\input{psfig}
\documentclass[referee]{aa}
\usepackage{epsfig}
\newcommand\eq[1]{Eq.~(\ref{#1})}

\newcommand\rfrac[2]{\left(\frac{#1}{#2}\right)}
\newcommand{\sub}[1]{_{\mbox{\scriptsize#1}}}

\newcommand\ee{\end{equation}}
\newcommand\be{\begin{equation}}
\newcommand\eea{\end{eqnarray}}
\newcommand\bea{\begin{eqnarray}}
\newcommand\eqa{\!\!\!&=&\!\!\!}

\newcommand\mone{^{-1}}
\newcommand\mtwo{^{-2}}
\newcommand\mthree{^{-3}}
\newcommand\mfour{^{-4}}
\newcommand\mhalf{^{-1/2}}

\newcommand\half{^{1/2}}
\newcommand\threehalf{^{3/2}}

\newcommand\msun{M_\odot}

\newcommand\del{{\mbox{\boldmath $\nabla$}}}
\newcommand\bfk{{\bf k}}

\newcommand\bfx{{\bf x}}
\newcommand\bfy{{\bf y}}

\newcommand\pa{\partial}
\newcommand\pdif[2]{\frac{\pa #1}{\pa #2}}

\newcommand\lsim{\mathrel{\rlap{\lower4pt\hbox{\hskip1pt$\sim$}}
    \raise1pt\hbox{$<$}}}
\newcommand\gsim{\mathrel{\rlap{\lower4pt\hbox{\hskip1pt$\sim$}}
    \raise1pt\hbox{$>$}}}
\newcommand{\dc}{$\delta_c$}
\newcommand{\dl}{$\delta_l$}

\newcommand{\res}{$\Lambda$}
\newcommand{\mres}{\Lambda}

\newcommand{\lam}{$\lambda$}
\newcommand{\luno}{$\lambda_1$}

\newcommand{\ltre}{$\lambda_3$}
\newcommand{\muno}{\lambda_1}
\newcommand{\mdue}{\lambda_2}
\newcommand{\mtre}{\lambda_3}

\renewcommand{\u}{${\bf u}$}

\newcommand{\q}{${\bf q}$}
\newcommand{\mq}{{\bf q}}

\newcommand{\mpdfl}{P_{\delta_l}(\delta_l;\Lambda)}

\newcommand{\mimfr}{\Omega(<\mres)}

\newcommand{\dmfr}{$n(\mres)$}

\newcommand{\mincir}{\raise -2.truept\hbox{\rlap
{\hbox{$\sim$}}\raise5.truept
\hbox{$<$}\ }}
\newcommand{\magcir}{\raise -2.truept\hbox{\rlap
{\hbox{$\sim$}}\raise5.truept
\hbox{$>$}\ }}
\newcommand{\minmag}{\raise-2.truept\hbox{\rlap
{\hbox{$<$}}\raise 6.truept\hbox
{$>$}\ }}
\newcommand{\ba}{\begin{eqnarray}}
\newcommand{\ea}{\end{eqnarray}}
\newcommand{\brr}{\begin{array}}
\newcommand{\err}{\end{array}}
\newcommand{\ec}{\end{center}}

\newcommand{\bk}{\mbox{\bf k}}

\newcommand{\vel}{\,{\rm km\,s^{-1}}}

\newcommand{\etal}{{et al.}~}

\newcommand{\p}{\partial}
\newcommand{\f}{\frac}

\newcommand{\de}{\delta}
\newcommand{\ded}{\delta_{_D}}

\newcommand{\Lam}{\Lambda}
\newcommand{\fde}{\tilde{\delta}}

\newcommand{\fW}{\widetilde{W}}

\newcommand{\calW}{{\cal W}}

\newcommand{\lan}{\langle}
\newcommand{\ran}{\rangle}

\newcommand{\0}{\circ}

\title{Dark matter, density perturbations and structure formation.}
\author{A. Del Popolo\inst{1,2,3}}
\offprints{A. Del Popolo - {\it E-mail adelpop@unibg.it}}
\institute{$^1$ Dipartimento di Matematica, Universit\`{a} Statale
di Bergamo,
via dei Caniana, 2 - I 24129 Bergamo, ITALY \\
$^2$ Feza G\"ursey Institute, P.O. Box 6 \c Cengelk\"oy, Istanbul, Turkey \\
$^3$ Bo$\breve{g}azi$\c{c}i University, Physics Department,
80815 Bebek, Istanbul, Turkey}

\begin{document}
%\maketitle
\def\cl{\centerline}
\def\bk{\hfill\break}
\def\no{\noindent}

%{\bf \huge A brief history of the cosmological mass function theory.\\}
{\bf \huge On the cosmological mass function theory.\\}

{\bf A. Del Popolo}\inst{1,2}\\
$^1$ Dipartimento di Matematica, Universit\`{a} Statale
di Bergamo,
via dei Caniana, 2 - I 24129 Bergamo, ITALY \\
%$^2$ Feza G\"ursey Institute, P.O. Box 6 \c Cengelk\"oy, Istanbul, Turkey \\
$^2$ Bo$\breve{g}azi$\c{c}i University, Physics Department,
80815 Bebek, Istanbul, Turkey

%%%%%%%%%%%%%%%%%%%%%%\vspace*{2cm}

%%%%%%%%%%%%%%%%%\baselineskip 0.8cm
\vspace{2.0cm}

\large {\bf Abstract} 

\bigskip
\bigskip

%\begin{flushleft}
%\begin{abstract}
%{\bf \large Abstract\\}
This paper provides, from one side, a review of the theory of the cosmological mass function from a theoretical point of view, starting from
the seminal paper of 
%\cite{pre} 
Press \& Shechter (1974) to the last developments (Del Popolo \& Gambera (1998, 1999), Sheth \& Tormen 1999 (ST), Sheth, Mo \& Tormen 2001 (ST1), Jenkins et al. 2001 (J01), Shet \& Tormen 2002 (ST2),  Del Popolo 2002a, Yagi et al. 2004 (YNY)), and from another side some improvements on the multiplicity function 
models in literature.
For what concerns this second aspect, I compare the numerical multiplicity function given in Yahagi, Nagashima \& Yoshii (2004),  with the theoretical multiplicity function obtained in the present paper by means of the excursion set model and an improved version of the barrier shape obtained in Del Popolo \& Gambera (1998), which implicitly takes account of total angular momentum acquired by the proto-structure during evolution and of a non-zero cosmological constant. 
I show that the multiplicity function obtained in the present paper, is in better agreement with Yahagi, Nagashima \& Yoshii (2004) simulations than other previous models (Sheth \& Tormen 1999; Sheth, Mo \& Tormen 2001; Sheth \& Tormen 2002; Jenkins et al. 2001). 
The multiplicity function of the present paper gives a good fit to simulations results as the fit function proposed by Yahagi, Nagashima \& Yoshii (2004), but differently from that it was obtained from a sound theoretical background.
Then, I calculate the mass function evolution in a $\Lambda$CDM model by means of the previous model. 
%by means of the excursion set model and an improved version of the barrier  
%shape obtained in Del Popolo \& Gambera (1998), 
I compare the result with Reed et al. (2003) (R03), who used a high resolution $\Lambda$CDM numerical simulation to calculate the mass function
of dark matter haloes down to the scale of dwarf galaxies, back to a redshift of
fifteen. I show that the mass function obtained in the present paper, gives similar predictions to the Sheth \& Tormen mass function but it does not show the
overprediction of extremely rare objects shown by the Sheth and Tormen mass function.
The results confirm previous findings that the simulated halo mass function can be described solely by the variance of the mass distribution, 
and thus has no explicit redshift dependence.
I moreover show that the PS-like approach together with the ellipsoidal model introduced in Del Popolo (2002b) gives a better description of the theoretical mass function.  

%\end{abstract}
%\end{flushleft}
\bigskip

{\it Subject headings:} clusters of galaxies --- cosmology: theory ---
dark matter --- galaxies: clustering --- large-scale structure of the
universe

\normalsize
\vspace{2cm}

%\begin{flushleft}
\section{\Large Introduction}
%\end{flushleft}

The universe that we observe appears quite clumpy and inhomogeneous on a spatial scale of $ \simeq 200 h^{-1}$ Mpc. Beyond that scale mass clumps appear to be homogeneously distributed. We observe massive clumps such as galaxies, groups of galaxies, clusters of galaxies, super-clusters (the most massive one among the hierarchy of structures) fill the space spanning a wide range of mass scales. The respective mass ranges characterizing these systems, approximately, are $\simeq 10^{9}-10^{10} M_{\odot}$, $\simeq 10^{11} M_{\odot}$, $\simeq 10^{12}-10^{14} M_{\odot}$, and $\simeq 10^{15} M_{\odot}$ where $M_{\odot}$ is the solar mass. All these objects, combinedly, form the structure what is known as the Large Scale Structure (LSS) in the universe. One of the most fundamental challenges in present universe is to understand the formation and evolution of the LSS. In order to understand the LSS, we need to have a theoretical framework within which predictions for structure formation can be made. 

The leading idea of all structure formation theories is that structures
was born from small perturbations in the otherwise uniform distribution
of matter in the early Universe, which is supposed to be, in great part,
dark (matter not detectable through light emission).
%One of the central problems in modern Cosmology is the study of the physics of the Dark Matter. 

With the term Dark Matter cosmologists indicate an hypothetic material component of the universe which does not emit directly electromagnetic radiation (unless it decays in particles having this property (Sciama 1990, but also see Bowyer et al. 1999)).\\
Dark matter, cannot be revealed directly, but nevertheless it is necessary to postulate its existence in order to explain the discrepancies between the observed dynamical properties of galaxies and clusters of galaxies  
and the theoretical predictions based upon models of these objects assuming that the only matter present is the visible one.
The original hypotheses on Dark Matter go back to measures performed by Oort (1932) of the surface density of matter in the galactic disk, which was obtained through the study of the stars motion in direction orthogonal to the galactic plane. The result obtained by Oort, which was after him named ``Oort Limit", gave a value of $ \rho = 0.15 M_{0} pc^{-3}$ for the mass density, and a mass, in the region studied, superior to that present in stars.
Nowadays, we know that the quoted discrepancy is due to the presence of HI in the solar neighborhood. Other studies (Zwicky 1933; Smith 1936) showed the existence of a noteworthy discrepancy between the virial mass of clusters (e.g. Coma Cluster) and the total mass contained in galaxies of the same clusters.
These and other researches from the thirties to now, have confirmed that a great part of the mass in the universe does not emit radiation that can be directly observed. \\

The study of Dark Matter has as its finality the explanation of formation of galaxies and in general of cosmic structures. For this reason, in the last decades, the origin of cosmic structures has been ``framed" in models in which Dark Matter constitutes the skeleton of cosmic structures and supply the most part of the mass of which the same is made.\\

The mass distribution of cosmic structures, i.e. the number of objects per unit
volume and unit mass interval, is commonly called mass function or
multiplicity function; it will be referred to as MF throughout the
text.  
The determination of the cosmological MF is a difficult, not fully
solved problem, both from the theoretical and the observational point
of view. An analytical exact prediction of the MF is hampered, even in
the simplest cosmological models, by the fact that highly non-linear
gravitational dynamics is involved in the formation of high density
objects; it is well known that the gravitational collapse problem has
never been exactly solved, except in the case of simple symmetries
(spherical planar). Large N-body simulations can be used to determine
the MF of simulated halos; however such simulations, besides being
time-expensive and limited in resolution, provide a numerical estimate
of the final solution of the problem without directly shedding light
on the difficult problem of gravitational collapse. Approximate
analytical arguments, while being of limited validity, can provide
useful and fully-understood solutions, which can then be compared to
the results of N-body simulations.

There is general consensus in setting the birth date of the MF theory
in 1974, when the seminal paper of Press \& Schechter (hereafter PS)
was published (the same PS formula can be found in
Doroshkevich 1967). That paper proposed a heuristic procedure, based
on linear theory, to obtain the distribution of the masses of
collapsed clumps.  That work inspired the fit for the galaxy
luminosity function proposed by Schechter (1976), but received a
limited attention for more than a decade. A real explosion of
attention to the MF theory started in 1988, when the first large
N-body simulations started to reveal a surprising adherence of their
results with the PS formula.  Many authors tried to extend the PS
procedure in many directions, or proposed different, alternative or
complementary procedures (Bond, Cole, Efstathiou \& Kaiser (1991) (BCEK)). These last years are witnessing a new wave
of interest, which has not yet been exhausted (see for example 
ST, Jenkins et al. 2001 (J01), ST1, ST2, Del Popolo 2002a, Yahagi et al. 2004 (YNY), Del Popolo 2005).

As mentioned before, the original PS work was developed within a model
in which structures grow from small seeds, either Poisson distributed
or set on a perturbed lattice. The extension of the PS work to more
general and ``standard'' cosmological settings was due to Efstathiou,
Fall \& Hogan (1979), who limited their analysis to power-law spectra
and an Einstein-de Sitter background.  The first application of PS
to a CDM spectrum was made by Schaeffer \& Silk (1985). They
``discovered' the PS MF (the procedure is described without any
reference to the PS paper), complete with its unjustified fudge factor
2, and criticized it in some interesting points; in particular, they
tried to model, on purely geometrical grounds, objects which do not
collapse spherically, like pancakes and filaments.

Starting from 1988, many authors extended the PS approach in many
directions, trying to understand why it appeared to work, in spite of
its heuristic and not fully satisfactory derivation.  The situation in
those years is reviewed in Lucchin (1989).  Just one year before,
Kashlinsky (1987) tried to determine the 2-point correlation function
of structures, collapsed according to the PS prescription.
Martinez-Gonzalez \& Sanz (1988a) performed similar calculations with
a different method, while Martinez-Gonzalez \& Sanz (1988b) attempted
to determine the luminosity function of galaxies of different
morphological types by means of an approach which was intermediate
between the PS and the peak one.  Lucchin \&
Matarrese (1988) formulated a PS MF for the case of non-Gaussian
perturbations.  Lilje (1992) and Lahav et al. (1991)
extended the PS result to open Universes and to flat Universes with a
cosmological constant.  Zhan (1990) changed the PS procedure by taking
into account the correction to the background density given by the
initial density contrast; as a matter of fact, such a correction is
negligible if the initial time is small.  Schaeffer \& Silk (1988a)
used the PS procedure to justify the presence of some small-scale
power in the HDM cosmology; in fact, the use of Gaussian smoothing
causes some large-scale power to be spread toward small scales.  Again
Schaeffer \& Silk (1988a,b), and Occhionero \& Scaramella (1988)
applied the PS formula to get many cosmological predictions on
collapsed structures. BCEK
solved the cloud-in-cloud problem using the 
``excursion set approach''.
Lacey \& Cole (1994), introduced the merging histories concept, which gives an important piece of information in the
formation history of dark matter objects. Del Popolo \& Gambera (1998), ST, ST1, ST2, Del Popolo (2002a), (2005), showed that 
the non-sphericity of collapse has important consequences on the MF.

Press and Schechter were the first to performed N-body simulations to
test the validity of their formula. They found some encouraging
agreement, but their simulations were limited to 1000 bodies, a very
small number to reach any firm conclusion.  Efstathiou, Fall \& Hogan
(1979) performed similar simulations, with the same number of point
masses, obtaining the same conclusions as PS.

Later, Efstathiou et al. (1988) compared the results of larger (32$^3$
P3M, scale-free power spectra) N-body simulations to the PS formula:
their dynamical range in mass was large enough to test the knee of the
MF.  The surprising result was that the PS formula nicely fitted their
abundances of simulated halos (as found by means of a percolation
friend-of-friend algorithm).  Further comparisons with N-body
simulations were performed by Efstathiou \& Rees (1988), Narayan \&
White (1988), Carlberg \& Couchman (1989), Carlberg (1990), BCEK, Brainerd \& Villumsen (1992), White, Efstathiou \& Frenk
(1993), Ma \& Bertschinger (1994), Jain \& Bertschinger (1994), Gelb
\& Bertschinger (1994), Katz, Quinn, Bertschinger \& Gelb (1994),
Lacey \& Cole (1994), Efstathiou (1995), Klypin \& Rhee (1994), Klypin
et al. (1995), Bond \& Myers (1996b), Governato et al. (1998), YNY.  Most authors reported the PS
formula to fit well their N-body results; nonetheless, all the authors
agree in stating the validity of the PS formula to be only
statistical, i.e.  the existence of the single halos is not well
predicted by the linear overdensity criterion of PS (see in particular
BCEK).

There are however some exceptions to this general agreement: Brainerd
{\&} Villumsen (1992) reported their MF, based on a CDM spectrum, to
be very similar to a power-law with slope $-2$, different from the PS
formula both at small and at large masses.  Jain \& Bertschinger
(1994), Gelb \& Bertschinger (1994) and Ma \& Bertschinger (1994)
noted that, to make the PS formula agree with their simulations (based
on CDM or CHDM spectra), it is necessary to lower the value of the
\dc\ parameter as redshift increases.  The same thing was found by
Klypin et al. (1995), but was interpreted as an artifact of their
clump-finding algorithm. More recent simulations seem to confirm this trend
(Governato et al. 1999; YNY)).

Lacey \& Cole (1994) extended the comparison to N-body simulations to
the predictions for merging histories of dark-matter halos; they found again a good agreement between theory and
simulations. This fact is noteworthy, as merging histories contain
much more detailed information about hierarchical collapse.

It is opportune to comment on two important technical points about
such comparisons.  First, the \dc\ parameters used by different
authors as ``best fit'' values range from the 1.33 of Efstathiou \&
Rees (1988) to the 1.9 found (in a special case) by Lacey \& Cole
(1994). The precise value of the \dc\ parameter is influenced by the
shape of the filter used to calculate the PS, Gaussian filters
requiring lower \dc\ values. Recent simulations tend to give
$\delta_c\simeq 1.5$ (e.g. Klypin et al. 1995) or $\delta_c=1.69$
(e.g. Lacey \& Cole 1994).  If \dc\ changes with time, a value 1.5
could be good at high redshifts, lowering to 1.7 at low redshifts.

Second, the numerical MF deeply depends on the way halos are picked up
from simulations. Typical algorithms, such as the friend-of-friend or
DENMAX, are parametric, i.e. they contain free parameters. For
instance, the frequently used friends-of-friends algorithm defines as
structures those clumps whose particles are separated among them by
distances smaller than a percolation parameter $b$ times the mean
interparticle distance. A heuristic argument, based on spherical
collapse, suggests a value of 0.2 for $b$ (with this percolation
parameter the mean density contrast of halos is about 180, which is
the expected density contrast of a virialized top-hat perturbation).
Obviously, the use of different $b$ parameters leads to different MFs.
In practice, what is obtained in this case is not ``the'' MF, but the
``friend-of-friend, $b$=0.2'' MF.  Then, the numerical MF contains
some hidden parameters, which, together with the \dc\ parameter (and
the mass associated to the filter in the Lacey \& Cole (1994) paper),
makes such comparisons more similar to parametric fits, rather than to
comparisons of a theory to a numerical experiment.\\

The present paper is organized as follows: section 2 contains some tools needed to formulate the MF theory:
it is focused on simple models for structure formation (density perturbation growth, 
spherical collapse model, Zel'dovich approximation, ellipsoid collapse model, etc.).

Section 3 contains a review of the theoretical MF problem and describes the theoretical works (such as the 
excursion set model)
which have tried to extend the validity of the original Press \&
Schechter work, or have proposed alternative procedures.

In the same section, it is reviewed the building of a MF theory, based on more 
realistic approximations for gravitational collapse than the spherical collapse. 
Finally, it is reviewed the work of Del Popolo \& Gambera (1998), ST, ST1, ST2, Del Popolo (2002a), (2005), who showed that 
the non-sphericity f collapse has important consequences on the MF. 
Section 4 is devoted to prospects and conclusions.

\section{\Large Theoretical bases of MF theory}

\subsection{ \bf Background cosmology}

The simplest cosmological model that describes, in a sufficient coherent manner, the evolution of the universe, from $ 10^{-2}
s$ after the initial singularity to now, is the so called {\it Standard Cosmological Model} (or Hot Big Bang model). It is based upon the Friedmann-Robertson-Walker (FRW) metric, which is given by:
\begin{equation}
ds^{2} = c^{2} d t^{2} -a(t)^{2}\left[\frac{d r^{2} }{1 - k r^{2}}+
r^{2} (d \theta^{2} +sin{\theta}^{2} d \phi^{2} )\right]
\end{equation}
where c is the light velocity, a(t) a function of time, or a scale factor called ``expansion parameter", t is the time coordinate, r, $\theta $ and  $ \phi $ the comoving space coordinates. The evolution of the universe is described by the parameter a(t) and it is fundamentally connected to the value $\rho$ of the average density.\\
The equations that describe the dynamics of the universe are the Friedmann's equations (Friedman, 1924) that we are going to introduce in a while. These equations can be obtained starting from the equations of the gravitational field of Einstein (Einstein, A., 1915): 
\begin{equation}
R_{ik}-\frac{1}{2} g_{ik} R =-\frac{8 \pi G}{ c^{4}} T_{ik}
\end{equation}
where now, $ R_{ik} $ is a symmetric tensor, also known as Ricci tensor, which describes the geometric properties of space-time, $ g_{ik} $ is the metric tensor, R is the scalar curvature, $T_{ik}$ is the energy-momentum tensor.\\
These equations connect the properties of space-time to the mass-energy. In other terms they describe how space-time is modeled by mass. Combining Einstein equations to the FRW metric leads to the dynamic equations for the expansion parameter, a(t). These last are the Friedmann equations:
\begin{equation}
d( \rho a^{3} )= -p d( a^{3} )
\end{equation}
\begin{equation}
\frac{1}{a^{2}} \dot{a}^{2} +\frac{k}{a^{2}} = \frac{8 \pi G }{3} \rho
\end{equation}
\begin{equation}
2\frac{\ddot{a}}{a} + \frac{\dot{a}^{2}}{a^{2}} +\frac{k}{a^{2}} =
-8 \pi G \rho
\end{equation}
where p is the pressure of the fluid of which the universe is constituted, k is the curvature parameter and a(t) is the scale factor connecting proper distances $ { \bf r }  $  to the comoving ones $ {\bf x} $ through the relation 
$ {\bf r}=a(t) {\bf x} $. Only two of the three Friedmann equations are independent, because the first connects density, $\rho$ to the expansion parameter a(t). The character of the solutions of these equations depends on the value of the curvature parameter, $k$, which is also determined by the initial conditions by means of Eq. 3. 
The solution to the equations now written shows that if $\rho$ is larger than $ \rho_{c} = \frac{3 H^{2} }{ 8 \pi G } = 1.88* 10^{-29} g/cm^{3} $ (critical density, which can be obtained from Friedmann equations putting $t=t_0$, $k=0$, and $ H= 100 km /s Mpc $), space-time  has a closed structure ($k=1$) and equations shows that the system go through a singularity in a finite time. This means that the universe has an expansion phase until it reaches a maximum expansion after which it recollapse. If $ \rho < \rho_{c} $, the expansion never stops and the universe is open $k=-1$ (the universe has a structure similar to that of an hyperboloid, in the two-dimensional case).
If finally, $ \rho = \rho_{c} $ the expansion is decelerated and has infinite duration in time, $k=0$, and the universe is flat (as a plane in the two-dimensional case). The concept discussed can be expressed using the parameter $ \Omega = \frac{\rho}{\rho_{c}} $. In this case, the condition $ \Omega = 1 $ corresponds to $k=0$, $ \Omega <1 $ corresponds to $k=-1$, and $ \Omega > 1 $ corresponds to $k=1$. Fig. 1 plots the evolution of the expansion parameter for $k=-1,0,1$.

In a flat model, the FRW equation becomes:

\be \frac{\dot{a}^2}{a^2}=\frac{8}{3}\pi G\bar{\rho}, \label{eq:frw_flat} \ee

\noindent whose solution is:

\be a(t)=(t/t_0)^{2/3}. \label{eq:a_ev_flat} \ee

In the case of open models with no cosmological constant: $\Omega<1$,
$\Lambda=0$, we can write:

\begin{equation} 
\frac{\dot{a}^2}{a^2}=\frac{8}{3}\pi G\bar{\rho}\left(1+
\left(\Omega_0^{-1} -1\right) a\right), \label{eq:frw_open} 
\end{equation}
\noindent 
and the $a(t)$ evolution can be expressed through the following
parametric representation:

\begin{eqnarray} 
a(\eta)&=&\frac{\Omega_0}{2(1-\Omega_0)}({\rm cosh} \eta -1) \\
     t(\eta)&=&\frac{\Omega_0}{2H_0(1-\Omega_0)^{3/2}}({\rm sinh}\eta -\eta).
\nonumber \label{eq:a_ev_open} 
\end{eqnarray} 
In the case of flat models with positive cosmological constant: $\Omega<1$,
$\Lambda\neq 0$, $\Omega+\Lambda/3H_0^2 = 1$, we can write:
\begin{equation} 
\frac{\dot{a}^2}{a^2}=\frac{8}{3}\pi G\bar{\rho}\left(1+
\left(\Omega_0^{-1} -1\right) a^3\right), \label{eq:frw_cosm} 
\end{equation} 
\begin{equation} 
a(t) = \left(\Omega_0^{-1}-1\right)^{-1/3}{\rm sinh}^{2/3}
\left(\frac{3}{2}\sqrt{\frac{\Lambda}{3}} t\right). \label{eq:a_ev_cosmo} 
\end{equation}

The value of $ \Omega $ can be calculated in several ways. The most common methods are the dynamical methods, in which the effects of gravity are used, and kinematics methods sensible to the evolution of the scale factor and to the space-time geometry.\\
% (BIBLIOGRAFIA). \\

\begin{figure}
%\centerline{\hbox{
%%%%%%%%%%%%%%%%%%%%%%%\psfig{file=fig.eps,width=1.2cm}
%}}
\caption[]{Evolution of the scale factor.}
\end{figure}

\subsection{\bf Perturbations evolution.}

Density perturbations in the components of the universe evolve with time. In order to get the evolution equations for $\delta$ in Newtonian regime, it is possible to use several models. In our model, we assume that gravitation dominates on the other interactions and that particles (representing galaxies, etc.) move collisionless in the potential $\phi$ of a smooth density function (Peebles 1980). 

The distribution function of particles for position and momentum is given by:
\begin{equation}
d N = f( {\bf x}, {\bf p},t) d^{3}x d^{3} p
\end{equation}
and density:
\begin{equation}
\rho({\bf x} , t )= m a^{-3} \int d^{3} p f({\bf x}, {\bf p}, t)=
\rho_{b}\left[1+ \delta({\bf x}, t)\right] \label{eq:densi}
\end{equation}
where m is the mass of a particle and $\rho_{b}$ the background density. Applying Liouville theorem to the probability density on a limited region of phase-space of the system we have that f verifies the equation:
\begin{equation}
\frac{\partial f}{\partial t }+\frac{{\bf p}}{m a^{2}}
\bigtriangledown f - m \bigtriangledown \phi \frac{
\partial f}{\partial {\bf p}} = 0 \label{eq:liou}
\end{equation}
The distribution function f that appears in the previous equations cannot be obtained from observations. It is possible to measure moments of f (density, average velocity, etc.). We want now to obtain the evolution equations for $\delta$. For this reason, we start integrating Eq. (\ref{eq:liou}) on $ {\bf p} $ and after using Eq. (\ref{eq:densi}), we get:
\begin{equation}
a^{3} \rho_{b} \frac{\partial \delta}{\partial t}+
\frac{1}{a^{2}} \bigtriangledown \int {\bf p} f d^{3} p =0 \label{eq:tre}
\end{equation}
If we define velocity as: 
\begin{equation}
{\bf v} = \frac{\int \frac{{\bf p}}{m a}f d^{3} p}{\int f d^{3} p}
\end{equation}
and introduce it in Eq. (\ref{eq:tre}) we get:
\begin{equation}
\rho_{b} \frac{\partial \delta}{\partial t}+
\frac{1}{a} \bigtriangledown (\rho {\bf v} )= 0
\end{equation}
We can now multiply Eq. (\ref{eq:liou}) for $ {\bf p} $ and integrate it on the momentum:
\begin{equation}
\frac{\partial}{\partial t} \int p_{\alpha} f d^{3} p +
\frac{1}{m a^{2}} \partial_{\beta} \int p_{\alpha}
p_{\beta} f d^{3} p + a^{3} \rho ( {\bf x}, t) \phi_{, \alpha} =0
\end{equation}
this last in Eq. (\ref{eq:tre}) leaves us with:
\begin{equation}
\frac{\partial^{2} \delta }{\partial t^{2}} + 2\frac{\dot{a}}{a}
\frac{\partial \delta}{\partial t} = \frac{1}{a^{2}}
\bigtriangledown \left[(1+\delta ) \bigtriangledown \phi\right]+
\frac{1}{\rho_{b} m a^{7}} \partial_{\alpha} \partial_{\beta}
\int p_{\alpha} p_{\beta} \phi d^{3} p
\end{equation}
and finally using 
\begin{equation}
< v_{\alpha} v_{\beta} > = \frac{\int f p_{\alpha} p_{\beta} d^{3} p }
{m a^{2} \int f d^{3} p}
\end{equation}
the equation for the evolution of overdensity becomes:
\begin{equation}
\frac{\partial^{2} \delta }{\partial t^{2}} + 2\frac{\dot{a}}{a}
\frac{\partial \delta}{\partial t} = \frac{1}{a^{2}}
\bigtriangledown \left[(1+\delta ) \bigtriangledown \phi\right]+
\frac{1}{a^{2}} \partial_{\alpha} \partial_{\beta}
\left[(1+\delta) < v^{\alpha} v^{\beta} >\right]
\end{equation}
(Peebles 1980). The term $ < v_{\alpha} v_{\beta} > $ is the tensor of anisotropy of peculiar velocity. This is present in the gradient and then it behaves as a pressure force. If we consider an isolated and spherical perturbation, it is possible to assume that initial asymmetries does not grow up and so we can suppose, in this hypothesis that $ < v_{\alpha} v_{\beta} > = 0 $. 

When the density contrast is much smaller than one, $\delta\ll 1$, and
peculiar velocities, $v$, are small enough to satisfy $(vt/d)^2\ll \delta$,
where $t$ is the cosmological time and $d$ is the coherence length of
the matter field, one can obtain a linear theory for perturbations evolution as follows. 
%, the system of equations (\ref{eq:poisson_old} --
%\ref{eq:continuity_old}) can be linearized, leading to the equation:

In this case 
%and with the linearity assumption $ \delta << 1 $,
we have:
\begin{equation}
\frac{\partial^{2} \delta}{\partial t^{2}} + 2\frac{\dot{a}}{a}
\frac{\partial \delta}{\partial t } = 4 \pi G \rho_{b} \delta
\end{equation}
This equation in an Einstein-de Sitter universe ($\Omega = 1$,
$\Lambda = 0 $) has the solutions:
\begin{equation}
\delta_{+}= A_{+}({\bf x}) t^{\frac{2}{3}} \hspace{1cm}
\delta_{-} ({\bf x},t) = A_{-} ({\bf x}) t^{-1}
\end{equation}
The perturbation is then done of two parts: a growing one (which shall be denoted with $b(t)$), becoming more and more important with time, and a decaying one
becoming negligible with increasing time, with respect to the growing one.

The solutions for the growing modes, relative to the three background models, 
are reported:

$\Omega=1$:

\be b(t)=a(t). \label{eq:flat_growingmode} \ee

$\Omega< 1$, $\Lambda=0$: it is useful to use the time variable

\be \tau = (1-\Omega(t))^{-1/2} = \sqrt{(a(\Omega_0^{-1}-1))^{-1}+1}.
\label{eq:tau_def} \ee

\noindent Then:

\be b(\tau) = \frac{5}{2(\Omega_0^{-1}-1)}\left( 1+3(\tau^2-1) \left(
1+\frac{\tau}{2} \ln \left( \frac{\tau-1}{\tau+1} \right)\right)\right).
\label{eq:open_growingmode} \ee

\noindent 
Note that this $b(t)$ function saturates to the value
$5/2(\Omega_0^{-1}-1)$ at large times.

$\Omega<1$, $\Lambda\neq 0$, flat: it is useful to use the time
variable

\be h = {\rm coth}(3t\sqrt{\Lambda/3}/2). \label{eq:htime_def} \ee

\noindent Then,

\be b(\tau) = h \int_h^\infty (x^2(x^2-1)^{1/3})^{-1}dx . 
\label{eq:cosm_growingmode} \ee

\noindent
Growing modes are normalized so as to give $b(t)\simeq a(t)$ at early
times, and $a(t_0)=1$.

In the MF theory, collapse time estimates are often based on an
extrapolations of the linear regime to density contrasts of order one.
It is then convenient to define the quantity:

\be \delta_l\equiv \delta(t_i)/b(t_i). \label{eq:deltal_def} \ee

\noindent 
This is the initial density contrast linearly extrapolated to the
time at which $b(t)=1$, which, in an Einstein-de Sitter background, is
the present time; it will be used in the next sections.

Before concluding this section, we want to find an expression for the velocity field in the linear regime. 
Using the equation of motion $ {\bf p} = m a^{2} {\bf \dot{x}} $, $ \frac{d {\bf p}}{d t} = -m \bigtriangledown \phi $ and the proper velocity of a particle, $ v = a \bf{\dot{x}} $, verify the equation:
\begin{equation}
\frac{d {\bf v}}{d t} + {\bf v} \frac{\dot{a}}{a} =
-\frac{\bigtriangledown \phi}{a} = G \rho_{b} a
\int d^{3} x \delta({\bf x}, t) \frac{{\bf x} -{\bf x'}}{\left|{\bf x}-
{\bf x'}\right|}
\end{equation}
Supposing that ${\bf v} $ is a similar solution for the density, $ {\bf v} = {\bf V_{+}({\bf x}, t)} t^{p}$, we get:
\begin{equation}
v^{\alpha} = \frac{H a}{4 \pi} \frac{\partial}{\partial x^{\alpha}}
\int d^{3} x' \frac{\delta ( {\bf x'}, t)}{\left|{\bf x'}-
{\bf x}\right|}
\end{equation}
(Peebles 1980). 
This solution is valid just as that for $ \delta $ in the linear regime. At time $ t = t_{0} $ this regime is valid on scales larger than $ 8 h^{-1} Mpc $.\\ 

\subsection{\bf The spectrum of density perturbation. }

In order to study the distribution of matter density in the universe it is generally assumed that this distribution is given by the superposition of plane waves independently evolving, at least until they are in the linear regime (this means till the overdensity $ \delta = \frac{\rho -\overline{\rho} }{\overline{\rho}}<<1 $). 
Let we divide universe in cells of volume $V_u$ and let we impose periodic conditions on the surfaces. If we indicate with 
$ \overline{\rho} $ the average density in the volume and with $ \rho({\bf r})$ the density in $ {\bf r} $, it is possible to define the density contrast as: 
\begin{equation}
\delta( {\bf r} ) = \frac{\rho( {\bf r}) - \overline{\rho}}{\overline{\rho}}
\end{equation}
This quantity can be developed in Fourier series:
\begin{equation}
\delta({\bf r}) = \sum_{{\bf k}} \delta_{{\bf k}}
exp(i{\bf k} {\bf r}) = \sum_{{\bf k}} \delta_{{\bf k}}
exp(-i{\bf k} {\bf r} ) \label{eq:sovra}
\end{equation}
( Kolb e Turner 1990), where $ k_{x} = \frac{2 \pi n_{x}}{l} $ (and similar conditions for the other components) and for the periodicity condition $ \delta(x,y,L) = \delta(x,y,0)$ (and similar conditions for the other components). 
Fourier coefficients $ \delta_{{\bf k}}$ are complex quantities given by: 
\begin{equation}
\delta_{{\bf k}} = \frac{1}{V_{u}} \int_{V_{u}} \delta({\bf r} )
exp(-i{\bf k}{\bf r}) d {\bf r}
\end{equation}
For mass conservation in $V_u$ we have also $ \delta_{{\bf k}=0}=0 $ while for reality of  $ \delta({\bf r}) $,  $ \delta_{{\bf k}}^{\ast} = \delta_{-{\bf k}} $.
If we consider n volumes, $V_u$, we have the problem of determining the distribution of Fourier coefficients $ \delta_{{\bf k}} $ and that of  $ \left|\delta\right| $.
We know that the coefficients are complex quantities and then $ \delta_{{\bf k}} = \left|\delta_{{\bf k}}
\right| exp (i \theta_{{\bf k}})$. If we suppose that phases are random, in the limit $ V_{u} \rightarrow \infty $ 
it is possible to show that  
we get 
$ \left|\delta\right|^{2} =
\sum_{{\bf k}} \left|\delta_{{\bf k}}\right|^{2} $. The Central limit theorem leads us to conclude that the distribution for 
$ \delta $ is Gaussian: 
\begin{equation}
P ( \delta ) \propto exp(\frac{-\delta^{2}}{\sigma^{2}} ) \label{eq:gau}
\end{equation}
(Efstathiou 1990).
The quantity $\sigma$ that is present in Eq. (\ref{eq:gau}) is the variance of the density field and is defined as: 
\begin{equation}
\sigma^{2} = < \delta ^{2} > = \sum_{{\bf k}} <
\left|\delta_{{\bf k}}\right|^{2} > = \frac{1}{V_{u}}
\sum_{{\bf k}}  \delta_{k}^{2}
\end{equation}
This quantity characterizes the amplitude of the inhomogeneity of the density field. If $ V_{u} \rightarrow \infty$, we obtain the more usual relation:
\begin{equation}
\sigma^{2} = \frac{1}{\left(2 \pi\right)^3} \int P(k) d^{3} k =
\frac{1}{2 \pi^{2}} \int P(k) k^{2} d k
\end{equation}
The term $ P( k ) = < \left|\delta\right|^{2}> $ is called ``Spectrum of perturbations". It is function only of k because the ensemble average in an isotropic universe depends only on r. A choice often made for the primordial spectrum is $ P( k) = A k^{n} $ which in the case $ n = 1 $ gives the scale invariant spectrum of Harrison-Zeldovich. An important quantity connected with the spectrum is the two-points correlation function $ \xi({\bf r}, t) $. It can be defined as the joint probability of finding an overdensity $\delta$ in two distinct points of space:
\begin{equation}
\xi({\bf r}, t) = < \delta( {\bf r},t )
\delta({\bf r}+{\bf x}, t) > \label{eq:corr}
\end{equation}
(Peebles 1980), where averages are averages on an ensemble obtained from several realizations of universe. Correlation function can be expressed as the joint probability of finding a galaxy in a volume $ \delta V_{1} $ and another in a volume 
$ \delta V_{2} $ separated by a distance $ r_{12} $:
\begin{equation}
\delta ^{2} P = n_{V}^{2} [1+ \xi(r_{12} )] \delta V_{1} \delta V_{2}
\end{equation}
where $ n_{V} $ is the average number of galaxies per unit volume. The concept of correlation function, given in this terms, can be enlarged to the case of three or more points.\\
Correlation functions have a fundamental role in the study of clustering of matter. If we want to use this function for a complete description of clustering, one needs to know the correlation functions of order larger than two (Fry 1982). By means of correlation functions it is possible to study the evolution of clustering. The correlation functions are, in fact, connected one another by means of an infinite system of equations obtained from moments of Boltzmann equation which constitutes the BBGKY (Bogolyubov-Green-Kirkwood-Yvon) hierarchy (Davis e Peebles 1978). 
This hierarchy can be transformed into a closed system of equation using closure conditions. Solving the system one gets information on correlation functions.\\
In order to show the relation between perturbation spectrum and two-points correlation function, we introduce in 
Eq. (\ref{eq:corr}), Eq. (\ref{eq:sovra}), recalling that $ \delta_{{\bf k}}^{\ast} = \delta( -{\bf k} )$ and taking the limit $ V_{u} \rightarrow \infty$, the average in the Eq. (\ref{eq:corr}) can be expressed in terms of the integral: 
\begin{equation}
\xi ( {\bf r} ) = \frac{1}{( 2 \pi ) ^{3}}
\int |\delta( {\bf k} )|^{2} exp( -i {\bf k} {\bf r} ) d^{3} k
\end{equation}
This result shows that the two-point correlation function is the Fourier transform of the spectrum. In an isotropic universe, it is $ |{\bf r}| = r $ and then $ | {\bf k} | =k $ and the spectrum can be obtained from an integral on $ |{\bf k}| = k $. Then correlation function may be written as:
\begin{equation}
\xi( r ) = \frac{1}{2 \pi^{2} }
\int k^{ 2} P (k ) \frac{sin( k r )}{k r} d k
\end{equation}
During the evolution of the universe and after perturbations enter the horizon, the spectrum is subject to modulations because of physical processes characteristic of the model itself (Silk damping (Silk 1968) for acollisional components, free streaming for collisional particles, etc.). These effects are taken into account by means of the transfer function
$ T(k;t) $ which connects the primordial spectrum $ P( k; t_{p} )$ at time $t_p$ to the final time $t_f$:
\begin{equation}
P(k;t_{f}) = \left[\frac{b(t_{f})}{b(t_{p})}\right]^{2}
T^{2}(k;t_{f}) P( k;t_{p})
\end{equation}
where b(t) is the law of grow of perturbations, in the linear regime. In the case of CDM models the transfer function is:
\begin{equation}
 T( k ) = \left\{1+\left[ak+(bk)^{1.5}+(ck)^{2}
\right]^{\nu}\right\}^{\frac{-1}{\nu}}
\end{equation}
(Bond e Efstathiou 1984), where $a=6.4 (\Omega h^{2} )^{-1}Mpc $; $b=3.0 (\Omega h^{2} )^{-1} Mpc$;
$ c=1.7 (\Omega h^{2} )^{-1} Mpc $; $ \nu=1.13$.
It is interesting to note that Eq. (\ref{eq:gau}) is valid only if $ \sigma << 1 $, since $ \left|\delta\right| \leq 1 $. This implies than non-linear perturbations, $ \sigma >> 1 $, must be non-Gaussian.
In fact when the amplitudes of fluctuations grow up, at a certain point modes are no longer independent and start to couple giving rise to non-linear effects that change the spectrum and correlation function (Juskiweicz et al 1984 ).
There are also some theories (e.g., cosmic strings (Kibble \& Turok 1986)) in which even in the linear regime perturbations are not Gaussian.

\subsection{\bf Spherical Collapse}

Linear evolution is valid only if $ \delta << 1 $ or similarly, if the mass variance, $\sigma$, is much less than    
unity. When this condition is no longer verified (e.g., if we consider scales smaller than $8h^{-1}$ Mpc), it is necessary to develope a non-linear theory. In regions smaller than $8h^{-1}$ Mpc
%of this size 
galaxies are not a Poisson distribution but they tend to cluster.   
If one wants to study the properties of galactic structures or clusters of galaxies, it is necessary to introduce a non-linear theory of clustering. A theory of this last item is too complicated to be developed in a purely theoretical fashion. The problem can be faced, by using N-Body simulations of the interesting system, assuming certain approximations that simplifies it, Spherical Collapse Model, 
Zel'dovich approximation (Zel'dovich 1970).  
%The approximations are often used to furnish the initial data to simulations. 
%In the simulations, a large number of particles are randomly distributed in a sphere, in the points of a cubic grid, in order to eliminate small scale noise. The initial 
%spectrum is obtained perturbing the initial positions by means of a superposition of plane waves having random distributed phases and wave vector (West et al. 1987).
%Obviously, the universe is considered in expansion (or comoving coordinates are used), and then the equation of motion of particles are numerically solved. For what concerns 
%the analytical approximations one of the most used is that of Ze`ldovich (1970). 
This last gives a solution to the problem of the grow of perturbations in an universe with $p=0$ not only in the linear regime but even in the mildly non-linear regime.

Spherical symmetry is one of the few cases in which gravitational
collapse can be solved exactly (Gunn \& Gott 1972; Peebles 1980).  In
fact, as a consequence of Birkhoff's theorem, a spherical perturbation
evolves as a FRW Universe with density equal to the mean density
inside the perturbation.

The simplest spherical perturbation is the top-hat one, i.e. a
constant overdensity $\delta$ inside a sphere of radius $R$; to avoid
a feedback reaction on the background model, the overdensity has to be
surrounded by a spherical underdense shell, such to make the total
perturbation vanish. The evolution of the radius of the perturbation
is then given by a Friedmann equation.

The evolution of a spherical perturbation depends only on its initial
overdensity. In an Einstein-de Sitter background, any spherical 
overdensity
reaches a singularity (collapse) at a final time:

\begin{equation} t_c=\frac{3\pi}{2}\left(\frac{5}{3}\delta(t_i)\right)^{-3/2} t_i.
\label{eq:spherical_coll} \end{equation}

\noindent
By that time its linear density contrast reaches the value: 

\begin{equation} \delta_l(t_c)=\delta_c=\frac{3}{5}\left(\frac{3\pi}{2}\right)^{3/2}
\simeq 1.69. \label{eq:delta_c_sph}\end{equation}

\noindent
In an open Universe not any overdensity is going to collapse: the
initial density contrast has to be such that the total density inside
the perturbation overcomes the critical density. 
%This can be
%quantified (not exactly but very accurately) as follows: the growing
%mode saturates at $b(t)=5/2(\Omega_0^{-1}-1)$, 
So, a perturbation
ought to satisfy $\delta_l>1.69\cdot 2(\Omega_0^{-1}-1)/5$ to be able
collapse.

Of course, collapse to a singularity is not what really happens in
reality. It is typically supposed that the structure reaches virial
equilibrium at that time. In this case, arguments based on the virial
theorem and on energy conservation show that the structure reaches a
radius equal to half its maximum expansion radius, and a density
contrast of about 178. In the subsequent evolution the radius and the
physical density of the virialized structure remains constant, and its
density contrast grows with time, as the background density decays.
Similarly, structures which collapse before are denser than the ones
which collapse later.

Spherical collapse is not a realistic description of the formation of
real structures; however, it has been shown (see Bernardeau 1994 for
a rigorous proof or Valageas 2002a,b) that high peaks ($> 2\sigma$) follow spherical
collapse, at least in the first phases of their evolution. However, 
a small systematic departure from
spherical collapse can change the statistical properties of collapse
times.

Spherical collapse can describe the evolution of underdensities. A
spherical underdensity is not able to collapse (unless the Universe is
closed!), but behaves as an open Universe, always expanding unless its
borders collide with neighboring regions. At variance with
overdensities, underdensities tend to be more spherical as they
evolve, so that the spherical model provides a very good approximation
for their evolution.

%\begin{flushleft}
\subsection{\bf Zel'dovich approximation.}
%\end{flushleft}
%\vspace*{5.0mm}

Most MF theories proposed in the
literature are based at best on spherical collapse,
which neglects tides which are the relevant dynamical interaction. 
Spherical top-hat collapse is a truly local dynamical approximation:
the fate of a spherical perturbation is determined just by its initial
overdensity. In other words, the dynamical role of the whole Universe
outside the perturbation (according to Birkhoff's theorem) is assumed to
be negligible.  It is possible to construct
a mixed Eulerian-Lagrangian system from the evolution equations of
fluid elements by
decomposing the tensor of (Eulerian) space derivatives of the peculiar
velocity \u\ into an expansion scalar $\theta$, a shear tensor
$\sigma_{ab}$ and a vorticity tensor $\omega_{ab}$. 
In this way, the following evolution
equation for the density contrast can be obtained (see, e.g., Ellis
1971; here the growing mode $b(t)$ is used as time variable):

\be \frac{d^2\delta}{db^2}+4\pi G\bar{\rho}\frac{b}{\dot{b}^2}
\frac{d\delta}{db}  = \frac{4}{3} \left(\frac{d\delta}{db}\right)^2 
\frac{1}{(1+\delta)} + (1+\delta) \left(4\pi G\bar{\rho}
\frac{b}{\dot{b}^2} \delta + 2\sigma^2-2\omega^2\right) 
\label{eq:delta_evol} \ee

\noindent 
Here $\sigma^2=\sigma_{ab}\sigma_{ab}/2$ and $\omega^2=\omega_{ab}
\omega_{ab}/2$ (note that $\sigma^2$ in this context is not the
mass variance).    According to linear theory, any vortical
mode is severely damped in the early gravitational evolution, and this
remains true during the mildly non-linear regime, up to "orbit crossing OC \footnote{In the dynamical evolution of cold matter, 
it can happen that two mass elements get to the same point. This event is called "orbit crossing"} (at OC
vorticity couples with the growing mode; see Buchert 1992). Then it is
reasonable to assume vanishing vorticity in the present framework. On
the other hand, the shear does not vanish in general: it provides the
link between the mass element and the rest of the Universe.

The evolution equation for the shear reads as follows:

\be \frac{{d\sigma}_{ab}}{db} + \frac{2}{3}\vartheta\sigma_{ab} + 
\sigma_{ac}\sigma_{cb} + 4\pi G\bar{\rho}\frac{b}{\dot{b}^2}\sigma_{ab} - 
\frac{2}{3}\sigma^2 \delta_{ab} = - 4\pi G\bar{\rho}\frac{b}{\dot{b}^2} 
E_{ab}. \label{eq:shear_evol}  \ee

\noindent 
The tensor $E_{ab}$, 
represents the tidal interactions between the mass element and the rest
of the Universe. Then, tides are the relevant dynamical interaction
neglected by spherical collapse.

It is useful to find which is the simplest way to introduce tides in
the evolution of a mass element. The simplest, realistic approximation of gravitational evolution in
the (mildly) non-linear regime is the Zel'dovich approximation, that I am going to summarize.

In this approximation, one supposes to have particles with initial position given in Lagrangian coordinates $ {\bf q} $.
The positions of particles, at a given time t, the
Lagrangian-to-Eulerian mapping is written as follows:
%are given by: 
\begin{equation}
{\bf x}={\bf q}+b(t){\bf p(q)}
\end{equation}
where ${\bf x}$ indicates the Eulerian coordinates, ${\bf p(q) } $ describes the initial density fluctuations and $b(t)$ describes their grow in the linear phase and it satisfies the equation:
\begin{equation}
\frac{d^{2}b}{dt^{2}}+2a^{-1}\frac{db}{dt}\frac{da}{dt}=4 \pi G
\rho b
\end{equation}
The equation of motion of particles, according to the quoted approximation, is given by: 
\begin{equation}
{\bf v} = \dot{a} {\bf q} + \dot{b} {\bf p}({\bf q} )
\end{equation}
The peculiar velocity of particles is given by:
\begin{equation}
{ \bf u}=\frac{d{\bf x}}{dt} =\frac{db}{dt}{ \bf p(q)}
\end{equation}
while the density of the perturbed system is given by:
\begin{equation}
\rho({\bf q},t)=\overline{\rho}
\left|\frac{\partial q_{j}}{\partial x_{k}}\right| =
\overline {\rho} \left| \delta_{jk} + b(t)
\frac{\partial p_{k} }{\partial q_{j} }\right|^{-1} \label{eq:cas}
\end{equation}

Developing the Jacobian present in Eq. (\ref{eq:cas}) at first order in $ b(t) {\bf p(q)} $, one obtains:
\begin{equation}
\frac{\delta \rho}{\overline{\rho}}\approx -b(t)
\bigtriangledown_{{ \bf q}}{\bf p(q)}
\end{equation}
This equation can be re-written, separating the space and time dependence, as in the equation for $ { \bf u} $, and writing:
\begin{equation}
b(t)=t^{\frac{2}{3}} \hspace{1.0cm} {\bf p(q)}=
\sum_{{\bf k}}i\frac{{\bf k}}{\left|{\bf k}\right|^{2}}A_{{\bf k}}
exp(i{\bf k q})
\end{equation}
in the form: 
\begin{equation}
\frac{\delta \rho}{\overline{\rho}}=
\sum_{{ \bf k}} A_{{\bf k}}t^\frac{2}{3} exp(i{\bf k q})
\end{equation}
(Efstathiou 1990), that leads us back to the linear theory. 
In other words, Ze`ldovich approximation is able to reproduce the linear theory, and is also able to give a good approximation in regions with $ \frac{\delta \rho}{\overline{\rho}}>>1$. Using the expression for $ p(q) $, the Jacobian in Eq. (\ref{eq:cas}) is a real matrix and symmetric that can be diagonalized. 
With this $ p(q) $ the perturbed density can be written as:
\begin{equation}
\rho({\bf q},t)=\frac{\overline{\rho}}{(1-b(t)\lambda_{1}(q))
(1-b(t)\lambda_{2}(q))(1-b(t)\lambda_{3}(q))} \label{eq:panc}
\end{equation}
where $ \lambda_{1} $, $ \lambda_{2} $, $ \lambda_{3} $ are the three eigenvalues of the Jacobian, describing the expansion and contraction of mass along the principal axes. From the structure of the last equation, we notice that in regions of high density Eq. (\ref{eq:panc}) becomes infinite and the structure of collapse in a pancake, in a filamentary structure or in a node, according to values of eigenvalues. Some N-body simulations (Efstathiou e Silk, 1983) tried to verify the prediction of 
Ze`ldovich approximation, using initial conditions generated using a spectrum with a cut-off at low frequencies. 
The results showed a good agreement between theory and simulations, for the initial phases of the evolution 
($ a(t)=3.6 $). Going on, the approximation is no more valid starting from the time of shell-crossing. After shell-crossing, particles does not oscillate any longer around the structure but they pass through it making it vanish. 
This problem has been partly solved supposing that particles, before reaching the singularity they sticks the one on the other, due to a dissipative term that simulates gravity and then collects on the forming structure.  
This model is known as ``adesion-model" (Gurbatov et al. 1985). \\
Summarizing, Zel'dovich approximation gives a description of the transition between linear and non-linear phase. It is expecially used to get the initial conditions for N-body simulations. The main problem with Zel'dovich and perturbative Lagrangian
approximations is that they break down after OC.  Many authors have
then tried to develop approximations which avoid OC or make particles
oscillate around pancakes (see Sahni
\& Coles 1996 for a complete review).
It is interesting to note that
the Zel'dovich approximation is
the first term of a perturbative series; the perturbed quantity is not
the density, as in Eulerian perturbation theory, but the displacement
of the particles from the initial position.
%AGGIUNGERE CHE E' UNA APPROSSIMAZIONE LINEARE NELLO SPOSTAMENTO.\\

\subsubsection{Collapse time in Zel'dovich approximation}

In the Ze'ldovich approximation, the density contrast $\delta$, the expansion $\theta$
and the shear $\sigma_{ab}$ evolve as follows:

\bea J(\mq,t) & = & (1-b(t)\lambda_1)(1-b(t)\lambda_2)(1-b(t)\lambda_3)
\label{eq:varie_zel_evol}\\
\theta(\mq,t) & = & - \frac{\lambda_1}{1-b(t)\lambda_1} - 
\frac{\lambda_2}{1-b(t)\lambda_2}-\frac{\lambda_3}{1-b(t)\lambda_3}\nonumber\\
\sigma_{ab}(\mq,t)  & = & {\rm diag}\left(\frac{\lambda_1}{1-b(t)\lambda_1}-
\frac{\theta}{3}, \frac{\lambda_2}{1-b(t)\lambda_2}-\frac{\theta}{3},
\frac{\lambda_3}{1-b(t)\lambda_3}-\frac{\theta}{3}\right)\nonumber\\
\delta(\mq,t)  & = & ((1-b(t)\lambda_1)(1-b(t)\lambda_2)
(1-b(t)\lambda_3))^{-1} -1.
\nonumber \eea

When $b(t)=1/\lambda_1$, caustic formation takes place: the Jacobian
determinant vanishes, and all the other quantities go to infinity.  It
is then quite reasonable, from the point of view of the mass element,
to define such instant as the collapse time. Hereafter collapse will always be defined as the OC event.
Note also that, after OC, the Zel'dovich evolution along the second and third
axes is not really meaningful, as Zel'dovich does not work after OC.
It is then possible to give collapse time estimates for any mass
element.  Initial conditions are ``locally'' given by the three
$\lambda$ eigenvalues, but the evolution is not physically local, as
initial conditions contain non-local information about tides

It is useful to define the following variables (M98):

\bea x & = & \lambda_1 - \lambda_2 \label{eq:x_and_y} \\
     y & = & \lambda_2 - \lambda_3, \nonumber \eea

\noindent 
and to use the growing mode $b(t)$ as time variable.  Moreover, it is
possible to consider regions with linear initial density contrast
$\delta_l=1$ or $-1$, as all the other cases can be obtained by a
simple rescaling of $b$. The Zel'dovich collapse time is finally given by:

\be b_c^{Zel'dovich} = \frac{3}{\delta_l+2x+y} \label{eq:zel_bc} \ee

Fig. 2 shows the collapse time curves $b_c(x,y)$ for $\delta_l=1$
and $-1$. A problem is soon apparent: in the spherical case, when
$x=y=0$, the collapse time is $3/\delta_l$, instead of the exact
$1.69/\delta_l$ value. This discrepancy is due to the fact that the Zel'dovich approximation 
is an exact solution (before OC of course) in
one dimension (see, e.g., Shandarin \& Zel'dovich 1989), and is then
able to describe the collapse of pancake-like structures, while it
severely underestimates the collapse rate in spherical symmetry.

A way to overcome this problem is to try some simple {\it ansatze} for
the ``true'' shape of the collapse time curve. In practice, a truly
realistic collapse time curve will depend not only on the $\lambda$
eigenvalues, but also on the values of the density (or potential)
field in all the points of (Lagrangian) space. 

A first way to change the Zel'dovich prediction is to
force it not to assume values larger than the spherical one (M98):

\be b_c^{an1}=\min \{ b_c^{Zel'dovich}, 1.69/\delta_l\}. \label{eq:ansatza} \ee

\noindent 
This $b_c$ curve is shown in Fig. 3a for $\delta_l=1$; it has a
plateau, of height 1.69, at small $x$ and $y$ values.  On the other
hand, it is unlikely that all quasi-spherical collapses have exactly
the same $b_c$ value; a systematic trend of lower $b_c$ values at
nonvanishing $x$ and $y$ values is more realistic.  Such a $b_c$ curve
can be modeled as the intersection of the Zel'dovich prediction with a
slightly inclined plane which reaches 1.69 at $x=y=0$ (M98):

\be b_c^{an2}= \min \{ b_c^{Zel'dovich}, 1.69/\delta_l-\epsilon(2x+y) \}. 
\label{eq:ansatzb} \ee

\noindent
This curve, with $\epsilon=0.2$, is shown in Fig. 3b.  Monaco (1995)
contains a more complete discussion of such {\it ansatze}.  A further
possibility, examined in Monaco (1995), is to insert ``by hand'' in
Eq.  (\ref{eq:delta_evol}) the shear as given by Zel'dovich (Eq.
\ref{eq:varie_zel_evol}), and then solve the equation numerically; the
resulting collapse time is very similar to that of Fig. 3a, but the
transition from spherical to Zel'dovich regime is smooth.

An interesting conclusion, is
that the reasonable systematic trend shown in Fig. 3b does influence
the large-mass part of the mass function, moving it toward large
masses, even though spherical collapse is recovered for very large
overdensities (which are characterized by small $x$ and $y$
values). In other words, the fact that large, rare fluctuations
asymptotically follow spherical collapse does not guarantee that the
``spherical'' PS MF (with $\delta_c=1.69$) is recovered at large masses.

\begin{figure}
\begin{center}\hbox{
\epsfig{file=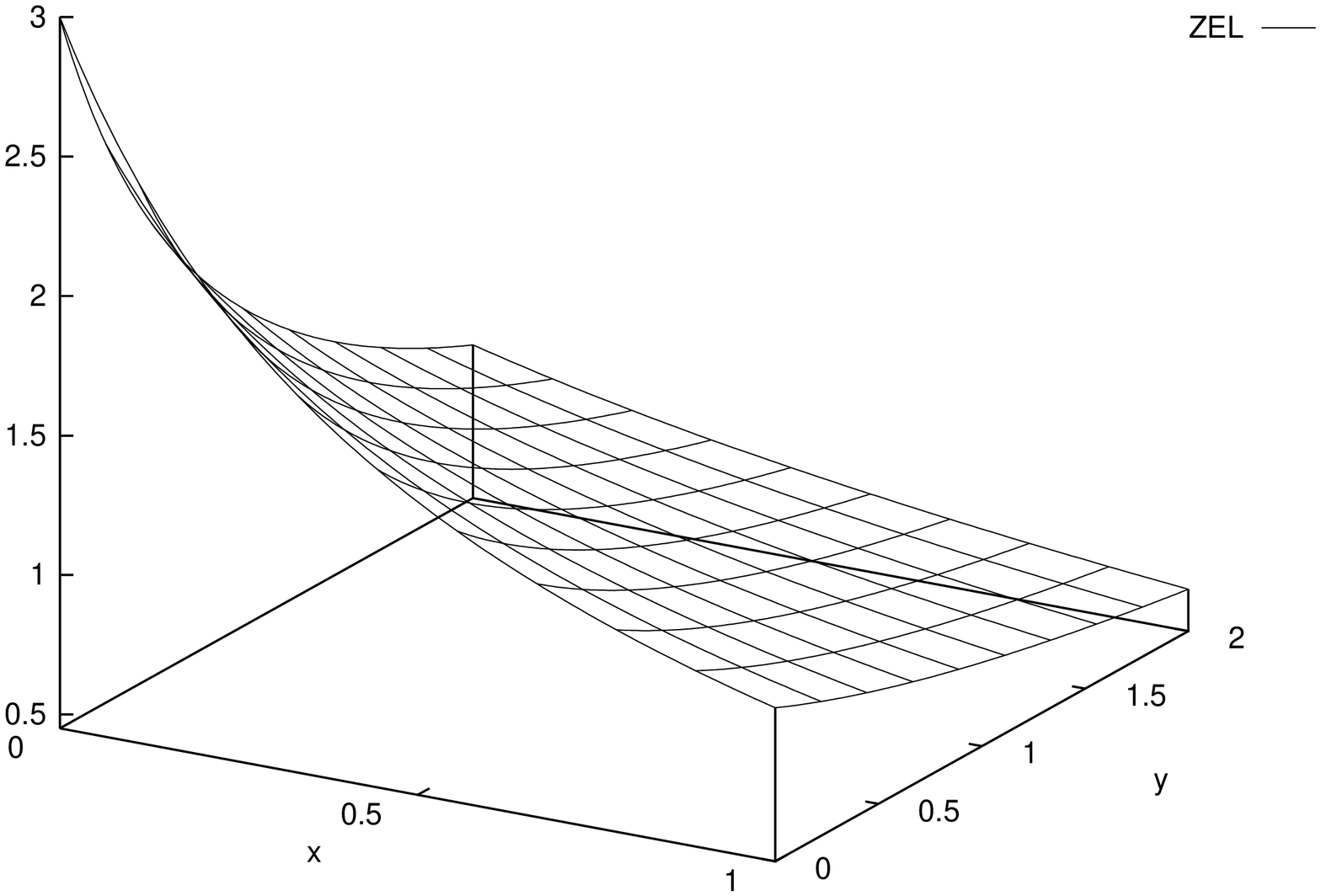,width=6cm,angle=-90}
\epsfig{file=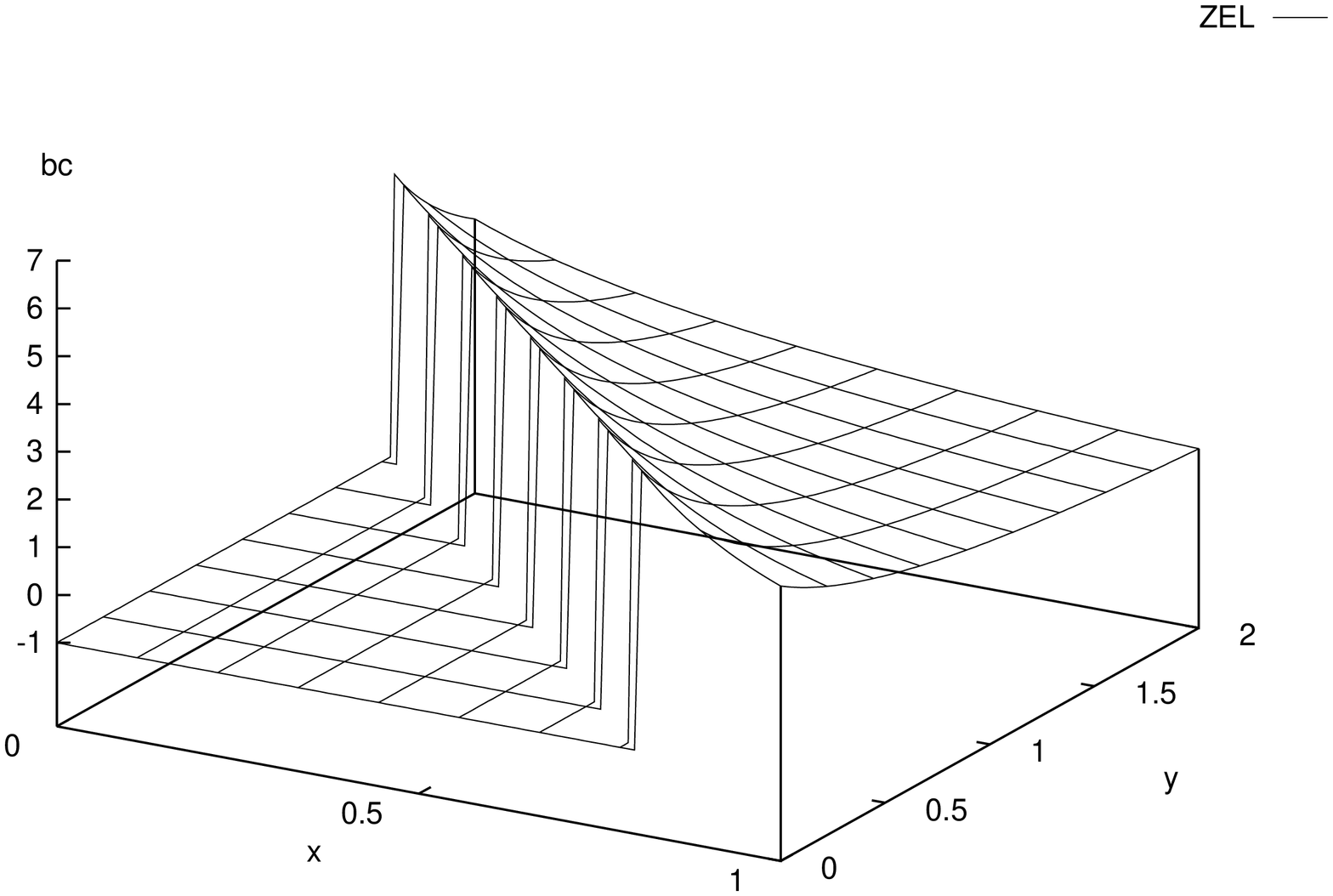,width=6cm,angle=-90}}
\caption{Collapse times with Zel'dovich approximation. Figures taken by M98.}
\end{center}
\end{figure}

\begin{figure}
\begin{center}
\hbox{
\epsfig{file=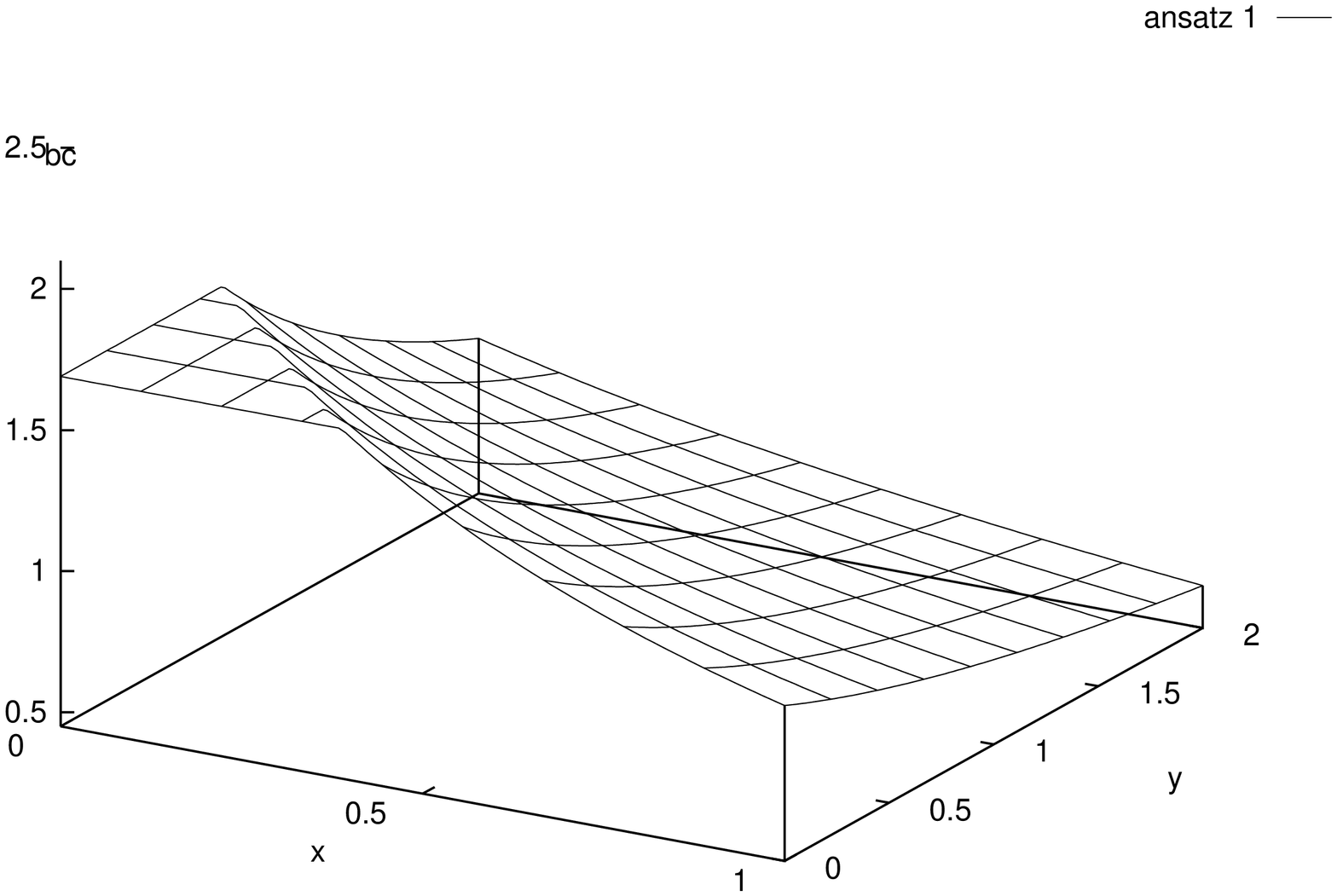,width=6cm,angle=-90}
\epsfig{file=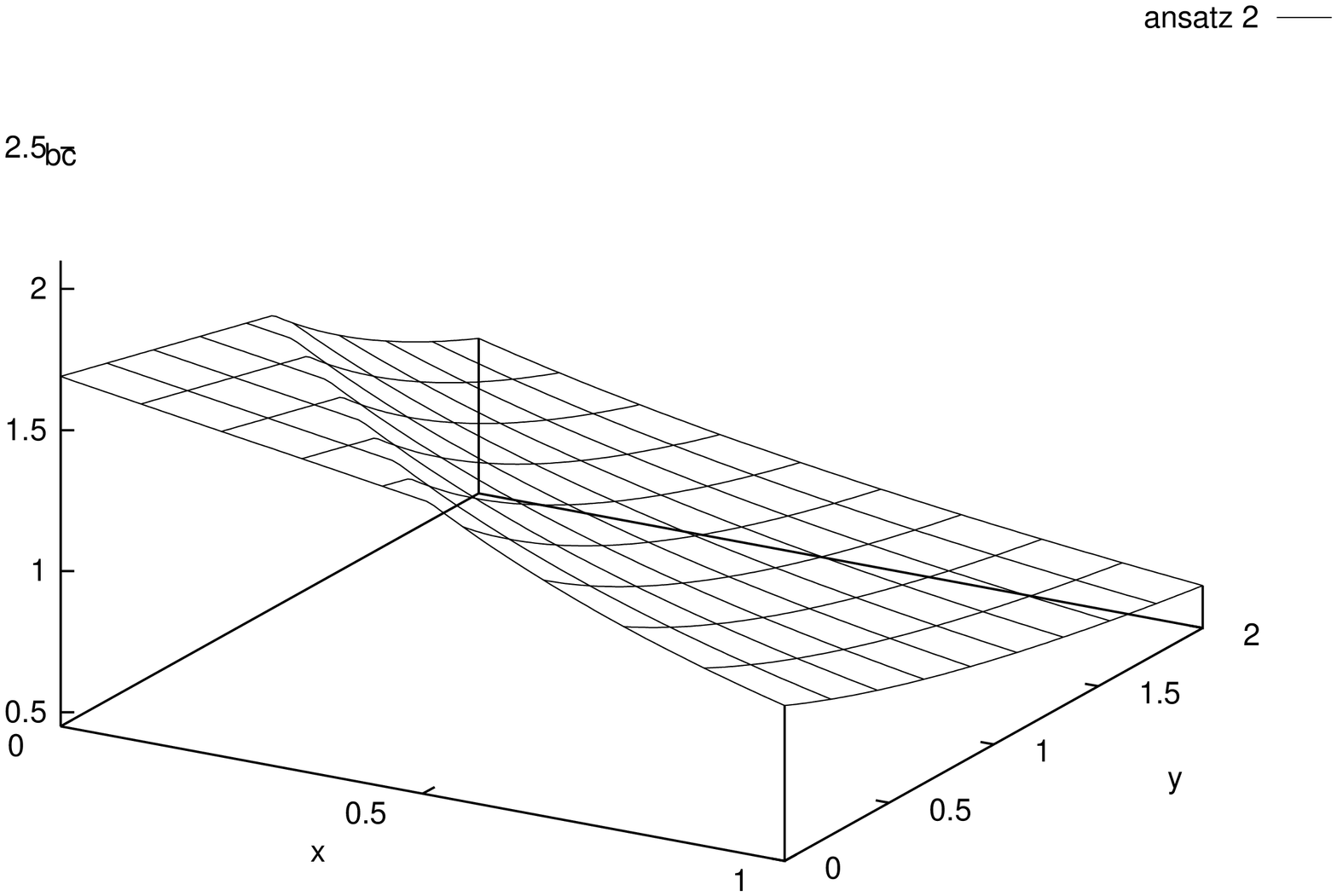,width=6cm,angle=-90}
}
\caption{Collapse times with the two {\it ansatze}. Figures taken by M98}
\end{center}
\end{figure}

\subsection{\bf Ellipsoidal collapse}

The convenience in using the homogeneous ellipsoid collapse model
resides in the fact that it can easily be solved by means of a
numerical integration of a system of three second-order ordinary
differential equations.  One of the advantages of spherical symmetry
is that, because of Birkhoff's theorem, it is possible to introduce in a
background metric a perturbation without influence the rest of the
Universe, provided any positive perturbation is compensated for by an
(outer) negative one, such to make the total mass perturbation
vanish. This is necessary to ensure the self-consistency of the
problem: the background has to evolve as if it were unperturbed.  This
reasoning is not valid any more when introducing a triaxial
perturbation in an unperturbed background: this is going to influence
the background, through non-linear feedback effects.  To use
ellipsoidal collapse in a cosmological context, the correct strategy
is not to try to insert an ellipsoid in a uniform background, but to
extract an ellipsoid from a perturbed FRW Universe.

In the following, we shall describe the ellipsoidal model and how to find solutions to the 
equations, both numerically and analytically.

%{\bf 
The dynamical variables of ellipsoidal collapse are the three axes
$a_i(t)$ of the ellipsoid; they are normalized as the scale factor:
$a_i(t)=a(t)$ if the ellipsoid is a sphere with null density contrast.
Their evolution equations are:

$$ \frac{d^2a_i}{da^2} - (2a(1+(\Omega_0^{-1}-1)a))^{-1}
\frac{da_i}{da} + (2a^2(1+(\Omega_0^{-1}-1)a))^{-1}a_i$$ 
\be\times \left[ \frac{1}{3} +\frac{\delta}{3} + \frac{b'_i}{2} \delta + 
\lambda'_{vi} \right]=0 \label{eq:ellips_open} \ee

\noindent in the open case (the Einstein-de Sitter case can be obtained 
by setting $\Omega_0=1$), while in the flat case with cosmological
constant they are:

$$ \frac{d^2a_i}{da^2} - \frac{1-2(\Omega_0^{-1}-1)a^3}{2a(1+(\Omega_0^{-1}
-1)a^3)}\frac{da_i}{da} + (2a^2(1+(\Omega_0^{-1}-1)a))^{-1}a_i $$ 
\be\times \left[ \frac{1}{3} + \frac{\delta}{3} + \frac{b'_i}{2} \delta + 
\lambda'_{vi} \right]=0. \label{eq:ellips_cosm} \ee

\noindent 
Note that this time the scale factor $a(t)$ has been used as time
variable.  The density contrast $\delta$ is:

\be \delta = \frac{a^3}{a_1a_2a_3} - 1,\label{eq:ellips_delta} \ee

\noindent 
while the quantities $b'_i$ and $\lambda'_{vi}$ are defined as:

\be b'_i=\frac{2}{3} [a_i a_j a_k R_D(a_i^2,a_j^2,a_k^2)-1]\;\;\;\;\;
i\neq j\neq k \label{eq:bprime} \ee

\noindent (where the $R_D$ is the Carlson's elliptical integral

\be R_D(x,y,z) = \frac{3}{2}\int_0^\infty \frac{d\tau}{(\tau+x)^{1/2}
(\tau+y)^{1/2}(\tau+z)^{3/2}}, \label{eq:carlson} \ee

\noindent 
which can be calculated by means of the routine given by Press \&
Teukolsky (1990)) and

\be \lambda'_{vi} = -\frac{a}{a_0} \left(\frac{\delta}{3} - 
a_0 \lambda_i \right). \label{eq:lambdavi} \ee

Initial conditions can be set by imposing that the $a_i$ evolve
according to Zel'dovich approximation at early times:

\be a_i \simeq a(1-a\lambda_i) \label{eq:ellips_incond} \ee
\be \frac{d a_i}{da} \simeq \frac{1}{a} (a_i(a)-a^2 \lambda_i)\; . \ee

These three coupled second-order ordinary differential equations, Eqs.
(\ref {eq:ellips_open}) or (\ref{eq:ellips_cosm}), can be integrated
by means of standard routines, as the Runge-Kutta one given in Press
et al. (1992).  The numerical integration has to be pushed to the
singularity, when at least one axis vanishes (and the density
diverges).  To do this, it is useful to use logarithmic variables, to
have more controlled variations from quasi-homogeneity to collapse.
Moreover, the integration can be divided into two parts: the first is
stopped at decoupling, defined as the instant at which the density
starts to increase, while in the second part the density is used as
time variable, and the integration is pushed up to large density
values (Monaco 1997a). The overall precision of the numerical
integration is better than 1\% for the spherical collapse, but becomes
about 8\% for pancake-like collapses. It is to be noted that, in any
case, the first collapsing axis is that corresponding to \luno, the
largest \lam\ eigenvalue.

Fig. 4 shows the collapse ``times'' $b_c$ of initially overdense
(\dl=1) and underdense (\dl=$-$1) ellipsoids in an Einstein-de Sitter
Universe (in this case $b_c=a_c$). Spherical collapse is obviously
recovered at $x=y=0$, while quasi-spherical collapses reasonably show
a systematic departure from spherical collapse, as in Fig. 3a.
The large-shear behavior is similar but not identical to that
predicted by Ze'ldovich approximation (ZEL); at variance with what happens with quasi-spherical
collapses, in this range ZEL tends to underestimate the collapse time.
Fig. 5 shows the $b_c$ curve for ellipsoids in an open Universe;
this time \dl=3 has been chosen, to allow all the ellipsoids to
collapse. As expected, this curve is nearly identical to the one shown
in Fig. 3a, apart from an obvious rescaling. Notably, numerical
calculations of collapsing ellipsoids in open Universes are affected
by larger errors than that quoted above if the ellipsoid takes a long
time to collapse.

%}

\begin{figure}
\begin{center}
\hbox{
\epsfig{file=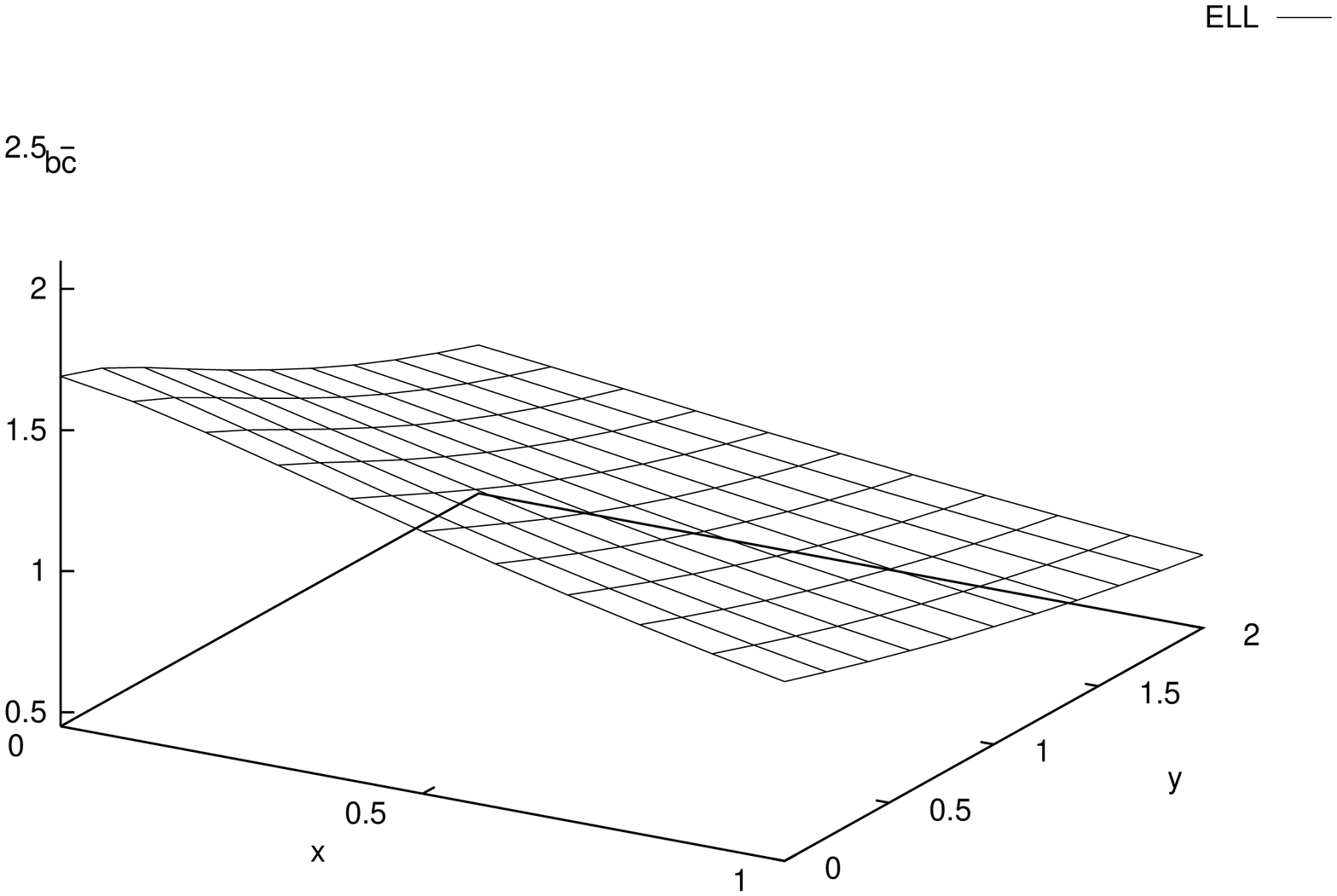,width=6cm,angle=-90}
\epsfig{file=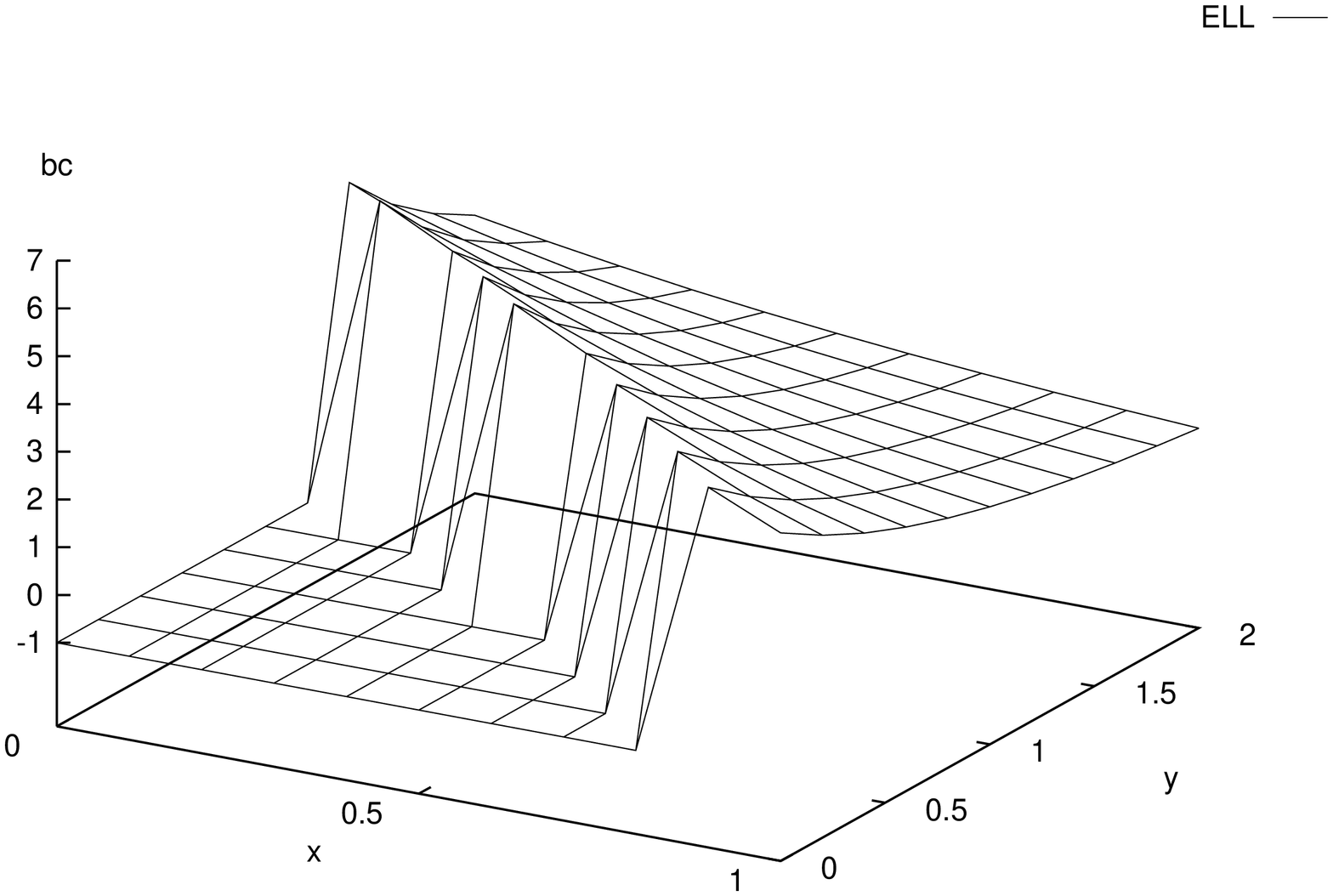,width=6cm,angle=-90}
}
\caption{Collapse times with Ellipsoidal model. Figures taken by M98.}
\end{center}
\end{figure}

\begin{figure}
\begin{center}
\epsfig{file=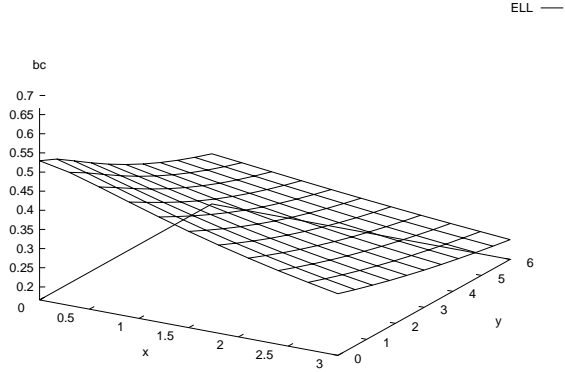,width=6cm,angle=-90}
\caption{Collapse times with Ellipsoidal model, open Universe. Figures taken by M98.}
\end{center}
\end{figure}

An analytic solution to the collapse model equations can be obtained as follows.
For the unisolated ellipsoid,
%obtained by adding the force due to
%the potential given by Eq. ~(\ref{eq:quad}) into
%Eq. ~(\ref{eq:WS}). Assuming
%that the principal axes of the external tidal tensor are always
%oriented along the principal axes of the mass tensor,
the evolution equations reduces to three equations for the three
semiaxes of the ellipsoid and are given by (Watanabe 1993;
van de Weygaert 1996):
\begin{eqnarray}
\frac{d^2a_{\rm i}}{dt^2}&=&-2\pi G\left\{ \rho _{\rm e}(\alpha
_{\rm i}-\gamma b_{\rm i})+\left[ \frac 23-(\alpha _{\rm i}-\gamma
b_{\rm i})\right] \rho _{\rm b}\right\} a_{\rm i} = \nonumber \\
& & -2\pi G\left[
\rho _{\rm e}\alpha _{\rm i}+\left( \frac 23-\alpha _{\rm
i}\right) \rho _{\rm b}\right] a_{\rm i}-\nonumber \\
& & 2 \pi G \gamma
\left(-b_{\rm i}\right) \left(\rho_{\rm e}-\rho_{\rm b}
\right)a_{\rm i}
%=-2\pi G\left[ \rho _{\rm e}\alpha _{\rm i}+\left( \frac 23-\alpha
%_{\rm i}\right) \rho _{\rm b}\right] a_{\rm i}-E_{\rm ij} a_{\rm
%j}
\label{eq:WSM}
\end{eqnarray}
where:
\begin{equation}
\gamma =\frac 3{2\pi }\frac Q\delta, \hspace{0.5cm} {\bf
b}=(-\beta ,\beta -1,1) \label{eq:bb}
\end{equation}
and 
where $\rho_{\rm b}$ is the density of the background universe, and $\rho_{\rm e}$ the
density within the ellipsoid. The coefficients $\alpha_{\rm i}$
are given by:
\begin{equation}
\alpha _{\rm i}=a_1a_2a_3\int_0^\infty \frac{d\lambda }{\left(
a_{\rm i}^2+\lambda \right) \left[ \left( a_1^2+\lambda \right)
\left( a_2^2+\lambda \right) \left( a_3^2+\lambda \right) \right]
^{\frac 12}} \label{eq:alpi}
\end{equation}

Assuming that the external structures, giving rise to the
tidal field, are at a large distance from the ellipsoid (see
Eisenstein \& Loeb 1995), the amplitude of the
external quadrupole force is assumed to increase with the linear
growth rate (Ryden 1988; Watanabe 1993; Eisenstein \& Loeb 1995),
$D(t)$ (this last quantity is given in Peebles 1980):
\begin{equation}
Q(t)=Q_0 \frac{D(t)}{D_0}
\label{eq:quadru}
\end{equation}
Here the subscript ``0" means that the corresponding quantity is
calculated at the present epoch and $D(t)=R_{\rm b}(t)$, for an
Einstein-de Sitter (hereafter EdS) universe.
In an Einstein-de Sitter universe, Eq. (\ref{eq:quadru}), 
$Q(t)=Q_0 \frac{R_{\rm b}(t_0) (t/t_0)^{2/3}}{R_{\rm b}(t_0)}$ reduces to 
$Q(t_0)=Q_0 $
at time $t_0$.

In order to have an estimate of the value of $Q_0$,
for a cluster interacting with a neighboring one, one can use the simple
model in Watanabe (1993), considering a cluster which has a
neighboring cluster with a mean density contrast $<\delta> \simeq
3$, a comoving separation $(0,0,x_3)$, and a comoving size $\Delta
x_3=x_3/3$, the $Q_{33}$ quadrupole component is given by:
\begin{equation}
Q_{33} \simeq \frac{8}{9} \pi <\delta> \left(\frac{\Delta
x_3}{x_3}\right)^3 \simeq 0.3 \label{}
\end{equation}
%Assuming
The previous estimate corresponds to a cluster interacting with a
neighbor having a mass excess comparable to that of the Virgo
cluster, and a separation three times its size.

%{\bf 
Another way of estimating $Q_0$ is by using the anisotropy of the velocity field in the 
LSC from data of Lilje, Yahil \& Jones (1986). If we indicate with $Q_{\rm v0}$ the component of the 
largest absolute value of the anisotropic velocity, one gets:
$Q_0 \Omega_0^{0.6}=\frac{4 \pi}{3} Q_{\rm v0}$ (Watanabe 1993). Since Lilje, Yahil \& Jones (1986) deduced a value of 
$Q_{\rm v0} \sim 0.1-0.2$ at the distance of the Local Group from Virgo, we have that
$Q_0 \Omega_0^{0.6} \sim 0.4-0.8$. 

\begin{figure*}
\label{Fig. 2} \centerline{\hbox{
(a)
\psfig{figure=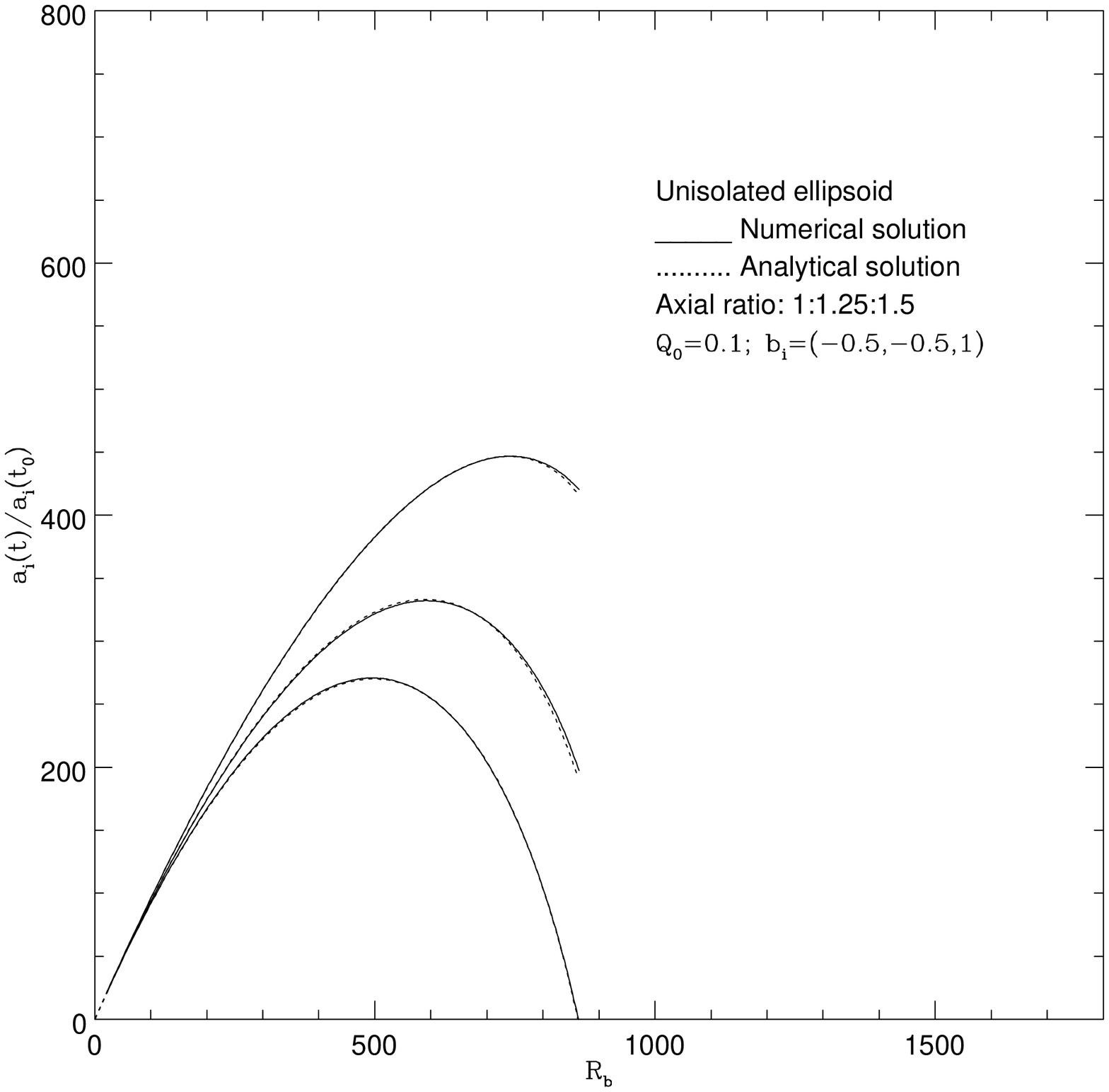,width=8.5cm} (b)  
\psfig{figure=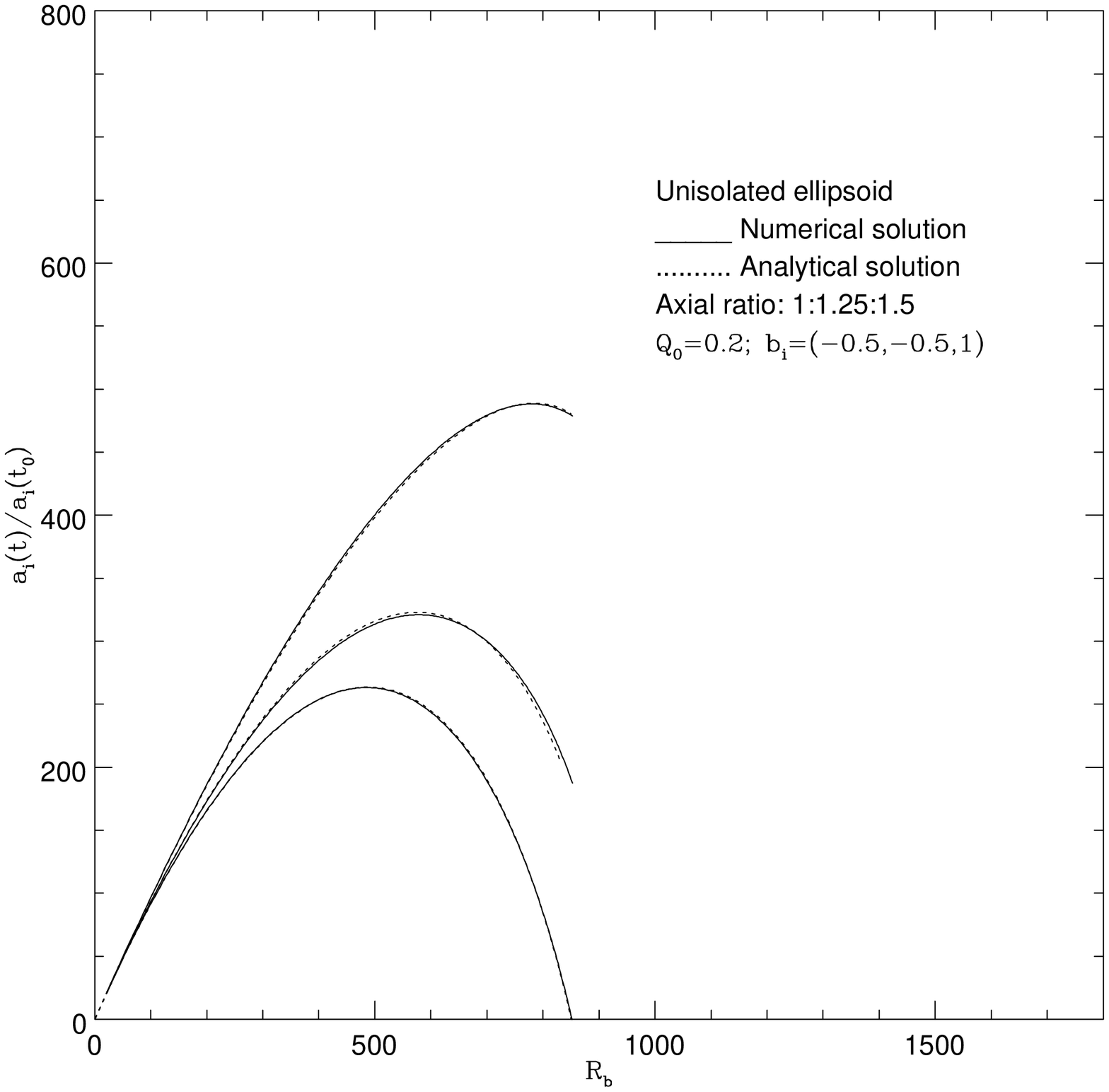,width=8.5cm}  
}}
%\caption{
\end{figure*}
\begin{figure*}
\label{Fig. 2} \centerline{\hbox{
(c)
\psfig{figure=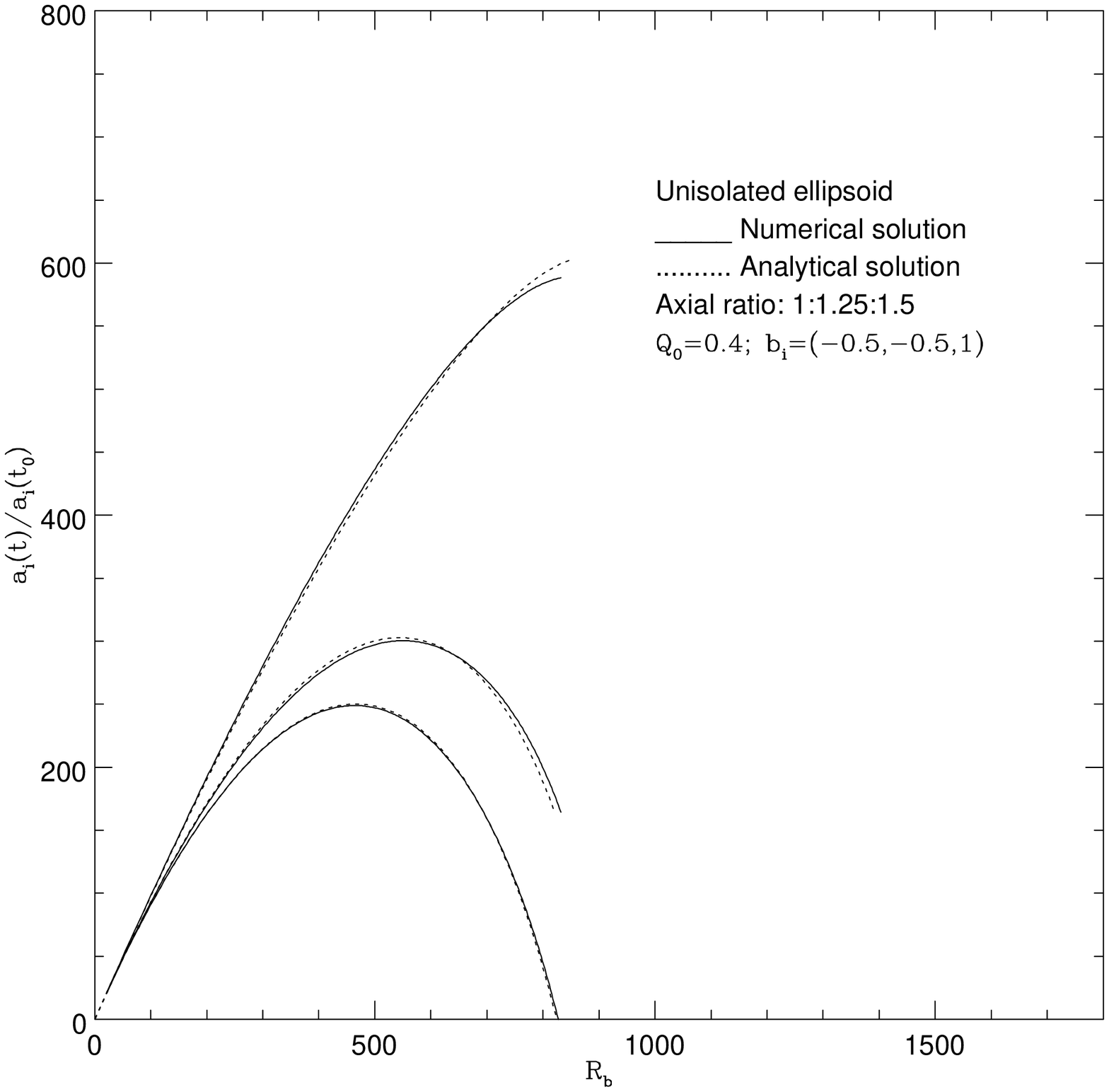,width=8.5cm}  }}
%{\bf Fig. 6} 
\caption{
Evolution of {\it unisolated} homogeneous ellipsoidal
perturbations in an EdS universe with $H_0=50 {\rm km/s/Mpc}$, 
$\rho_{\rm e}/\rho_{\rm b}=1.003$, axial ratio is $1:1.25:1.5$, 
$b_{\rm i}=(-0.5,-0.5,1)$, and $Q_0=0.1$ (a), $Q_0=0.2$ (b), $Q_0=0.4$ (c). Figures taken from Del Popolo (2002b).
}
\end{figure*}

It is
possible to obtain an analytical solution of Eq.
(\ref{eq:WSM}), describing the evolution of an {\it unisolated}
ellipsoid as shown in Del Popolo (2002b) In this case the solution can be written in the form:
\begin{equation}
\frac{a_1(t)}{a_1(t_i)}=R_{\rm b}-\frac 32 {\tilde \alpha_1}\left(
R_{\rm b}-R_{\rm e}\right)-d \times R_{\rm b}^{\left(\frac{2+3
c_1}{2}\right)} \left(1-\frac{3 {\tilde \alpha_1}}{2}\right)
\label{eq:predd1}
\end{equation}

\begin{equation}
\frac{a_{\rm 2}(t)}{a_{\rm 2}(t_i)}=R_{\rm b}-\frac 32 {\tilde
\alpha _{\rm 2}}\left( R_{\rm b}-R_{\rm e}\right)
\label{eq:predd2}
\end{equation}

\begin{equation}
\frac{a_{\rm 3}(t)}{a_{\rm 3}(t_i)}=R_{\rm b}-\frac 32 {\tilde
\alpha _{\rm 3}}\left( R_{\rm b}-R_{\rm e}\right)
\label{eq:predd3}
\end{equation}
where 
$R_{\rm b}$ is the scale
factor of the background universe, and $R_{\rm e}$ that of the ellipsoid.

For ellipsoids having the initial axial ratio $1:a_2:a_3$, with $a_1
\ge 1.25$ and $a_2 \ge 1.5$, we now have that $c_1=1.23$,
$d=6\times10^{-7}$ and:
\begin{equation}
{\tilde \alpha _{\rm 1}}=\alpha_1+0.0672
\left(\frac{b_1}{b_2}\right)^{0.15} b^{0.6}_3 \label{}
\end{equation}
\begin{equation}
{\tilde \alpha _{\rm 2}}=a_{10}^{0.07}a_{20}^{-0.06}a_{30}^{-0.01}
\left[\alpha_2+0.031 \left(\frac{b_2}{b_1}\right)^{0.5} b^{0.95}_3
\right] \label{}
\end{equation}
\begin{equation}
{\tilde \alpha _{\rm 3}}=1.002
a_{10}^{0.1}a_{20}^{-0.035}a_{30}^{-0.065} \left(\alpha_3-0.063
a_{30}^{0.09} b^{0.95}_3 \right) \label{}
\end{equation}
where $b_{\rm i}$ was defined in Eq. (\ref{eq:bb}).

\begin{figure*}
\label{Fig. 3} \centerline{\hbox{
(a)
\psfig{figure=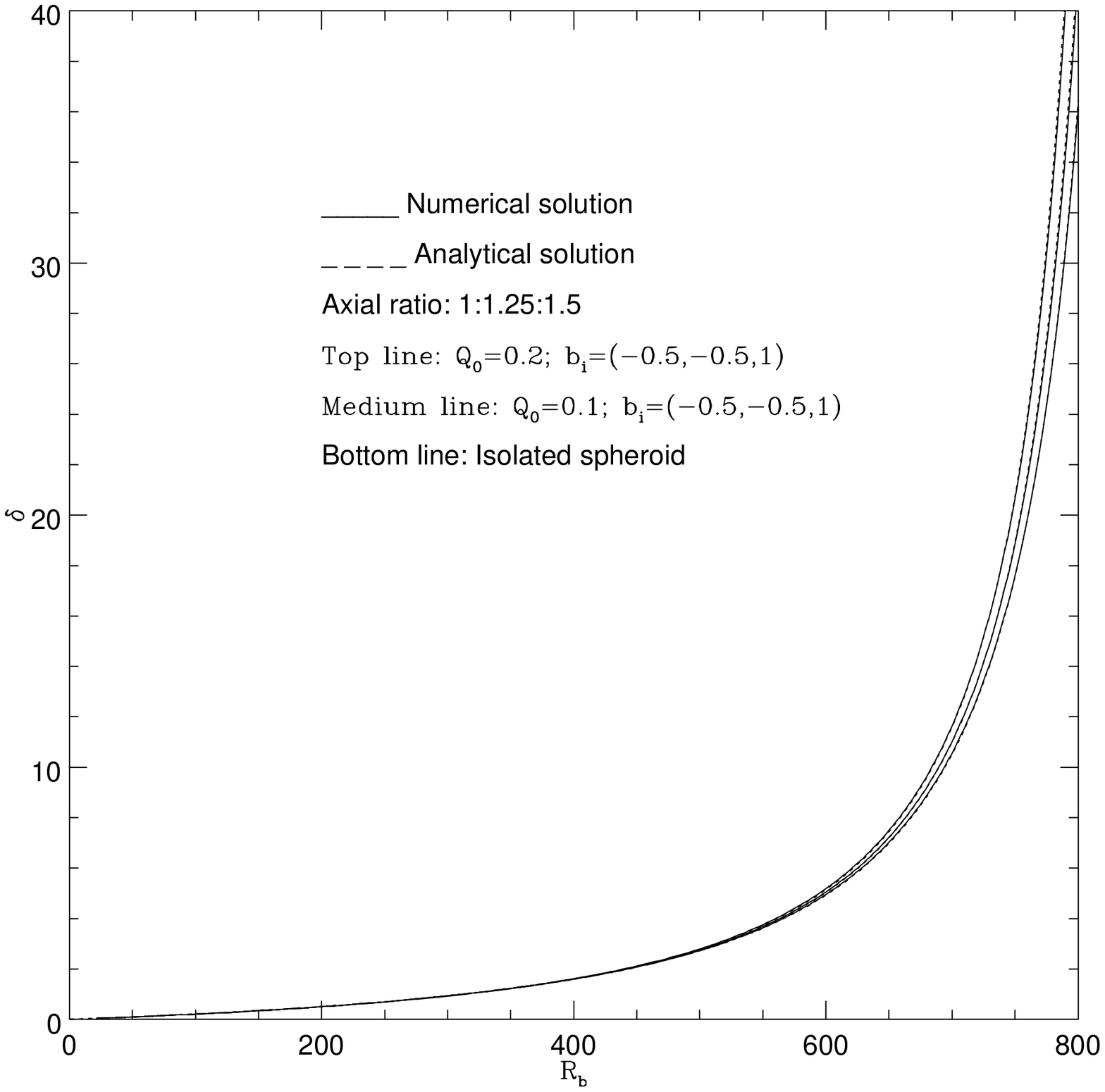,width=8.7cm} (b)  
\psfig{figure=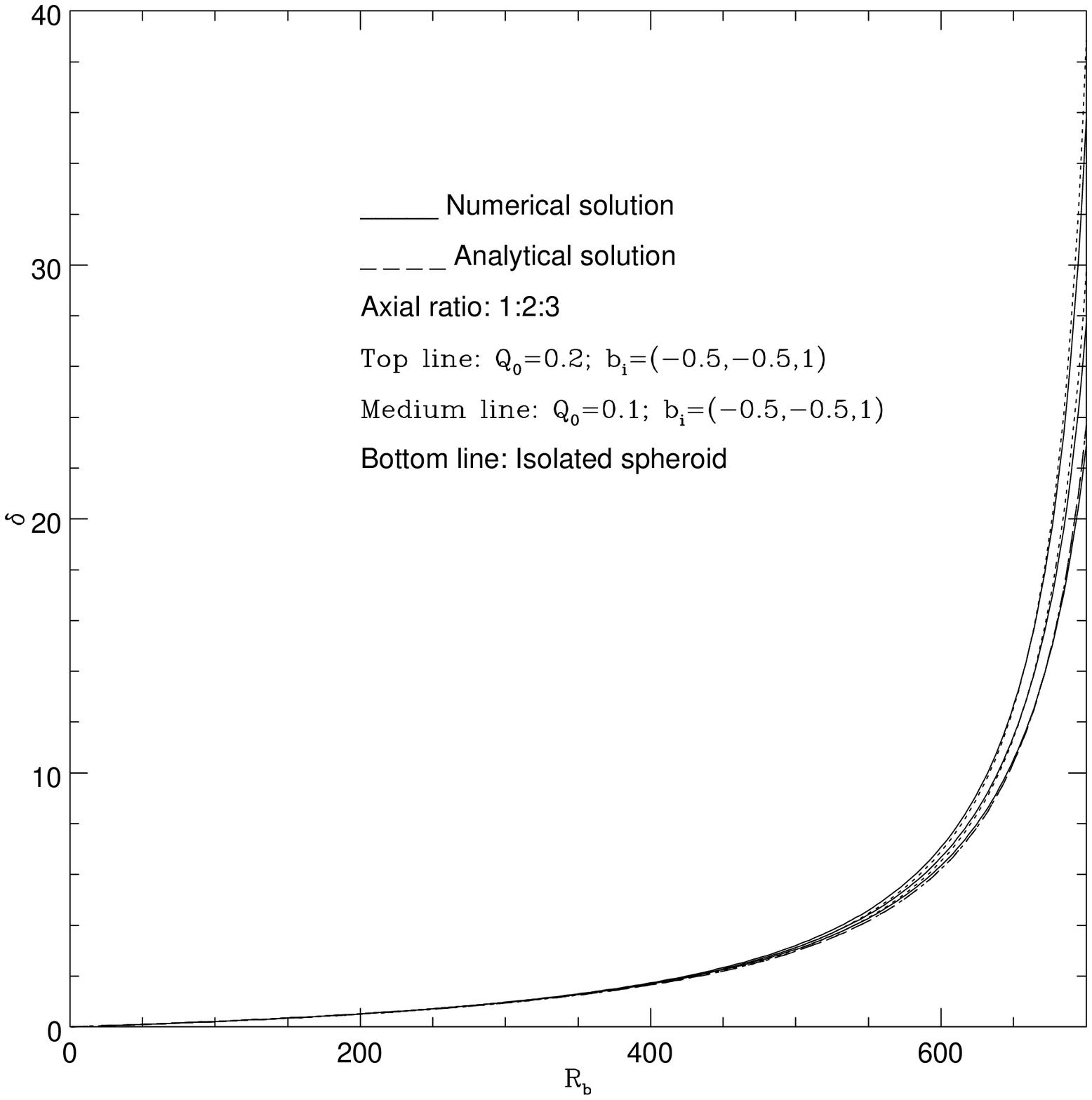,width=8.7cm}  
}}
%{\bf Fig. 7} 
\caption{The evolution of the density contrast. The axial
ratio of the ellipsoid is $1:1.25:1.5$ (a) and $1:2:3$ (b), the lines from bottom to
top represent the case of an {\it isolated} ellipsoid ($b_{\rm
i}=(0,0,0)$), $Q_0=0.1$, $b_{\rm i}=(-0.5,-0.5,1)$, and $Q_0=0.2$, $b_{\rm i}=(-0.5,-0.5,1)$,
respectively. Figure taken from Del Popolo (2002b).
%\caption{
}
\end{figure*}
\begin{figure*}
%\label{Fig. 4} 
\label{Fig. 4} \centerline{\hbox{
\psfig{figure=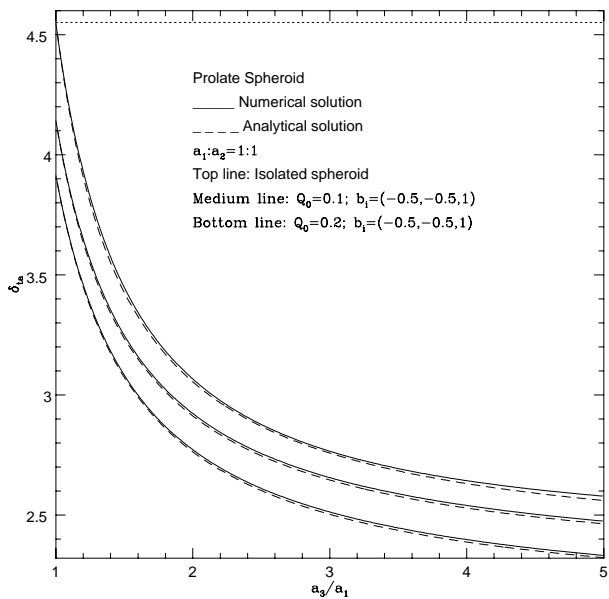,width=8.7cm}  
}}
\caption{
%{\bf Fig. 8} 
The density contrast at turnaround for a prolate
spheroid for several values of the longest axis, $a_3$, (the other
two axes have fixed value $a_1:a_2=1:1$). The solid lines, from
top to bottom, represent numerical results for the density
contrast for an {\it isolated} spheroid ($b_{\rm i}=(0,0,0)$), and
for {\it unisolated} spheroids with $Q_0=0.1$, $b_{\rm i}=(-0.5,-0.5,1)$ and
$Q_0=0.2$, $(-0.5,-0.5,1)$, respectively. The dashed lines represent the
approximate solution (Eq. (37)). The upper
dotted line represents the value of the density contrast at
turnaround for a spherical perturbation. Figure taken from Del Popolo (2002b).}
\end{figure*}

\begin{figure}
\label{Fig. 9} \centerline{\hbox{
\psfig{figure=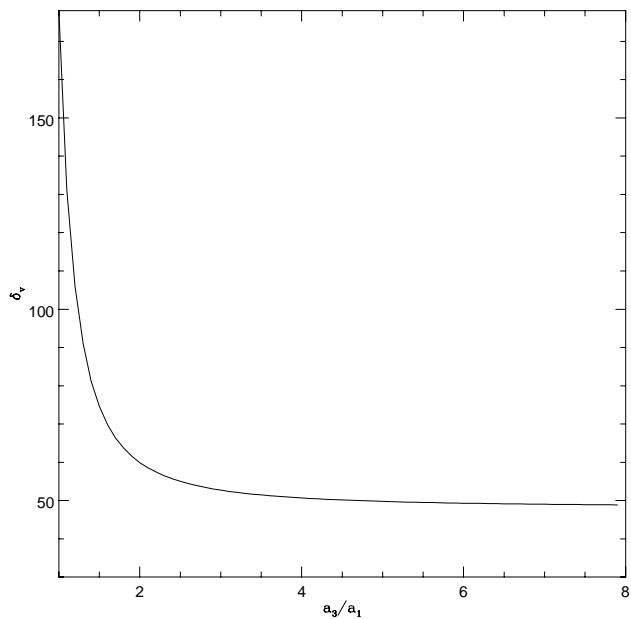,width=9cm}  }}
\caption{
%{\bf Fig. 8} 
Density contrast at virialization. The solid line
refers to an ${\it isolated}$ prolate spheroid. Figure taken from Del Popolo (2002b).
}
\end{figure}

\begin{figure}
\label{Fig. 10} \centerline{\hbox{
\psfig{figure=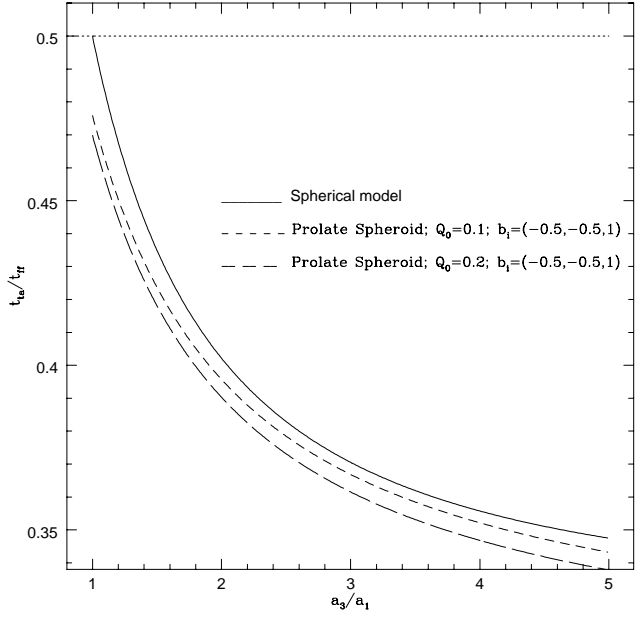,width=9cm}  }}
\caption{
%{\bf Fig. 9} 
Turnaround epoch for a prolate spheroid. The solid,
short-dashed and long-dashed lines, represent respectively the
time of turnaround for an {\it isolated} spheroid and for
spheroids having $Q_0=0.1$, $b_{\rm i}=(-0.5,-0.5,1)$ and $Q_0=0.2$, $(-0.5,-0.5,1)$. The
upper dotted line represents the value of the density contrast at
turnaround for a spherical perturbation. Figure taken from Del Popolo (2002b).
}
\end{figure}

In the
case of prolate spheroids, with axial ratio $1:1:a_3$ and $1 \leq
a_3 \leq 5$, a better approximation to the $\alpha_{\rm i}$ is:
\begin{equation}
{\tilde \alpha _{\rm 1}}=\alpha_1+0.037 \left(
\frac{a_{30}}{a_{10}}
\right)^{0.35}
\left(\frac{b_1}{b_2}\right)^{0.15} b^{0.6}_3 \label{}
\end{equation}
\begin{equation}
{\tilde \alpha _{\rm 2}}=\left(
\frac{a_{10}}{a_{30}}
\right)^{0.01}
\left[\alpha_2+0.031
\left(
\frac{a_{10}}{a_{30}}
\right)
\left(\frac{b_2}{b_1}\right)^{0.5} b^{0.95}_3
\right] \label{}
\end{equation}
\begin{equation}
{\tilde \alpha _{\rm 3}}=1.002
\left(
\frac{a_{20}}{a_{30}}
\right)^{0.065} \left(\alpha_3-0.063
a_{30}^{0.09} b^{0.95}_3 \right) \label{}
\end{equation}

In the case of an {\it unisolated} ellipsoid the length of the uncollapsed axes at collapse can be 
obtained by means of Eq. (\ref{eq:predd2})-(\ref{eq:predd3}):
\begin{equation}
\frac{a_{\rm 3}(t_{\rm c})}{a_{\rm 2}(t_{\rm c})}=\frac{a_{\rm 3}(t_{\rm i})}{a_{\rm 2}(t_{\rm i})}
\frac{R_{\rm b}-\frac 32 \tilde \alpha
_{\rm 3}\left( R_{\rm b}-R_{\rm e}\right)}{R_{\rm b}-\frac 32 \tilde \alpha
_{\rm 2}\left( R_{\rm b}-R_{\rm e}\right)}
\end{equation}

The evolution of the density contrast can be calculated using the
usual definition:
\begin{equation}
\delta =\frac{\rho _{\rm e}-\rho _{\rm b}}{\rho _{\rm
b}}=\frac{\rho _{\rm e0}}{\rho _{\rm
b0}}\frac{a_{10}}{a_1}\frac{a_{20}}{a_2}\frac{a_{30}}{a_3}\left(
\frac{R_{\rm b}}{R_{\rm b0}}\right) ^3-1 \label{eq:denss}
\end{equation}

If $x(t)=x_{\rm o} X(t)$,
$y(t)=y_{\rm o} Y(t)$ and $z(t)=z_{\rm o} Z(t)$, are the principal
axes ($x_{\rm o}$, $y_{\rm o}$ and $z_{\rm o}$ are the initial
values of the axes), the overdensity of the ellipsoid is the same
used previously, the initial conditions are $X=Y=Z=R_{\rm b}=R_{\rm
e}=1$ at $t=t_0$ and as before the initial velocity is equal to
the Hubble velocity at $t_0$ (representing the initial time). The
parametric equations satisfied by $R_{\rm e } (t)$ are:
\begin{equation}
R_{\rm e }=\frac{1}{2 \delta} \left(1-\cos(\vartheta)\right),
\hspace{0.5cm} \frac{t}{t_0}=\frac{3}{4
\delta^{3/2}}\left(\vartheta-\sin(\vartheta)\right) \label{eq:tim}
\end{equation}
% ???
%%($\delta$ evolve o no???)
%
while, since our background is an EdS universe, $R_{\rm b}(t)
\propto t^{2/3}$.

It is easy to find that the density contrast
%density
%at turnaround, which
is given by:
\begin{eqnarray}
\delta_{\rm i} &=&\frac{R_{\rm b}^3}{XYZ}-1= f_1^2 
\left( \vartheta\right) \nonumber\\
& &
[\left( 1-\frac{3{\tilde \alpha} _1}2\right)
f_1^{\frac 23}\left( \vartheta \right) +\frac 34{\tilde \alpha}
_1f_2\left( \vartheta \right) \nonumber\\
& &
-d\left( 1-\frac{3{\tilde \alpha}
_1}2\right) \frac{f_1^{\frac{2f}3}\left( \vartheta
\right) }{\delta ^{f-1}}]^{-1} \times \nonumber \\
& & \left[ \left( 1-\frac{3{\tilde \alpha} _2}2\right) f_1^{\frac
23}\left( \vartheta \right) +\frac 34{\tilde \alpha} _2f_2\left(
\vartheta \right) \right]^{-1} \nonumber\\ 
& &
\left[ \left( 1-\frac{3{\tilde
\alpha} _3}2\right) f_1^{\frac 23}\left( \vartheta \right) +\frac
34{\tilde \alpha} _3f_2\left( \vartheta \right) \right]^{-1} -1
\label{eq:denstu}
\end{eqnarray}
where $f=\frac{2+3c_1}{2}$, $f_1(\vartheta)=\frac{3}{4}
\left(\vartheta-\sin(\vartheta)\right)$ and
$f_2(\vartheta)=1-\cos(\vartheta)$. The density contrast at
turn-around is obtained by calculating $\delta_{\rm i}(\vartheta_{\rm
ta})$, where the parameter $\vartheta_{\rm ta}$ at the turnaround epoch is
given solving the equation:
\begin{equation}
\frac 2{3 {\tilde\alpha_1} }=\frac{dfR^{f-1}_{\rm b}+\frac{\sin
(\vartheta_{\rm ta} )}{f_2(\vartheta_{\rm ta} )}f_1^{\frac
13}-1}{dfR^{f-1}_{\rm b }-1} \label{eq:time}
\end{equation}
%Eqs. (\ref{eq:denstu}), (\ref{eq:time}) reduces to BS
%Eqs. (76) and (72) for $d=0$, ${\tilde\alpha_1}=\alpha_1$,
%${\tilde\alpha_2}=\alpha_2$, ${\tilde\alpha_3}=\alpha_3$.
%\footnote{Some equations in BS contains some
%typographical misprints, for example Eq.
%(76)  and Eq. (80).} 
Eq. (\ref{eq:denstu}) yields the
familiar value $\delta=\left(3 \pi/4\right)^2$  in the spherical
case,
($d=0$,
${\tilde\alpha_1}=\alpha_1={\tilde\alpha_2}=\alpha_2={\tilde\alpha_3}=\alpha_3=2/3$).
In general, in order to obtain the density contrast at turnaround,
one has first to solve Eq. (\ref{eq:time}) for $\vartheta$
for an arbitrary axial ratio and substitute the value in
Eq. (\ref{eq:denstu}). The time of turn-around can be
calculated by:
\begin{equation}
t=\frac{3 t_0}{4
\delta^{3/2}}\left(\vartheta-\sin(\vartheta)\right)=\frac{t_{\rm
ff}}{2 \pi} \left(\vartheta-\sin(\vartheta)\right) \label{}
\end{equation}
where $t_{\rm ff}$ is the free-fall time:
\begin{equation}
t_{\rm ff}=\frac{3 \pi}{2 \delta^{3/2}} t_0
\label{}
\end{equation}

Similarly it is possible to calculate the density contrast at collapse and collapse time.
%CALCOLARE delta al tempo di collasso

%The good approximation to quantities n ellipsoidal collapse are plotted in Figs. 6-10.
Figs. 6-10 plots the comparison of the analytical model with numerical simulations for several 
interesting quantities. 

Fig. 6 shows the evolution of {\it unisolated} homogeneous ellipsoidal
perturbations in an EdS universe with $H_0=50 {\rm km/s/Mpc}$, 
$\rho_{\rm e}/\rho_{\rm b}=1.003$, axial ratio is $1:1.25:1.5$, 
$b_{\rm i}=(-0.5,-0.5,1)$, and $Q_0=0.1$ (a), $Q_0=0.2$ (b), $Q_0=0.4$ (c).

Fig. 7, plots the evolution of the density contrast. The axial
ratio of the ellipsoid is $1:1.25:1.5$ (a) and $1:2:3$ (b), the lines from bottom to
top represent the case of an {\it isolated} ellipsoid ($b_{\rm
i}=(0,0,0)$), $Q_0=0.1$, $b_{\rm i}=(-0.5,-0.5,1)$, and $Q_0=0.2$, $b_{\rm i}=(-0.5,-0.5,1)$,
respectively.

Fig. 8, shows the density contrast at turnaround for a prolate
spheroid for several values of the longest axis, $a_3$, (the other
two axes have fixed value $a_1:a_2=1:1$). The solid lines, from
top to bottom, represent numerical results for the density
contrast for an {\it isolated} spheroid ($b_{\rm i}=(0,0,0)$), and
for {\it unisolated} spheroids with $Q_0=0.1$, $b_{\rm i}=(-0.5,-0.5,1)$ and
$Q_0=0.2$, $(-0.5,-0.5,1)$, respectively. The dashed lines represent the
approximate solution. The upper
dotted line represents the value of the density contrast at
turnaround for a spherical perturbation.

Fig. 9, is the density contrast at virialization, while Fig. 10 represents
the turnaround epoch for a prolate spheroid. The solid,
short-dashed and long-dashed lines, represent respectively the
time of turnaround for an {\it isolated} spheroid and for
spheroids having $Q_0=0.1$, $b_{\rm i}=(-0.5,-0.5,1)$ and $Q_0=0.2$, $(-0.5,-0.5,1)$. The
upper dotted line represents the value of the density contrast at
turnaround for a spherical perturbation.

\subsection{\bf Collapse time and the MF}

In the previous subsections, we have studied the evolution of different mass elements and 
the calculation of collapse time. In fact, the problem of finding an expression for the MF is strictly connected 
to that of finding realistic estimates of collapse
times of generic mass elements. Knowing this last, by means of statistical methods (as we shall see later) it is possible 
to obtain an expression for the MF. 
In the case of the spherical model (SPH), Zel'dovich approximation and ellipsoidal model, 
%the collapse times \bc\ have been calculated for every
%point: spherical (hereafter SPH), Zel'dovich (ZEL), second-order (2ND)
%third-order (3RD) and ellipsoidal (ELL) ones.  SPH, ZEL and ELL
collapse times can be calculated analytically: in the case of spherical collapse (SPH) is simply
1.69/\dl, in the case of Ze'ldovich approximation (ZEL) is 1/\ltre, while the ellipsoidal collapse (ELL) can be calculated by finding the
smallest positive root of Eq. (3.24) of M98
%(\ref{eq:ell_collapse}), 
and by correcting for quasi spherical collapses as in Eq. (3.27) of M98
or using the analytical model of section 2.6.  
%(\ref{eq:ell_corr})???????????????????.  
Second (2ND) and third (3ND) order 
\footnote{The Zel'dovich approximation is
the first term of a perturbative series; the perturbed quantity is not
the density, as in Eulerian perturbation theory, but the displacement
of the particles from the initial position. 
The problem of the evolution of a self-gravitating fluid can be
reformulated in terms of equations for the displacement field.
The equations for the displacement field have been found by
several authors: Buchert (1989), Bouchet et al. (1995) and Catelan
(1995). The Lagrangian system, (see Eqs. 1.42 and 1.43 in M98), can be
perturbatively solved for small displacements. This has been done by
Buchert (1989), Moutarde et al. (1991), Bouchet et al. (1992), Buchert
(1992), Buchert \& Ehlers (1993), Lachi\`eze-Rey (1993a,b), Buchert
(1994), Bouchet et al. (1995) and Catelan (1995); see also Bouchet
(1996) and Buchert (1996) for reviews. The first order solution, for
suitable initial conditions (as given by the linear growing mode), is
the well-known Zel'dovich (1970) approximation.
The perturbative series has been calculated up to third order (from this comes the notation 3RD).
}
collapse times have been found by
looking for the first instant at which $J<0$, then using conventional
root-finding algorithms to find the accurate value. 
%As a matter of
%fact, it is possible that the Jacobian determinant $J$ returns
%positive just after having become negative; a very small number of
%such events, with an incidence of a few times 10$^{-5}$, were missed
%by the algorithm used in the calculations.

%\section{Inverse collapse times}

Collapse times present the disadvantage that the passage from
collapse to non-collapse takes place through infinity. It is more
convenient to define the quantity $F(\mq;\mres)$ as the inverse
collapse time of the point \q, when the initial field is smoothed at a
resolution (mass variance) \res:

\be F(\mq;\mres)=\frac{1}{b_c(\mq;\mres)} \label{eq:inverse_coll} \ee

\noindent 
$F$ is the basic dynamical quantity needed to construct the MF. In the
SPH case $F$ is simply proportional to the linear density contrast,
$F=\delta_l/1.69$; in the Zel'dovich (ZEL) case $F$ is just equal to
\luno.

The quantity $F$ is obviously a non-Gaussian process, and it is not a
simple non-linear function of a Gaussian field (such as, for instance,
a lognormal distribution): it is a complicated non-linear and non-local
functional of the whole initial Gaussian perturbation field. 
%As shown in Chapter 2, 
Its 1-point PDF is the minimal amount of statistical
information needed to construct the MF.
%; this point will be further discussed in \S 4.4.

\begin{figure}
\begin{center}
\epsfig{file=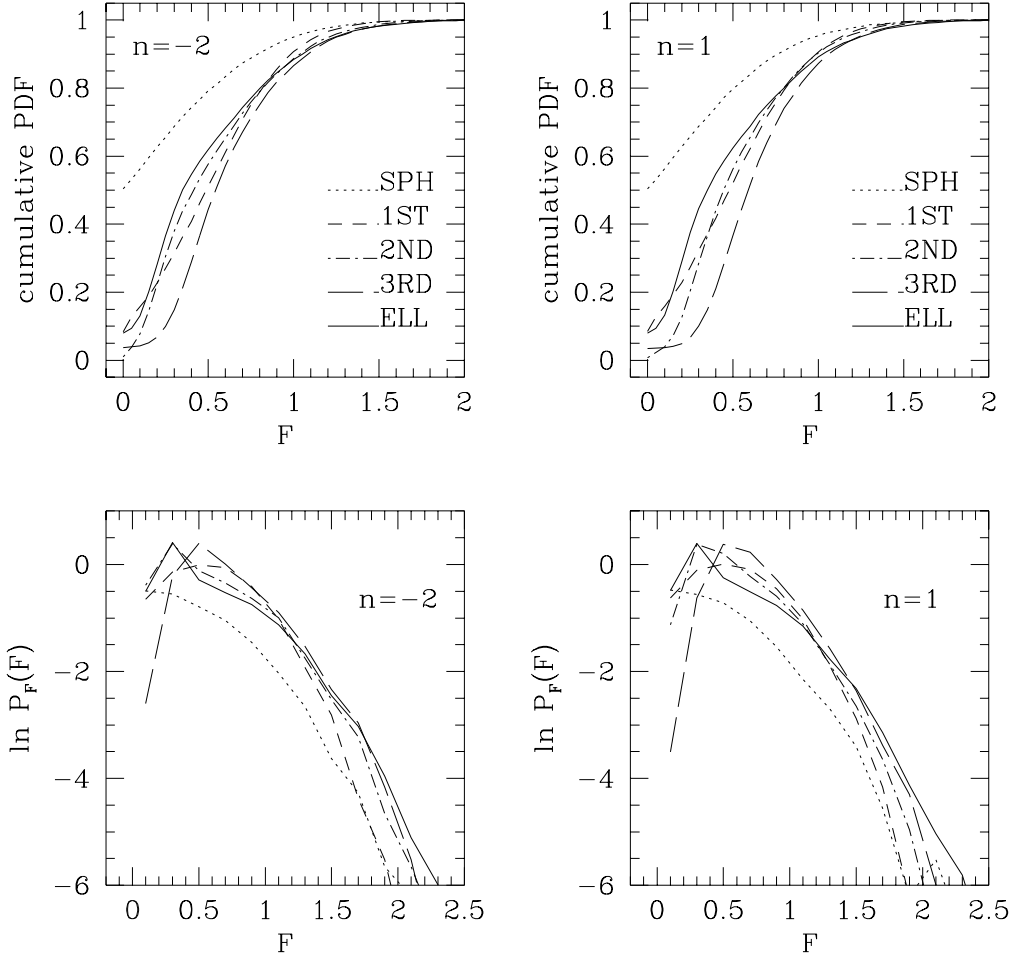,width=14cm}
\caption{Cumulative and differential PDFs of $F$. Figure taken from M98.}
\end{center}
\end{figure}

The 1-point PDFs for the $F$ processes, relative to different
dynamical predictions, can be estimated by means of the Monte Carlo
realizations (see M98).  
%described in last section. 
Fig. 11 
%3.10????? 
shows such PDFs for
two different spectral indexes, namely $n=-2$ and 1. Both cumulative
and differential curves (the latters in logarithm scale) are shown.
%:
%cumulative curves, being binning-free, are less noisy and directly
%show the total fraction of mass which lies in a certain range of $F$,
%while differential curves better represent the behavior of the various
%PDFs, especially at large $F$ values. 

In the following, the ELL and 3RD predictions will be considered, and
the PDFs will be mediated over four the spectral indexes, namely
$n=-2$, $-1$, 0 and 1.

\begin{figure}
\centerline{\hbox{
\psfig{file=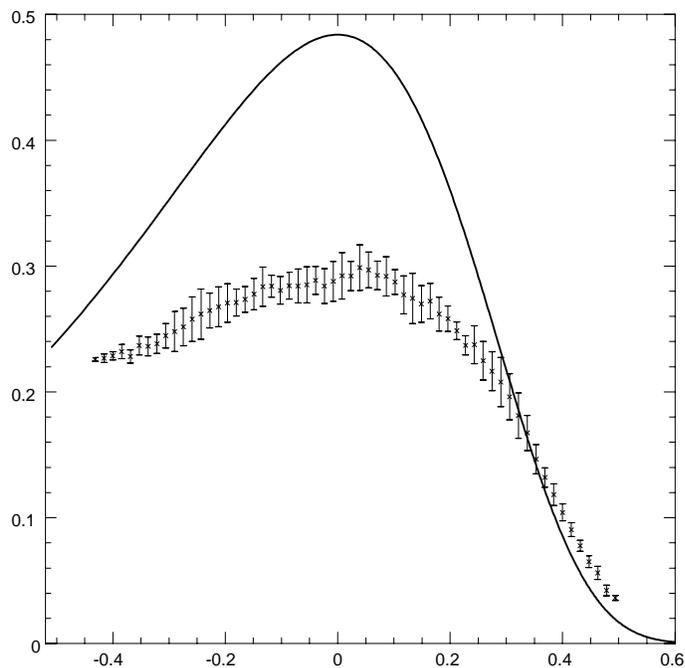,width=10cm}
}}
\caption[]{Comparison of PS multiplicity function with simulations.  
In the plot the solid line represents the multiplicity function
obtained by PS, solid line, and the numerical multiplicity function obtained by YNY. Figure taken from Del Popolo (2005).
}
\end{figure}

%\begin{figure}
%\begin{center}
%\epsfig{file=fig3_12.ps,width=7cm}
%\caption{$F$-PDFs for  ELL and 3RD, with Gaussian and complete fits.}
%\end{center}
%\end{figure}

Summarizing, the MF problem can be formulated in terms of a non-Gaussian process
$F$, representing the inverse collapse time of a generic mass
element. If linear theory with a threshold (or spherical collapse) is
considered, $F$ is proportional to the linear density contrast, and is
then Gaussian. More realistic estimates of collapse times lead to $F$
processes which have a different statistical behavior, even in the
rare event tail.

%\begin{flushleft}
\section{\Large Mass function.}
%\end{flushleft}

\subsection{\bf PS MF}

%One of the most important quantities in cosmology is 
The mass function or multiplicity function can be described by the relation:
\begin{equation}
dN = n( M ) dM
\end{equation}
that is the number of objects per unit volume, having a mass in the range $ M $ ed $ M + dM $. The multiplicity function can also be used to define the luminosity function after having fixed the ratio $ \frac{ M}{ L} $.
Obtaining the mass function starting from that of luminosity is complicated since the ratio $ \frac{ M }{ L } $
is known with noteworthy uncertainty and it is different for different objects and moreover the luminosity function for objects like galaxies depend on the morphological type  (Binggeli \& Tamman 1985). Finally trying to determine the luminosity function observatively is problematic (see for example G.Efstathiou, R.S.Ellis, B.A.Peterson 1988).\\
For the above reasons, the theoretical determination of the mass function is very important. 

The theoretical derivation of the mass function of gravitationally bound objects has been pioneered by Press \& Schechter (1974). 
%[Press W. \& Schechter P., 1974, ApJ, 187, 425]. 
To incorporate the dynamics of the structure, PS formalism adopted top-hat spherical model, according to which the collapse condition for forming massive objects is determined purely by its local average density. To make statistical predictions, however, the PS formalism assumes that initial density field is Gaussian and that massive objects form in the peaks of the density field.

 This theory is based upon these hypotheses:
\begin{itemize}
\item
The linear density field is described by a stochastic Gaussian field. The statistics of the matter distribution is Gaussian.
\item 
The evolution of density perturbations is that described by the linear theory. Structures form in those regions where the overdensity linearly evolved and filtered with a top-hat filter exceeds a threshold $ \delta_{c} $ ($ \delta_{c} =1.68 $, obtained from the spherical collapse model (Gunn \& Gott 1972)). 
\item 
for $ \delta \geq \delta_{c} $ regions collapse to points. The probability that an object forms at a certain point is proportional to the probability that the point is in a region with $ \delta \geq \delta_{c} $ given by:
\begin{equation}
P ( \delta , \delta_{c} )=
\int_{\delta_{c}}^{\infty}d \delta \frac{1}{\sigma
(2 \pi)^{\frac{1}{2}}}
 exp \left(-\frac{\delta^{2}}{2 \sigma^{2}}\right)
\end{equation}
The multiplicity function is given by:
\begin{equation}
\rho ( M, z )= - \rho_{0}\frac{ \partial P}{\partial M } dM =
n( M ) M dM \label{eq:ps}
\end{equation}
\end{itemize}
%To the previous assumptions it is necessary to 
If we add the conditions
$ \Omega=1 $, $ \left|\delta_{k}\right|^{2} \propto k^{n}$,
the Press-Schechter solution is autosimilar and has the form:
\begin{eqnarray}
\rho ( M, z) =\frac{ \rho}{\sqrt{2 \pi}}
\left(\frac{n+3}{3}\right) \left( \frac{M}{M_{\ast }}( z)
\right)^{\frac{n+3}{6}} \nonumber \\
\times exp \left[ -\frac{1}{2} \frac{M}{M_{\ast}}( z )^{
\frac{n+3}{3}}\right]\frac{dM }{M}
\end{eqnarray}
where $ M_{\ast}(z ) \propto (1+z)^{-\frac{6}{n+3}} $. 
Several are the problems of the theory: 
\begin{itemize}
\item 
{\bf Statistical problems:} in the limit of vanishing smoothing radii,
or of infinite variance, the fraction of collapsed mass, 
asymptotes to 1/2. This is a signature of
linear theory: only initially overdense regions, which constitute half
of the mass, are able to collapse. Nonetheless, underdense regions can
be included in larger overdense ones, or, more generally,
non-collapsed regions have a finite probability of being included in
larger collapsed ones; this is commonly called {\it cloud-in-cloud
problem}.  PS argued that the missing mass would accrete on the formed
structures, doubling their mass without changing the shape of the MF;
however, they did not give a true demonstration of that. Then, they
multiplied their MF by a ``fudge factor'' 2. Other authors (see
Lucchin 1989) used to multiply the MF by a factor $(1+f)$, with $f$
denoting the fraction of mass accreted by the already formed
structures.
\item
{\bf Dynamical problems:} the heuristic derivation of the PS MF
bypasses all the complications related to the highly non-linear
dynamics of gravitational collapse. Spherical
collapse helps in determining the $\delta_c$ parameter and in identifying
collapsed structures with virialized halos. However, the PS procedure
completely ignores important dynamical elements, such as the role of
tides and the transient filamentary geometry of collapsed structures.
Moreover, supposing that every structure virializes just after
collapse is a crude simplification: when a region collapses, all its
substructure is supposed by PS to be erased at once, while in
realistic cases the erasure of substructures is connected to the
two-body interaction of already collapsed clumps, an important piece
of gravitational dynamics which is completely missed by the PS
procedure.
\item
{\bf Geometrical problems:} to estimate the mass function from
the fraction of collapsed mass at a given scale it is
necessary to relate the mass of the formed structure to the resolution

In practice,
the true geometry of the collapsed regions in the Lagrangian space
(i.e. as mapped in the initial configuration) can be quite complex,
especially at intermediate and small masses; in this case a different
and more sophisticate mass assignment ought to be developed, so that
geometry is taken into account. For instance, if structures are
supposed to form in the peaks of the initial field, a different and
more geometrical way to count collapsed structures could be based on
peak abundances.
\end{itemize}

%{\bf
Improvements of the model are due to several authors, Heavens and Peacock (1989), Peacock \& Heavens (1990), BCEK, strictly connected to the lacking factor of 2 in the theory, which 
has been shown to be related to the so called "cloud in cloud" problem.
Heavens and Peacock (1989) took into considerations the underdense regions ($ \delta < \delta_{c} $).  
Filtering a density field,  $ \delta (x) $ on a given scale $ R_{f} $, one obtains a set of points exceding the threshold $ \delta_{c} $: this 
set is named "excursion set" (see next subsection) of the filtered field (Adler 1981). In this field is possible to identify objects using special criteria (see Apple \& Jones 1990). An object who had a given value $ \delta>\delta_{c} $, at a fixed $ R_{f} $, shall have a smaller value $ \delta=\delta_{c} $, for a larger $ R_{f} $. If $ R_{f} $ is 
furtherly increased it shall disappear under the threshold. As a consequence, objects belonging to the smaller scale of clustering shall be englobed in the larger one and as a consequence half of the mass is accounted.
Heavens and Peacock (1989) solved the problem taking into consideration also the underdense regions with  $ \delta < \delta_{c} $.  
They used, differently from PS, the relation:
\begin{equation}
p( >R_{f} ) = p_{G} ( \delta >\delta_{c} )+ 
\int_{-\infty}^{\infty} \frac{d p_{G} }{ d \delta} 
p_{up} (\delta_{c}, \delta ) d \delta
\end{equation}
where $ p_{G} $ is the Gaussian distribution used by PS. This relation divides the points into two classes: those going from $ \delta > \delta_{c} $ to $ \delta = \delta_{c} $, when $ R_{f} $ is increased, and they are associated to objects of radius $ >  R_{f} $, and points under the threshold having the probability $ p_{up} $ that in a following filtering, they have $ \delta >\delta_{c} $. Using the relation for $ p_{up} $ obtained by Heavens and Peacock (1989) one gets:
\begin{equation}
p ( >R_{f} ) = 2 p_{G}
\end{equation}
which solves the problem of the fudge factor of 2 in PS.

In next subsections a wider description of the improvements to the PS theory are summarized.

Despite all of its problems, the PS procedure for a long time proved successful, as
compared to N-body simulations, and a good starting point for all the
subsequent works on the subject (Efstathiou et al. 1988; Efstathiou \& Rees 1988, 
BCEK), 
%Brainerd \& Villumsen (1992), 
White, Efstathiou \& Frenk 1993,
%Ma \& Bertschinger (1994), 
Jain \& Bertschinger 1994, 
%Gelb \& Bertschinger (1994), Katz, Quinn, Bertschinger \& Gelb (1994),
Lacey \& Cole 1994, Efstathiou 1995, 
%Klypin \& Rhee (1994), Klypin
%et al. (1995), 
Bond \& Myers 1996b, Governato et al. 1999).

Although the analytical framework of the PS model has been greatly refined and extended (see next subsection), more recently, it has been shown that the PS mass function, while qualitatively correct, disagrees with the results of 
high resolution N-body simulations. In particular, the PS formula overestimates the abundance of haloes near the characteristic mass 
$M_{\ast}$ and underestimates the abundance in the high mass tail (Efstathiou et al. 1988; White, Efstathiou \& Frenk 1993; Lacey \& Cole 1994; Tozzi \& Governato 1998; Gross et al. 1998; Governato et al. 1999; YNY). The quoted discrepancy 
is not surprising since the PS model, as any other analytical model, should make several assumptions to get simple 
analytical predictions. In Fig. 12, it is plotted 
a comparison of PS multiplicity function with simulations.  
In the plot the solid line represents the multiplicity function
obtained by PS, solid line, and the numerical multiplicity function obtained by YNY. 
The plot shows what is well known, namely: 
%it is well known that 
the PS mass function, while qualitatively correct, disagrees with the results of 
N-body simulations. In particular, the PS formula overestimates the abundance of haloes near the characteristic mass 
$M_{\ast}$ and underestimates the abundance in the high mass tail (Efstathiou et al. 1988; 
%White, Efstathiou \& Frenk 1993; 
Lacey \& Cole 1994; Tozzi \& Governato 1998; Gross et al. 1998; Governato et al. 1999). 

%
%\begin{figure}
%\begin{center}
%\hbox{
%\epsfig{file=fig2_1a.ps,width=7cm}
%\epsfig{file=fig2_1b.ps,width=7cm}}
%\caption{PS mass function: (a) $n=-2$, (b) $n=1$.}
%\end{center}
%\end{figure}
%

\subsubsection{\bf The cloud-in-cloud problem}

%{\bf  DA MONACO

\begin{figure}
\begin{center}
\epsfig{file=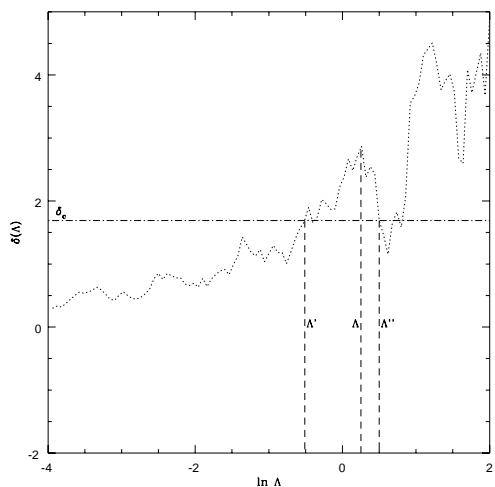,width=7cm}
\caption{The absorbing barrier problem. Figure taken by M98.}
\end{center}
\end{figure}

The cloud-in-cloud problem has origin in the following inconsistency
of the original PS procedure. A collapse prediction is given to any
point (of the Lagrangian space) {\it for any resolution}
\res; in other words, a whole trajectory in the $\delta_l$--\res\
plane is given to any point, as in Fig. 13. Such trajectories start
from 0 at \res=0 (vanishing resolution implies complete smoothing, and
then vanishing density contrast), then wander around the plane,
eventually upcrossing or even downcrossing the threshold line
$\delta_l=\delta_c$.  When a trajectory lies above the threshold, the
point is assumed to be part of a collapsed region of radius $R'\geq
R(\mres)$; it is clear that the size $R'(\mres')$ of the formed
structure is connected to the point $\mres'$ of first upcrossing of
the trajectory. On the other hand, when a trajectory downcrosses the
barrier at a resolution $\mres''$, the point is interpreted as not
included in any region of size $>R''(\mres'')$ ($R$ decreases with
increasing \res), which is clearly in contradiction with what stated
above.

To solve the cloud-in-cloud problem, a point whose trajectory has
experienced its first upcrossing of the threshold line has to be
considered as collapsed at that scale, regardless of its subsequent
downcrossings. This can be done as follows: an absorbing barrier is
put in correspondence with the threshold line, so as to eliminate any
downcrossing event (BCEK). Alternatively, a non-collapse
condition can be formulated as follows: a point is not collapsed at
\res\ if its density contrast {\it at any} $\mres'<\mres$ is below the
threshold (Peacock \& Heavens 1990).

The mathematical nature of the problem, and the resulting MF, strongly
depend on the shape of the filter.  For general filters, trajectories
are strongly correlated in \res, and then all the N-point correlations
of the process at different resolutions have to be known to solve the
problem. However, if the smoothing filter sharply cuts the density
field in the Fourier space, then independent modes are added as the
resolution changes, and the resulting trajectories are Gaussian random
walks. Such kind of filter is commonly called {\it sharp} $k$-{\it
space filter}; it will be referred to as SKS filter throughout the
text.

In the SKS case, the problem is suitably solved within the diffusion
framework proposed by BCEK (see next subsection).

%}

\subsection{\bf Excursion set approach}

%MODELLO BOND
%\section{Excursion set approach}

In this section, we review the formulation of the excursion set
approach, mostly following the treatment given by BCEK.
The terminology ``excursion set approach'' was introduced by BCEK, to indicate that the MF determination is based on the
statistics of those regions in which the linear density contrast \dl\
is larger than a threshold \dc\ (such regions are called excursion sets
in the theory of stochastic processes; see, e.g., Adler 1981).  The PS
procedure is clearly included in this approach.  This section presents
those works which are based on the excursion set approach.

\subsubsection{Langevin equation}

%Let us assume that the linear density fluctuations form a homogeneous
%and isotropic Gaussian random field $\delta({\bf x},z)$, with $\bf x$
%the Lagrangian comoving position and $z$ the redshift. In linear theory
%one can factor out the redshift dependence as $D(z)\de(\bfx)$, where
%$D(z)$ is the linear growth factor (e.g. Peebles 1980). With the
%normalization $D(z=0)=1$, $\de(\bfx)$ becomes the mass density
%fluctuation linearly extrapolated to the present epoch.

As previously described, the statistical properties of the Gaussian field $\de(\bfx)$ are
completely specified by the two-point function in Fourier space, which
is related to the power-spectrum $P(k)$ by $\lan \fde (\bfk_1)\, \fde
(\bfk_2)\ran = (2\pi)^3 \ded (\bfk_1+\bfk_2) P(k_1)$, where $\ded$
represents the Dirac delta function, and the brackets $\lan \cdot \ran$
denote ensemble averaging. Our Fourier transform convention is
$\fde(\bfk)=\int d\bfx\,\de(\bfx)\,{\rm e}^{i\bfk\cdot\bfx}$.

We want now to study the statistical properties of the density
fluctuation field at some resolution scale $R_f$. This is introduced by
convolving $\de(\bfx)$ by some filter function $W(|\bfx'-\bfx|, R_f)$,
\be
\de(\bfx, R_f)=\int d\bfx'\,W(|\bfx'-\bfx|, R_f)\,\de(\bfx')=
\f{1}{(2 \pi)^3}
\int d\bfk\, \fW(kR_f)\,\fde(\bfk)\,
{\rm e}^{-i\bfk\cdot\bfx}\;,
\label{filtra}\ee
where $\fW$ is the Fourier transform of the filter.  At each point
$\bfx$ the smoothed field represents the weighted average of
$\de(\bfx)$ over a spherical region of characteristic dimension $R_f$
centred in $\bfx$. The detailed properties of $\de(\bfx, R_f)$ clearly
depend upon the specific choice of window function.  The most commonly
used smoothing kernels are the top-hat filter $W_{TH}(|\bfx|,
R_f)=3\,\Theta(R_f-|\bfx|)/4\pi R_f^3$, where $\Theta(x)$ is the
Heaviside step function, and the Gaussian one $W_{G}(x,R_f)=(2\pi
R_f^2)^{-3/2}\exp(-x^2/2R_f^2)$.  Recently, for convenience of
analysis, top-hat filtering has been also applied in momentum space
$\fW_{SKS}(k, R_f)= \Theta (k_f-k)$, where $k_f=1/R_f$ and
$k_f=|\bfk_f|$. This kernel is generally called sharp $k$-space filter.
While it is easy to associate a mass to real space top-hat filtering
$M_{TH}(R_f)=4\pi\rho_b R_f^3/3$, there is always a bit of
arbitrariness in assigning a mass to the other window functions.  The
most common procedure is to multiply the average density by the volume
enclosed by the filter, obtaining $M_{G}(R_f)= (2\pi)^{3/2}\rho_b
R_f^3$ and $M_{SKS}(R_f)=6\pi^2\rho_b k_f^{-3}$ (Lacey \& Cole 1993 (LC93)).
An alternative procedure, originally introduced by Bardeen \etal
(1986) (BBKS), corresponds to the choice $M_{SKS}(R_f)= 4\pi \rho_b
R_{TH}^3/3$, where $R_{TH}$ is chosen by requiring
$\sigma^2_{SKS}(R_f)=\sigma^2_{TH}(R_{TH})$, and similarly for the
Gaussian filter. In this way one obtains good agreement with numerical
simulations of clustering growth (Lacey \& Cole 1994).
 
%In hierarchical bottom-up models of structure formation at each epoch
%mass fluctuations have typically undergone non-linear collapse up to
%some scale $R_f$, and the first objects that form are those of lower
%mass. Haloes of larger mass arise by the merging and accretion of
%subunits. The mass distribution deriving from this hierarchical halo
%formation process has been successfully modeled by Peacock \& Heavens
%(1990), Cole (1991), BCEK and Lacey \& Cole (1993).  

In order to mimic
the accretion of matter one consideres a full hierarchy of
decreasing resolution scales $R_f$ (Peacock \& Heavens
1990, Cole 1991, BCEK and LC93). The effect of varying $R_f$ can be
obtained by differentiating eq. (\ref{filtra})
\be
\label{lan}
{\partial \de(\bfx,R_f) \over \partial R_f}=
{1 \over (2\pi)^3} \int d\bfk\,\fde(\bfk)\,
{\partial \fW(kR_f) \over \partial R_f}\,
{\rm e}^{-i\bfk\cdot\bfx} \equiv \eta(\bfx, R_f) \;.
\ee
This has the form of a Langevin equation, which shows how an
infinitesimal change of the resolution scale $R_f$ affects the value of
the density fluctuation field $\delta(\bfx, R_f)$ in the given position
$\bfx$ through the action of the stochastic force $\eta(\bfx, R_f)$. In
the limit $R_f \to \infty$ one has $\de(\bfx;R_f)\to 0$, which can be
adopted as initial condition for our first-order stochastic
differential equation.  Thus, by solving it, we can associate to each
point $\bfx$ a trajectory $\delta(\bfx, R_f)$ obtained by varying the
resolution scale $R_f$. 
%Notice that, since eq. (\ref{lan}) is linear,
%$\eta(\bfx, R_f)$ is also a zero mean Gaussian random field, uniquely
%specified by its auto-correlation function
%
%\be
%\langle \eta(\bfx_1,R_{f1})\, \eta(\bfx_2,R_{f2}) \rangle = {1\over
%2\pi^2} \int_0^\infty dk_1\, k_1^2\, P(k_1)\, 
%{\p \fW(k_1R_{f1})\over \p R_{f1}}\,
%{\p \fW(k_1R_{f2}) \over \p R_{f2}}\,
%j_0(k_1r)\;,
%\label{eq3}
%\ee
%
%where $r=|\bfx_1-\bfx_2|$ and $j_0(x)$ is the zeroth order spherical
%Bessel function. 
Trajectories associated to different neighbouring
points will be statistically influenced by the correlation properties
of the force $\eta(\bfx, R_f)$, i.e. of the underlying Gaussian field
$\de(\bfx)$. On the other hand the coherence of each trajectory along
the $R_f$ direction depends exclusively on the analytic form of the
filter function and vanishes for the sharp $k$-space window
(BCEK). With such a filter, by decreasing the smoothing length one adds
up a new set of Fourier modes of the unsmoothed distribution to
$\delta(\bfx, R_f)$. For a Gaussian field this is completely
independent of the previous increments, and each trajectory
$\delta(\bfx, R_f)$ becomes a Brownian random walk.

In the case of sharp $k$-space filtering the notation greatly
simplifies if we use as time variable the variance of the filtered
density field, $\Lambda \equiv\sigma^2(k_f) = \lan \de(k_f)^2\ran=
(2\pi^2)^{-1} \int_0^{k_f}dk\, k^2\, P(k)$.  In such a case the
stochastic process reduces to a Wiener one, namely
\be \f{\p \de(\bfx, \Lambda)}{\p\Lambda} = \zeta(\bfx, \Lambda) \;,
\label{eq4}
\ee
with $\langle \zeta(\bfx, \Lambda) \rangle=0$ and
\be
\langle \zeta(\bfx, \Lambda_1) \,\zeta(\bfx, \Lambda_2) \rangle = \ded
(\Lambda_1-\Lambda_2)
\label{eq5}
\ee
[see eq. (17) for the spatial correlation]. In the following we will
adopt $\Lambda$ as time variable, unless explicitly stated.  The
solution of the Langevin equation (\ref{eq4}) in an arbitrary point of
space (the position $\bfx$ is here understood), with the initial condition
$\de(\Lambda=0)=0$, is simply $\de(\Lambda)=\int_0^\Lambda
d\Lambda^\prime \zeta(\Lambda ^\prime)$.  By ensemble averaging this
expression one obtains $\langle \de(\Lambda) \rangle= 0$ and $\langle
\de(\Lambda_1) \de(\Lambda_2) \rangle = {\rm min} (\Lambda_1,
\Lambda_2)$, which uniquely determine the Gaussian distribution of
$\de(\Lambda)$. 

%\subsection{Press-Schechter mass function}

As already told PS model is intrinsically
flawed by the cloud-in-cloud problem, namely the fact that a
fluctuation on a given scale can contain substructures of smaller
scales and the same fluid elements can be assigned, according to the PS
collapse criterion, to haloes of different mass.  Moreover, in a
hierarchical scenario, one expects to find all the mass collapsed in
objects of some scale, while the PS model can account only for half of
it: this problem is intimately related to the fact that in a Gaussian
field only half volume is overdense.  Press and Schechter faced the
problem by simply multiplying their result by a fudge factor of 2.  In
this section we review how the Langevin equations introduced above can
be used to extend the PS theory in such a way to solve both
problems.

The solution of the cloud-in-cloud problem has been given by Peacock \&
Heavens (1990), Cole (1991) and BCEK. Their approach consists in
considering at any given point the trajectory $\delta(R_f)$ as a
function of the filtering radius, and then determining the {\it
largest} $R_f$ at which $\delta(R_f)$ upcrosses the threshold
$t_f(z_f)$ corresponding to the formation redshift $z_f$.  This
determines the largest mass collapsed at that point, all sub-structures
having been erased. So, the computation of the mass function is
equivalent to calculating the fraction of trajectories that first
upcross the threshold $t_f$ as the scale $M$ decreases. The solution of
the problem is enormously simplified for Brownian trajectories, that is
for sharp $k$-space filtered density fields. In such a case one only
has to solve the Fokker-Planck equation for the probability density
$\calW(\de, \Lam)\,d\de$ that the stochastic process at $\Lam$ assumes
a value in the interval $\de, \de+d\de$,
\be
\f{\p \calW(\de, \Lam)}{\p\Lam} = 
\f{1}{2}\,\f{\p^2\calW(\de,\Lam)}{\p\de^2}\; ,
\label{fp}
\ee
with the absorbing boundary condition $\calW(t_f, \Lam)=0$ and initial
condition $\calW(\de, 0)=\de_D(\de)$. The solution is well known
(Chandrasekhar 1943)
\be
\calW(\de, \Lam; t_f)\,d\de =
{1\over \sqrt{2\pi\Lambda}} \left[ \exp \left(-{\delta ^2 \over 2\Lambda}
\right) - \exp \left(-{(\delta -2t_f)^2 \over 2\Lambda} \right)
\right]\,d\de \;.
\label{chandra}
\ee
Defining $\,S(\Lambda, t_f)= \int _{-\infty}^{t_f} \!d\delta\,{\cal
W}(\delta, \Lambda, t_f)\,$ as survival probability of the
trajectories, one obtains the density probability distribution of
first-crossing variances by differentiation
\be
{\cal P}_1 (\Lambda) =-{\p S(\Lam, t_f) \over \p \Lam} =
-{\p  \over \p \Lam}\int_{-\infty}^{t_f}\!d\de\, 
\calW(\de,\Lam; t_f) =
\left[- {1\over 2} {\partial \calW(\delta,\Lambda, t_f)
\over \partial \delta}
\right] _{-\infty}^{t_f}=
{t_f \over \sqrt {2\pi \Lambda^3} }
\exp \left( -{t_f^2 \over 2 \Lambda} \right)\;.
\label{Smir}
\ee
The function ${\cal P}_1(\Lambda) \,d\Lambda$ yields the probability
that a realization of the random walk is absorbed by the barrier in the
interval $(\Lambda, \Lambda+d\Lambda)$ or, by the ergodic theorem, the
probability that a fluid element belongs to a structure with mass in
the range [$M(\Lambda+d\Lambda), M(\Lambda)$]. Finally, the comoving
number density of haloes with mass in the interval $[M, M+dM]$
collapsed at redshift $z_f$ is
\be
\label{genps}
n(M,z_f)\, dM = {\rho_b \over M}\, {\cal P}_1(\Lambda) \,
\left| {d\Lambda \over dM} \right|\, dM \;.
\ee
Inserting the expression of ${\cal P}_1(\Lambda)$ of eq. (\ref{Smir})
in the latter equation one obtains the well-known PS expression for the
mass function
\be
\label{ps}
n(M,z_f)\, d M =
{\rho_b \,t_f(z_f) \over \sqrt {2\pi}}\,
{1\over M^2 \sqrt{\Lam(M)} } \,
\left| {d\ln \Lambda \over d\ln M} \right|\;
\exp \left( - {t_f(z_f)^2 \over 2 \,\Lambda (M)} \right) \,dM\;.
\ee
The original fudge factor of 2 of the PS approach is now naturally
justified. 

Previous investigations (e.g. Peacock \& Heavens 1990; BCEK) have shown
that only for sharp $k$-space filtering it is possible to write an
analytic formula for the mass function obtained from the excursion set
approach.  Numerical solutions of the cloud-in-cloud problem with
physically more acceptable smoothing kernels like Gaussian and top-hat
result in mass functions that are a factor of two lower in the
high-mass tail and have different small-mass slopes compared with the
PS formula.  The standard interpretation of this result is that the
excursion set method is not reliable for $M\ll M_\ast$, where $M_\ast$,
defined by $\Lambda(M_\ast)=t_f^2$, is the typical mass collapsing at
$z_f$.

\begin{figure}
%\centerline{\epsfxsize= 17 cm  \epsfysize= 8 cm \epsfbox{fig1.ps}}
%
\centering{
\vbox{
%\hbox{
\psfig{figure=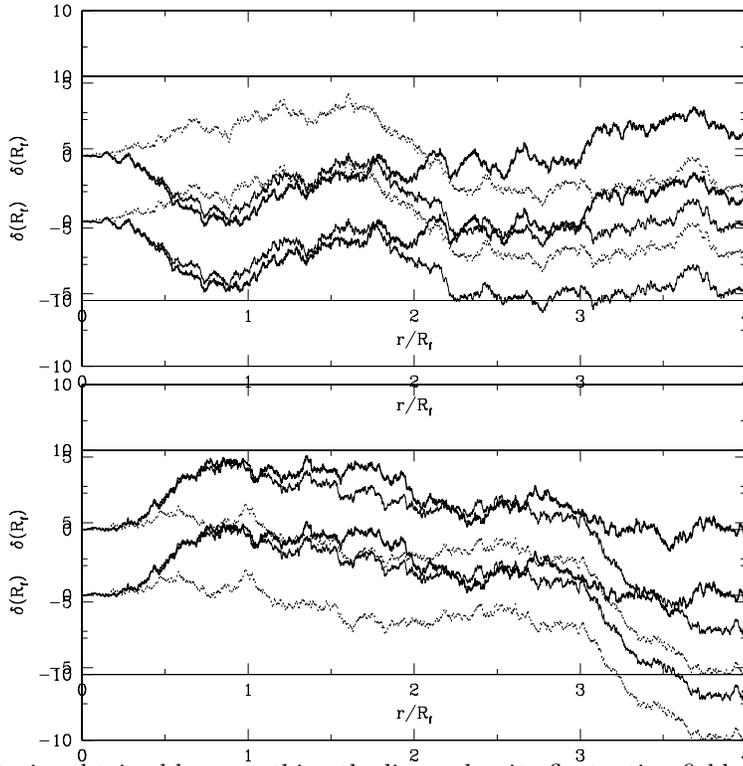,bbllx=21pt,bblly=472pt,bburx=568pt,bbury=693pt,clip=t,width=10cm}
\psfig{figure=fig1.ps,bbllx=21pt,bblly=165pt,bburx=568pt,bbury=422pt,clip=t,width=10cm}
%}
}
}
\caption{Examples of trajectories obtained by smoothing the linear
density fluctuation field with a series of sharp $k$-space filters
(with decreasing resolution scales $R_f=1/k_f$) at two points separated
by the distance $r$. Here we consider $r=4 \,{\rm Mpc}$ and a CDM power
spectrum (with density parameter $\Omega=1$ and present Hubble constant
$H_\0= 50 \vel\,{\rm Mpc}^{-1}$) linearly extrapolated to $\sigma_8=1$.
The {\it heavy continuous} line represents the trajectory associated to
a point $\bfx$ in Lagrangian space.  The trajectory associated to the
point $\bfy$ such that $ |\bfy - \bfx| = r$ is plotted with a {\it
light continuous} line.  The {\it dotted} line is obtained by: {\em i})
considering the trajectory associated to $\bfy$, {\em ii}) artificially
removing any correlation with the trajectory at $\bfx$ and {\em iii})
suitably rescaling the result.  In practice, the dotted line represents
a trajectory which is completely independent of the one associated to
$\bfx$, but which has the same statistical one-point properties.  This
is what is generally used in building up merger trees of dark matter
haloes.  
%However, the actual trajectories associated to neighbouring
%points are strongly correlated when the smoothing length $R_f$ is
%comparable to the physical separation between the points $r$.  The same
%trajectories tend to become less and less correlated as $R_f$ decreases. 
%It is interesting to note that a memory of the strong correlation existing for $r/R_f \ll 1$ remains 
%even when $r/R_f \gg
%1$. In fact, when $R_f$ is small compared to $r$, the real trajectory
%associated to $\bfy$ can be accurately approximated by adding a
%constant value to the independent trajectory so that it matches the
%random walk at $\bfx$ when $R_f \magcir\,r$.  In other words, since the
%trajectories are continuous functions of $R_f$, the value assumed at
%$R_f \sim r$ (i.e. at the end of the strongly correlated regime)
%represents the initial condition for the random walk at small $R_f$
%(i.e. in the independent regime).  This derives from a sort of natural
%peak-background split: by evaluating the density fluctuation field in
%two points separated by $r$ one samples the same background value
%(obtained by summing up all the Fourier components corresponding to
%wavelenghts larger than $r$) but different high frequency modes.
}
\label{trajectories}
\end{figure}
%

%...................

\subsubsection{\bf Merging histories}

Merging histories are an important piece of information in the
formation history of dark matter objects. They are a natural outcome
of MF theories: a mass point, found in an object of mass $M$ at a
given time, will be found in another, more massive object at a
subsequent time; from the conditional probabilities connected with
such events it is possible to construct the statistics of accretion
and merging histories of collapsed structures.  This was attempted
first by Carlberg (1990), whose results are in contradiction with more
recent works outlined in the following. Bower (1991) constructed
merging histories by using the PS formalism, obtaining the same
results as that obtained by means of the diffusion formalism, which
are now reported.

\label{sec:ps}
Suppose that we have smoothed the initial density distribution on a scale $R$
using some spherically symmetric window function $W_{M}(r)$, where $M(R)$ is
the average mass contained within the window function. There are various
possible choices for the form of the window function (c.f. LC93). We
use a real-space top-hat window function, $W_M(r) = \Theta(R-r)(4\pi
R^3/3)^{-1}$, where $\Theta$ is the Heaviside step function. In this case $M =
4\pi \rho_0 R^3/3$, where $\rho_0$ is the mean mass density of the
universe. The mass variance $S(M) \equiv \sigma^2(M)$ may be calculated from
\begin{equation}
\label{eqn:massvariance}
\sigma^2(M) = \frac{1}{2\pi^2} \int P(k) W^2(kR) k^2 {\rm d}k \, ,
\end{equation}
where $P(k)$ is the power spectrum of the matter density fluctuation, and
$W(kR)$ is the Fourier transform of the real space top-hat.

The ``excursion set'' derivation due to BCEK leads naturally to the
extended Press-Schechter formalism. The smoothed field $\delta(M)$ is a
Gaussian random variable with zero mean and variance $S$. The value of $\delta$
executes a random walk as the smoothing scale is changed. Adopting an ansatz
similar to that of the original Press-Schechter model, we associate the
fraction of matter in collapsed objects in the mass interval $M, M+{\rm d}M$ at
time $t$ with the fraction of trajectories that make their \emph{first
upcrossing} through the threshold $\omega \equiv \delta_c(t)$ in the interval
$S, S+{\rm d}S$. This may be translated to a mass interval through equation
(\ref{eqn:massvariance}). The threshold $\delta_c(t)$ corresponds to the
critical density at which a pertubation will separate from the background
expansion, turn around, and collapse. It is extrapolated using linear theory,
$\delta_c(t)=\delta_{c,0}/D_{\rm lin}(z)$, where $D_{\rm lin}(z)$ is the linear
growth function.  The halo mass function (here in the
notation of LC93) is then:
\begin{equation}
\label{eqn:ps}
f(S, \omega) {\rm d}S = \frac{1}{\sqrt{2\pi}} \frac{\omega}{S^{3/2}} 
\exp{\left[-\frac{\omega^2}{2S}\right]} {\rm d}S \, .
\end{equation}

The \emph{conditional} mass function, the fraction of the trajectories in halos
with mass $M_1$ at $z_1$ that are in halos with mass $M_0$ at $z_0$ ($M_1 <
M_0$, $z_0 < z_1$) is 
\begin{eqnarray}
\label{eqn:flc}
\lefteqn{f(S_1, \omega_1 \mid S_0, \omega_0) {\rm d}S_1 =} 
\nonumber \hspace{1truecm}\\
 & &  \frac{1}{\sqrt{2\pi}} \frac{(\omega_1-\omega_0)}{(S_1-S_0)^{3/2}}
\exp{\left[-\frac{(\omega_1-\omega_0)^2}{2(S_1-S_0)}\right]} {\rm d}S_1
\,.
\end{eqnarray}
The probability that a halo of mass $M_0$ at redshift $z_0$ had a
progenitor in the mass range $(M_1, M_1+{\rm d}M_1)$ is given by (LC93):
\begin{eqnarray}
\label{eqn:Nlc}
\lefteqn{\frac{{\rm d}P}{{\rm d}M_1}(M_1, z_1 \mid M_0, z_0) {\rm d}M_1 =}
\nonumber \hspace{2.5truecm}\\
& & \frac{M_0}{M_1} f(S_1, \omega_1 \mid S_0, \omega_0) 
\left| \frac{{\rm d}S}{{\rm d}M}\right| {\rm d}M_1\, ,
\end{eqnarray}
where the factor $M_0/M_1$ converts the counting from mass weighting to number
weighting.

We can now derive two more useful quantities. 
%that will be useful later. 
Given the mass
of a parent halo $M_0$ and the redshift step $z_0 \rightarrow z_1$, the
\emph{average} number of progenitors with mass larger than $M_l$ is:
\begin{equation}
\label{eqn:avgnumprog}
\bar{N} \equiv
\langle N_p (M \,\vert M_0) \rangle = \int_{M_l}^{M_0} \, {\rm d}M\,
\frac{M_0}{M}\, \frac{{\rm d}P}{{\rm d}M}(M,z_1 \vert M_0, z_0) 
\,.
\end{equation}
We can also readily calculate the average fraction of $M_0$ that dwelt
in the form of progenitor halos of mass $M>M_l$:
\begin{equation}
\label{eqn:faccbar}
\bar{f}_p =\int_{M_l}^\infty{\rm d}M\, \frac{{\rm d}P}{{\rm d}M}
(M,z_1 \vert M_0,z_0)\,,
\end{equation}
and define the complimentary quantity for the average fraction of $M_0$ that
came from ``accreted'' mass, $\bar{f}_{\rm acc} =1-\bar{f}_p$.

In order to describe the accretion history of an object, it suffices to lower
the barrier in a continuous way, and follow the position in \res\ of
the first upcrossing point: if this performs a discontinuous jump
(which happens when the trajectory goes down and then up again), the
object containing the mass point considered suffers a discontinuous
merging with a structure of comparable size, while if the point moves
continuously the object is just accreting material.  It is clear that,
if the trajectory is a random walk, the first upcrossing point will
always perform discrete jumps; continuous accretion will be recognized
only if an (arbitrary) minimum resolution step is fixed.

LC93 also proposed a Monte Carlo approach to simulate
ensembles of formation histories, based on realizing a large number of
random walks. Their Monte Carlo method for simulating merging
histories is commonly used to model the formation of virialized
galactic halos, in which gas dynamical is inserted ``by hand''. Such
Monte Carlo models of galaxy formation will be discussed in \S 5.1.2.
As a matter of fact, they found a weak inconsistency in their
formalism (a probability density going slightly negative), probably
caused by the simplicistic mass assignment. The same authors
(Lacey \& Cole 1994) compared their results to N-body simulations,
finding an overall satisfactory agreement.

Finally, Sheth (1995,1996) obtained a complete analytical description
for the merging histories of objects formed from a Poisson
distribution of seed masses, the problem analyzed by the original PS
paper. The resulting MF has turned up to be the same as the
distribution function proposed by Saslaw \& Hamilton (1984).

\subsection{\bf The statistics of the collapsed regions}

An idea which traces back to Doroshkevich (1970) is to suppose that
structures form in the peaks of the initial density field.  This idea
became a standard paradigm in the framework of biased galaxy
scenarios. Kaiser (1984) noted that high-level
peaks show an enhanced correlation with respect to the underlying
matter field, a fact which provided an explanation for the large
correlation length of clusters with respect to galaxies, and gave
freedom to tune the normalization of the CDM model to reproduce the
large-scale distribution of galaxies.  Peacock \& Heavens (1985) and
BBKS calculated a number of statistical expectation
values for the peaks of a Gaussian random field, as the number density
of peaks of given height. This quantity seemed suitable to determine a
peak MF, but two important problems, recognized by BBKS 
%(who did not attempt to determine an MF from the peak
%number density) 
hampered such a determination (see next subsection for details).

In order to get the peak MF, it is necessary to 
study the stochastic properties of the collapsed regions, defined as
regions where the linearly evolved density contrast $\delta(M,\bfx)$ exceeds
a threshold $\delta_c$. 
The correct choice of the
threshold $\delta_c$, insofar as it is well defined,
is a matter of debate. Many authors take the value $1.69$
inspired by the spherical collapse model. 
On the other hand, comparison
of the quasi-linear estimate of $n(>M)$ (described below) with estimates
from numerical simulations suggests a smaller value, Carlberg and Couchman
(1989) advocating $\delta_c=1.44$ and Efstathiou and Rees (1988) advocating
$\delta_c=1.33$ (but see also Brainerd and Villumsen (1992)
and Katz, Quinn and Gelb (1992)).
%
%In the collapsed regions, $\delta(M,\bfx)$ is more than $\nu$
%standard deviations above zero, where
%\bea \nu(z,M)\eqa \delta_c/\sigma(z,M) \label{nufirst} \\
%\eqa \delta_c (1+z)/\sigma_0(M) \label{nusec} \\
%\eqa \delta_c\frac{1+z}{1+z\sub{nl}(M)} \label{nudef} \eea
%We are working in the linear regime, corresponding to $\nu
%>\delta_c>1$.
%
The collapsed regions occupy a volume fraction $V$ given by the Gaussian
distribution,
\be \frac{{\rm d}V}{{\rm d}\nu}=\frac1{\sqrt{2\pi}} e^{-\nu^2/2}
\label{dgauss} \ee
leading to
\bea V(\nu)
\eqa\rm{erfc}(\nu/\sqrt2)/2\\
\eqa (2\pi)\mhalf\nu\mone e^{-\nu^2/2}
(1-\nu\mtwo+O(\nu\mfour) ) \label{gauss} \eea
%
%For $\nu=1,2,3,4$ the volume fraction is $V=.16,.023,.0013,.000031$. In
%practice one is not interested in values $\nu\gsim4$, because the collapsed
%regions are then too rare to be physically significant. The corresponding mass
%fraction is about $(1+\delta_c)=2$ to $3$ times bigger than the volume
%fraction.
%
To say more one needs to know the shape of the spectrum of perturbations. We shall list the
relevant results given by BBKS. They involve only two moments of the spectrum,
defined by
\bea
\langle k^{2}(M)\rangle
\eqa \sigma \mtwo(M) \int^\infty_0 k^{2} \exp(-k^2 R_f^2) P(k)
\frac{{\rm d}k}{k} \label{ksqm}
\\
\langle k^{4}(M)\rangle
\eqa \sigma \mtwo(M) \int^\infty_0 k^{4} \exp(-k^2 R_f^2) P(k)
\frac{{\rm d}k}{k} \eea
The quantity $\langle k^2 \rangle$ is the mean
of the $\nabla^2$ operator, ie., of
the quantity $\delta\mone\nabla^2\delta$.
Similarly, $\langle k^4 \rangle$ is the mean
of $\nabla^4$.

A relevant length
scale is defined by $R_*^2=3\langle k^2 \rangle/\langle k^4 \rangle$. For any
spectral index $n>-1$, it is easy to show that in the limit
of small filtering scale $R_f$,
\be \frac{R_*}{R_f}=\rfrac6{1+n} \half \ee
For the case of CDM with $.7<n<1$, the ratio is
in the range $1$ to $3$ for the entire range of cosmologically interesting
masses.

Another relevant length scale
is $\langle k^2 \rangle\mhalf$. On large filtering scales,
such that
$P(k)$is increasing fairly strongly at $k\mone\simeq R_f$,
the ratio $\langle k^2 \rangle\mhalf/R_f$ is close to 1.
As the scale is reduced it increases, but is $\lsim 10 $
for $M>10^6\msun$.

Finally, it is convenient to define the dimensionless parameter
\be \gamma(M)=\langle k^2 \rangle/\langle k^4 \rangle\half \label{gamm} \ee
It falls from about $.7$ to about $.3$ as $M$ decreases from $10^{15}\msun$ to
$10^6\msun$, for $.7<n<1$.

For sufficiently large $\nu$, each collapsed region is a sphere surrounding a
single peak of $\delta$. However, the departure from sphericity is
considerable in the cosmologically interesting regime. BBKS show that a
quantity $x\mone$, which is roughly the fractional departure from sphericity,
is well approximated by
\be
x=\gamma\nu+\theta(\gamma,\gamma \nu)
\ee
where
\be
\theta(\gamma,\gamma \nu)=\frac{3(1-\gamma^2)+(1.216-.9\gamma^4)
\exp[-\gamma/2(\gamma\nu/2)^2]}
{\left[3(1-\gamma^2)+.45+(\gamma\nu/2)^2\right]\half+\gamma\nu/2}
\ee
%The asphericity is plotted in Figure 10 for $M=10^8\msun$ and $10^{14}\msun$,
%for both $n=1$ and $n=.7$, and is seen to be $\gsim .3$ even at $\nu=4$
%and $10^{14} \msun$. 
We emphasize that this is the asphericity seen in
the linearly evolved, filtered density contrast. The asphericity in
the true, unfiltered density contrast will be bigger, and will increase
during collapse.

Three useful number densities are given by BBKS. First, the density $n_\chi$
of the Euler number of the surfaces bounding the collapsed regions is
\be
\frac12 n_\chi(\nu,\langle k^2\rangle )=
\frac{(\langle k^2 \rangle/3)\threehalf}{(2\pi)^2}
(\nu^2-1) e^{-\nu^2/2} \label{nchi}
\ee
Second, the number density of upcrossing points on these surfaces is
\bea
&& n\sub{up}(\nu,\langle k^2\rangle ,\gamma)
=\frac{(\langle k^2 \rangle/3)\threehalf}{(2\pi)^2}\times\nonumber\\
&&\left[ \nu^2-1+\frac{4\sqrt3}{5\gamma^2(1-5\gamma^2/9)\half}
e^{-5\gamma^2\nu^2/18}\right] e^{-\nu^2/2} \label{nup}
\eea
An upcrossing point on a surface of constant $\delta$ is defined as one where
$\del \delta$ points along some arbitrarily chosen reference direction.

The third  number density is $n\sub{peak}$, the number density of peaks
which are more than $\nu$ standard deviations high. BBKS give
expressions for $n\sub{peak}$, but they point out also that in the
cosmologically interesting regime it is quite well approximated by
$n\sub{up}$. We shall use this approximation in what follows. It suggests that
if a collapsed region contains several peaks, they are not buried deep inside
it; rather, the boundary of the region is presumably corrugated, wrapping
itself partially around each peak.

\subsubsection{\bf The number density $n(>M)$}

The main application of these results is to estimate the number density
$n(>M)$ of gravitationally bound systems with mass bigger than $M$, at a given
epoch before $z\sub{nl}(M)$ (where $nl$ stands for "non linear"). The systems
are supposed to be identifiable by looking at
the linearly evolved density contrast $\delta(M,\bfx)$. Each collapsed
region, defined as one in which $\delta(M,\bfx)>\delta_c$, is supposed to
contain one or more systems with mass bigger than $M$.

If each collapsed region is identified with a single system, then
$n(>M)=n\sub{coll}$. In general this recipe is useless for lack of an
expression for $n\sub{coll}$. A different prescription, which does lead to a
calculable expression, is to identify each peak within a collapsed region with
a different collapsed object,
\be n(>M)= n\sub{peak} \simeq n\sub{up} \label{peak} \ee
This estimate (usually without the simplifying second equality) is
widely used in the literature.
It is certainly
the same as the estimate $n(>M)=n\sub{coll}$ for large $\nu$, where we know
that there is just one peak per collapsed region. To what extent the
prescriptions are the same for lower $\nu$ is not known, because the number of
peaks per collapsed region is not known.

By means of the previous quoted theory 
%This quantity 
it seemed suitable to determine a
peak MF, but two important problems, recognized by BBKS, (who did not attempt to determine an MF from the peak
number density) hampered such a determination:  (i) the peak number density
was based on the initial field smoothed at a single scale, so the peak
MF suffered from the same cloud-in-cloud problem (the {\it
peak-in-peak} problem) as the PS one; (ii) it was not clear
which mass had to be assigned to a peak.

The first (i) problem can be easily described as follows.
If at some epoch the linearly evolved density contrast does have many
peaks within a collapsed region, an interesting
situation arises. At a somewhat earlier epoch, $\delta(M,\bfx)$ was smaller,
and a separate contour $\delta(M,\bfx)=\delta_c$ was wrapped around each
peak.
In other words, each peak of the linearly evolved density contrast,
filtered on scale $M$, was inside a single collapsed region, and presumably
represented a separate gravitationally bound system. At the later epoch when
the collapsed region encompasses many peaks of the linearly evolved density
contrast, we have a bigger gravitationally bound system. If the original
systems
survive, the identification of each peak with a separate system is correct,
but it misses the larger system which contains the original systems. Of
course, missing this one system does not affect the total count much, so if
this case is typical of collapsed regions containing many peaks the estimate
$n(>M)=n\sub{peak}$ is better than the estimate $n(>M)=n\sub{coll}$. If, on
the other hand, the original systems have merged, that identification is
wrong, and the whole of the collapsed region should be identified with just
one
gravitationally bound system. If this case is typical, the estimate $n(>M)
=n\sub{coll}$ would be better, if only we had a formula for it. Which
case is the more likely? A clue is provided by the observation made earlier,
that if there are several peaks in a collapsed region they typically seem to
lie near the surface of the region, a part of the surface wrapping itself
around each peak. This picture would suggest that the estimate
$n(>M)=n\sub{peak}$ is the more reasonable, the peaks of the linearly evolved
density contrast in a typical collapsed region representing structures which
have not existed long enough to merge.

Even if the determination of a peak MF has some problems, several authors tried to
%A peak MF can be 
formulate it in different ways, e.g. Colafrancesco et al. (1989). 
Let's 
Using the method of the last paper, let's denote by $N_{\rm pk}(\nu, M) d \nu$ the number density of peaks of heightbetween $\nu$ and 
$\nu$+$d \nu$ and
\begin{equation}
n_{\rm pk}= \int_{\nu_c}^{\infty} N_{\rm pk}(\nu, M) d \nu
\end{equation}

In order to obtain the mass multiplicity for objects belonging to a catalog of contrast $\delta_c$, one 
considers peaks on a hierarchy of resolution scales, M. Following the original PS choice, the mass of 
different objects is identified with the filtering mass $M$. The mass per unit volume $n(M) M dM$, is obtained by differentiating with respect 
$M$ the mass of the peaks of the filtered field with overdensity exceeding $\delta_c$, that is:

%Then, if $M_{\rm pk}(\delta_l,\mres)$ is the
%mass associated to the peak, and if the $M$ variable is used in place
%of \res\ (the uncertainty in the $\mres\rightarrow M$ relation can
%be absorbed in the $M_{\rm pk}$ definition) the fraction of collapsed
%mass can be written as:
%
%\be \mimfm = \frac{1}{\bar{\rho}} \int_{\delta_c}^\infty d\delta_l
%n_{\rm pk}(\delta_l,M) M_{\rm pk}(\delta_l,M), \label{eq:intmf_peak} \ee
%
%\noindent 
%where \dc\ is a density threshold for the peak. By using the same
%``golden rule'' as in the PS approach (Eq. \ref{eq:ps_plain}), one
%obtains:

\be n(M) M dM = (1+f) \left| \frac{d[n_{\rm pk}(\delta_c;M)M_{\rm pk}(\delta_c,M)]}
{dM} \right| dM.\label{eq:peak_plain} \ee
which s a generalization of PS Ansatz, where $1+f$ takes account of the secondary infall of matter initially in underdense regions, and 
$M_{\rm pk}$ is the average mass of peaks above $\delta_c$ (see Eq. 3-7 of Colafrancesco et al. 1989) 
calculated modelling the peak as an ellipsoid,
and estimated its mass by means of the volume inside the ellipsoid
surface with density larger that a given threshold. 

As a matter of fact, there is not a general agreement on the actual
validity of the peak paradigm. From the theoretical point of view,
structures are {\it not} predicted to form in the peaks of the initial
field; for instance, according to Zel'dovich approximation, structures
form in the peaks of the largest eigenvalue of the deformation
tensor. Then the peak paradigm can not be valid in general, except for
the highest peaks.  Some numerical simulations (Katz, Quinn \& Gelb
1993; van de Weygaert \& Babul 1994) have shown that galactic-size
peaks often disrupt or merge with larger structures, as a result of
tidal interactions with external structures. Manrique \&
Salvador-Sol\'e have argued that such results are due to the lack of
correction for the peak-in-peak problem. On the other hand, Bond \&
Myers (1996b) have found their peak-patch structures,
which account for the peak-in-peak problem, to represent well N-body
structures.\\

The excursion set and peak approaches are somewhat complementary.  In
fact, excursion sets are effective in determining the total fraction
of collapsed mass, and then the global normalization, but are not
accurate in deciding how the collapsed mass fragments into clumps,
i.e. to count the number of objects formed.  On the contrary, the peak
approach clearly determines the number of formed objects, but does not
determine the mass to be associated with the structures, and hence the
global normalization.

\subsubsection{\bf Relations between the peak model and PS formalism}

The peak model can be connected to PS formalism. As known Press and Schechter (1974) 
worked with the differential number
density,
\be \frac{{\rm d}n}{{\rm d}M}\equiv \frac{\rm d}{{\rm d}M} n(>M)
\label{dndm} \ee
At a given epoch, if the filtering mass $M$ is increased by an amount ${\rm
d}M$ then $\nu$ is increased by an amount ${\rm d}\nu$, and the volume
fraction occupied by the collapsed regions is reduced by an amount ${\rm d}V$
given by \eq{dgauss}. Press and Schechter suppose that the eliminated volume
consists of objects with mass between $M$ and $M+{\rm d}M$, corresponding to
the idealisation that filtering the density contrast on any mass scale $M$
cuts out precisely those objects with mass less than $M$ while leaving
unaffected objects with mass bigger than $M$. Ignoring the overdensity
$\simeq(1+\delta_c)$ of the collapsed regions this implies that the number
density ${\rm d}n$ of such objects is given by
\bea M\frac{{\rm d}n}{{\rm d}M}\eqa \left[M\frac{{\rm d}(R_f^2)}{{\rm d}M}
	\right] \frac{{\rm d}(\sigma^2(M))}{{\rm d}(R_f^2)} \frac{{\rm d}\nu}
	{{\rm d}(\sigma^2(M))} \frac{{\rm d}V}{{\rm d}\nu} \frac{{\rm d}n}
	{{\rm d}V} \\
\eqa \left[\frac{2R_f^2}{3} \right] \left[-\sigma^2(M)\langle k^2
\rangle\right]
	\left[-\frac{\nu}{2\sigma^2(M)} \right]
	\left[\frac1{\sqrt{2\pi}}e^{-\nu^2/2}\right]
	\left[\frac1{V_f}\right]\\
\eqa \frac{R_f^2 \langle k^2 \rangle}{3}\frac{1}{4\pi^2 R_f^3} \nu e^{-\nu^2/2}
	\label{psch} \eea

Press and Schechter multiplied this formula by a factor 2, so that when
integrated over all masses it would give the  total mass density of the
universe, rather than just the half corresponding to the regions of space
where the linearly evolved density contrast is positive. They thus arrived at
the estimate
\be n(>M)\simeq n\sub{ps}
\equiv\int_M^\infty
\frac{\langle k^2 \rangle^\prime }{6\pi^2 R_f^\prime }
\nu^\prime
e^{-\nu^{^\prime 2}/2} \frac{dM^\prime }{M^\prime } \label{nps} \ee
In this equation, $R_f^\prime =R_f(M^\prime )$, and similarly for
$\langle k^2\rangle ^\prime $ and $\nu^\prime $.
The factor 2 inserted by Press and Schechter is not justified (as previously described) by their
argument, because the linearly evolved density contrast has nothing to do with
reality in the non-linear regime $\sigma(M)>1$. On the other hand, the
neglected overdensity gives a factor $\simeq(1+\delta_c)=2$ to 3. Thus the
factor 2 goes in the right direction, and the Press-Schechter formula is
reasonably well founded theoretically. A somewhat different justification for
the formula has been given by Bond {\em et al} (1991b).

As told before, there are two alternative estimates
$n(>M)=n\sub{ps}$ and $n(>M)=n\sub{peak}$.
%, for $M=10^{10}\msun$ and for
%$M=10^{15}\msun$. 
A comparison of these two shows that the estimates agree to better than a factor 2
for $\nu\lsim 2$. Presumably, this indicates that in this regime the
assumptions underlying the two estimates are compatible, in that increasing
$M$ by a small amount cuts out portions of the collapsed regions which have
mass of order $M$ and are centred on peaks with height of order $\nu(M)$.

For large $\nu$, the Press-Schechter estimate falls below $n\sub{peak}$. This
can be understood analytically, from the expression
\be \frac{{\rm d}n\sub{peak}}{{\rm d}M}
	\simeq \frac{{\rm d}n\sub{up}}{{\rm d}M}
	=\pdif{n\sub{up}}{\nu}
	\frac{{\rm d}\nu}{{\rm d}M} + \pdif{n\sub{up}}{\langle k^2 \rangle}
	\frac{{\rm d}\langle k^2 \rangle}{{\rm d}M} + \pdif{n\sub{up}}{\gamma}
	\frac{{\rm d}\gamma}{{\rm d}M}
\ee
The first term dominates for large $\nu$, leading to the ratio
\be \frac{{\rm d}n\sub{ps}/{\rm d}M}{{\rm d}n\sub{peak}/{\rm d}M}=2\rfrac{3}
	{R_f^2\langle k^2\rangle} \threehalf\nu\mthree
\label{ratio} \ee
Apart from the factor 2, this is just the filter volume divided by the average
volume of a collapsed region.

As we saw earlier, the smallness of the size of the `collapsed regions' is an
artefact of the filtering. The conclusion is that for very rare fluctuations,
$n\sub{peak}$ is a better estimate that $n\sub{PS}$, the latter being
considerably too small (Thomas \& Couchman 1992). 
However, other sources of error are likely to be more important than the
difference between $n\sub{peak}$ and $n_{PS}$.

\subsection{\bf Dynamics and MF}

The models treated till now
%Both the excursion set and the peak approaches agree in 
identifies collapsed structures as those regions whose linear density contrast
exceeds some threshold. Many simplifications are used to obtain the mass function
%But linear theory is not suitable to follow
%the complicated behavior of collapsing matter; spherical collapse, on
%the other hand, 
which at the end leads to 
neglects important elements such as the role of tides.
Such simplifications could lead to oversimplified
and misleading arguments.  A number of authors have tried to insert
elements of realistic dynamics in the theoretical MF. Such attempts are reviewed in the following.

Some authors have inserted elements of realistic dynamics in the MF
problem by extending the original PS approach or the diffusion or the
peak one (Lucchin \& Matarrese 1988). This approach named {\it PS-like method} shall be analized in the next subsection. 
The peak-patch formalism by Bond \& Myers (1996a), described above, is
also characterized by a more realistic description of the dynamical
evolution of peaks, even though structures are always identified
through the peaks of the linear field.  Other determinations of the MF
were proposed by Henriksen \& Lachi\`eze-Rey (1990), where collapsed
regions were identified by means of correlated velocity structures,
and by Newman \& Wasserman (1990) and Bernardeau \& Schaeffer (1991),
who related (in quite different ways) the MF to the correlation
properties of the matter field.
Monaco (1995) constructed a MF in a PS-like approach, based
on realistic collapse time estimates, found by means of extensions of
the Zel'dovich approximation, and by the use of the homogeneous
ellipsoid collapse model. 

One of the problems with the PS approach, reported by Cavaliere et al.
(1991), is that it supposes matter clumps to instantaneously pass from
non-collapsed to collapsed, and to be immediately incorporated in a
larger clump. In other words, PS seems to imply vanishing time scales
for the formation and destruction of clumps. As a matter of fact, PS
simply does not contain any information on such timescales: the change
of the MF with time is a combination of {\it creation} of new clumps,
{\it destruction} of old clumps and {\it accretion} of mass onto
existing clumps.  Such terms cannot be
disentangled by means of the PS approach alone, without further
assumptions: for instance, the ``static'' procedure proposed by LC93, which is based only on statistics, cannot provide a
precise definition of formation time (it is arbitrarily, though
reasonably, defined as the time taken by a clump to double its mass).

Cavaliere et al. (1991) proposed a {\it dynamical} procedure, based on
creation and destruction time scales, to model the MF and an evolution
equation for $n(M,t)$.
%(without the accretion term).

%is given by:
%\be \frac{\partial n(M,t)}{\partial t} = \frac{n(M,t)}{\tau_+} - 
%\frac{n(M,t)}{\tau_-}. \label{eq:ccs_evol} \ee

Blain \& Longair (1993a,b) and Sasaki (1994) used a similar approach,
obtaining identical results. 
%They imposed that the destruction time
%scale has no characteristic time, then demonstrated that such time
%scale is mass-independent, and explicitly derived the formation time
%scale, by imposing Eq. (\ref{eq:ccs_evol}) to give the PS MF as a
%solution.  In particular, Blain \& Longair (1993) found a numerical
%solution, while Sasaki (1994) derived an analytical expression for the
%formation time scale.

A completely different approach to the MF was proposed by Silk (1978)
and Silk \& White (1978). Aggregation (and fragmentation) of collapsed
clumps of similar size can be described by means of an aggregation
kinetic equation (Smoluchowski 1916; Ernst 1986).
%, of the kind:
%
%$$ \frac{\partial n(M,t)}{\partial t} = \frac{1}{2} 
%\int_0^M K(M,M'-M,t) n(M',t)n(M-M',t)dM' $$ 
%\be - n(M,t)\int_0^\infty K(M,M',t)n(M',t)dM'. \label{eq:smoluchowsky} \ee
%
%\noindent 
The behavior of the MF given by Smoluchowsky equation has been
reviewed elsewhere (see, e.g., Lucchin 1989; Cavaliere, Colafrancesco
\& Menci 1991b; Cavaliere, Menci \& Tozzi 1994).

More recently, Shaviv \& Shaviv (1993, 1995) have analyzed the Smoluchowsky
equation (always in the gravitational context) in a different way; at
variance with Cavaliere and coworkers, they found their MF to depend
on initial conditions.  Another application of Smoluchowsky equation
in a cosmological context is due to Edge et al. (1990), to explain the
evolution of the X-ray luminosity function of galaxy clusters.

As a general remark, such kinetic approaches can describe those
aggregation events which take place between already collapsed clumps.
The direct hierarchical (first) collapse of structures remains well
described by a diffusion formalism, like the one proposed by BCEK. 

The first determination of a mass function, based on a self-consistent
realistic dynamical approximation, is probably due to the works on the
adhesion model. 

Some authors compared the predictions
of clump formation, as given by the adhesion model, to N-body
simulations, in 1D (Doroshkevich \& Kotok 1990; Williams et al. 1991)
and 2D (Nusser \& Dekel 1990; Kofman et al. 1992), finding
satisfactory agreement. 

Vergassola et al. (1994), attempted an analytical estimate of the MF
with adhesion, by using and extending a number of theorems
demonstrated by Sinai (see references in the cited paper).  They were
able to find the asymptotic dependences of the MF: it behaves
exactly like PS at large masses (the exact position of the typical
mass is not determined), but has a different slope at small
masses. However, they could not find the exact normalization of the
MF.

Lagrangian perturbations can be used to construct an MF fully based on
realistic dynamics. At variance with adhesion theory, this dynamical
approximation is of truncated type, i.e. small-scale power has to be
filtered out to avoid small-scale multi-stream regions. Moreover,
Lagrangian perturbations show interesting connections with the
homogeneous ellipsoid collapse model, which can also be used to give
reliable collapse time estimates, as in Monaco (1995). Such topics
have been addressed in Monaco (1997a,b).

\subsubsection{\bf PS-like approach}

In previous sections, we saw 
how finding realistic estimates of collapse
times of generic mass elements.  The problem of translating such
information into an expression for the MF is of purely statistical
nature. Such a statistical problem, in
the simple case in which the inverse collapse time $F$ is proportional
to the initial density contrast, has received much attention in the
scientific literature. Two main approaches were identified, namely the
excursion set and the peak ones; the first approach was shown to be
easier to manage than the second one, at the expense of a simplified
treatment of the geometry of collapsed regions in Lagrangian space,
while the peak approach, whose validity relies on the validity of the
peak hypothesis, better takes into account geometry, at the expense
of an increased complexity of the formalism, especially when trying to
include a proper treatment of the peak-in-peak problem.

In this section, the MF will be determined by means of a simplified PS-like,
single-scale approach, which consists of estimating the probability of
having initial conditions that make the mass element collapse.  

%%%%%%%%%%%%%%%%%%%%%%%%%%%%%%%%%%  4.1  %%%%%%%%%%%%%%%%%%%%%%%%%%%%%%%%%%

A first determination of the MF can be obtained by applying the same
statistical approach as in the original PS paper: the fraction of
collapsed mass is obtained by integrating, at a given fixed scale,
over all initial conditions which make a mass element collapse before
a given time.  This determination obviously suffers from the same
cloud-in-cloud problem as the PS one; however, 
most mass is predicted to finally collapse by realistic collapse time
estimates, so a PS-like MF is nearly normalized, by more than 90\%. It
is then natural to suspect that an MF obtained by means of the
absorbing barrier formalism can be not very different from the PS-like
one, as only a minor part of the mass, lying in the strongest
underdensities, has to be redistributed; this mass is not expected to
influence the MF in any interesting mass range. As shown by M98,
%In \S 4.2 and 4.3 it
%will be shown 
that under some conditions the PS-like and the absorbing
barrier MFs are very similar;  
the
simplified PS-like approach suffices in finding the main features of
the dynamical MF (M98).

\begin{figure}
\begin{center}
\epsfig{file=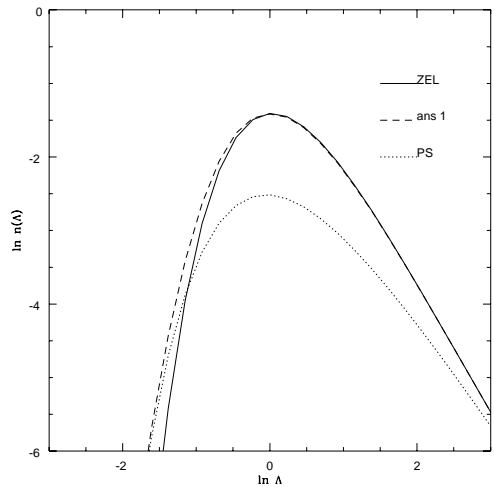,width=7cm}
\caption{PS-like $n(\mres)$ curves for Ze'ldovich approximation and ans\"atz 1, compared to the PS 
one. Figure taken from M98.}
\end{center}
\end{figure}

\begin{figure}
\begin{center}
\epsfig{file=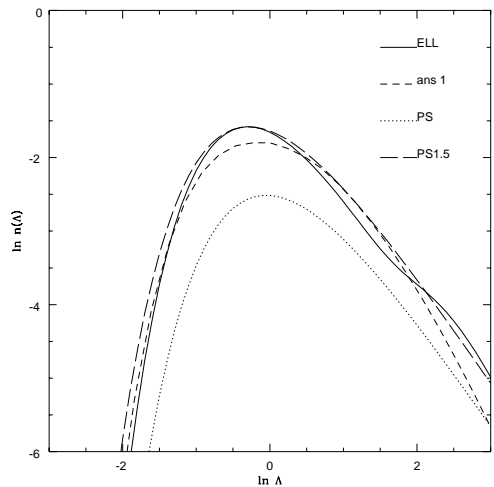,width=7cm}
\caption{PS-like $n(\mres)$ curves for ELL and ans\"atz 2, compared to 
the PS one and with a PS with \dc=1.5 and the fudge factor 2; see text. Figure taken from M98.}  
\end{center}
\end{figure}

%
%\begin{figure}
%\begin{center}
%\epsfig{file=fig4_3.ps,width=7cm}
%\caption{PS-like $n(\mres)$ curves for ELL and 3RD, compared to 
%a PS with \dc=1.5 and the fudge factor 2; see text.}
%\end{center}
%\end{figure}
%
%
%\begin{figure}
%\begin{center}
%\epsfig{file=fig4_7.ps,width=7cm}
%\caption{$n(\mres)$ curves for the ELL prediction.}
%\end{center}
%\end{figure}
%

Integration over initial conditions requires that initial conditions
are specified, and that their joint PDF is known. In the PS case,
where linear theory and spherical collapse were used, initial
conditions were simply provided by the initial density contrast.  In
the most general case, initial conditions are given by the value of the
density (or, equivalently, of the potential) at every point, so that a
direct integration is hard to perform. An intermediate case is
provided by the Zel'dovich approximation, and by the other
approximations which require the same initial conditions and ellipsoidal collapse. In this
case, the joint PDF of initial conditions, the \lam\ eigenvalues, is
known (Doroshkevich 1970):

$$ P_\lambda(\muno,\mdue,\mtre)d\muno d\mdue d\mtre = 
\frac{675\sqrt{5}}{8\pi\mres^3} \exp\left( -\frac{3}{\mres} \mu_1^2+
\frac{15}{2\mres}\mu_2\right) $$
\be \times (\muno-\mdue)(\muno-\mtre)(\mdue-\mtre)d\muno d\mdue d\mtre.
\label{eq:dorosh} \ee

\noindent 
\res\ is again the mass variance; $\mu_1=\muno+ \mdue+ \mtre$ and
$\mu_2=\muno\mdue+ \muno\mtre+ \mdue\mtre$ are principal invariants of
the Zel'dovich deformation tensor.
It is convenient to express this PDF in terms of the linear density
\dl\ and the $x$ and $y$ variables defined in Eq.  
(\ref{eq:x_and_y}); in this case the joint PDF is factorized into a
Gaussian for \dl\ and a joint PDF for $x$ and $y$:

$$ P(\delta_l,x,y) d\delta_ldxdy= \frac{1}{\sqrt{2\pi\mres}}\exp\left( 
-\frac{\delta_l^2} {2\mres} \right)d\delta_l\frac{225}{4} \sqrt{\frac{5}{2\pi}}
\frac{1}{\mres^{5/2}} \exp \left(-\frac{5}{2\mres} (x^2+xy+y^2)\right) $$
\be \times xy(x+y)dxdy=\mpdfl d\delta_l \times P_{x,y}(x,y;\mres)dxdy. 
\label{eq:dorosh_due} \ee

The fraction of collapsed mass can then be obtained as follows:

\be \mimfr = \int_0^\infty \!dx \!\int_0^\infty \!dy P_{x,y}(x,y;\mres)
\!\int_{\delta_c(x,y)}^\infty \! d\delta_l\, \mpdfl, \label{eq:integralold} \ee

\noindent 
where the function $\delta_c(x,y)$, which substitutes the \dc\
parameter of PS, is defined as the solution of the equation:

\be b_c(\delta_c,x,y)=b(t_0), \label{eq:deltac_def} \ee

\noindent
where $b(t_0)$ is the instant at which the MF is wanted (it will
usually be set equal to one). By writing the function $\delta_c$ as
$\delta_0-f(x,y)$, where $\delta_0$ is the spherical value 1.69 and
the positive function $f(x,y)$, vanishing at the origin, gives the
effect of the shear, it is possible to write the \dmfr\ function as:

\be n(\mres)=n_{PS}(\mres)\times {\cal I}(\mres), \label{eq:correction} \ee

\noindent 
where $n_{PS}(\mres)$ is the PS curve, 
%Eq. (\ref{eq:ps_ndires}), 
and
${\cal I}(\mres)$ is a correction term:

$$ {\cal I}(\mres) = \frac{1}{\mres} \int_0^\infty \!dx 
\!\int_0^\infty \!dy\, P_{x,y}(x,y) \exp \left( -\frac{1}{2\mres}f^2(x,y)+
\frac{\delta_0}{\mres} f(x,y) \right)$$
\be \times\left( 
1-\frac{1}{\delta_0} \left(f-x\frac{\partial f}{\partial x} -y 
\frac{\partial f}{\partial y} \right)\right). \label{eq:i_factor} \ee

\dmfr\ curves have been calculated for Zel'dovich, ellipsoidal model, and for the two {\it
ansatze} previously presented. 
%in \S 3.1????????????????; 
Details of the calculations are reported
by Monaco (1995). Fig. 15
%4.1??????????? 
presents the \dmfr\ curve for Zel'dovich and for
the first {\it ans\"atz}, Eq. (\ref{eq:ansatza}), in which spherical
collapse is recovered when Zel'dovich predicts a slower collapse. The
canonical PS \dmfr curve is shown for comparison; the PS curve has not
been multiplied by the fudge factor of two at this stage, in order to
compare the results of the PS-like procedures in the different cases,
with no guarantee of normalization (in Monaco 1995 all the curves were
multiplied by two).  It can be seen that the Zel'dovich curve underestimates
the number of large-mass\footnote{I freely use the word mass in this
context to indicate the large-mass (small \res) or small-mass (large
\res) part of the MF} objects, and gives more intermediate- and
small-mass objects, also thanks to its better normalization. The {\it
ans\"atz} curve reproduces the PS one at large masses, and reduces to
the Zel'dovich one at small masses, as expected.

Fig. 16
%4.2????????????? 
shows the ellipsoidal model prediction, in comparison with the second {\it
ans\"atz}, Eq. (\ref{eq:ansatzb}) (with $\epsilon=0.2$), the canonical
PS curve (without the fudge factor 2, as before) and a PS curve with
\dc=1.5, representative of the typical outcome of N-body simulations
(and then with the factor 2). Both the {\it ans\"atz}
curve and ellipsoidal model predict an overabundance of large-mass clumps with
respect to the canonical PS curve: it is again demonstrated that a
systematic displacement of collapse times from the spherical value
influences the large-mass tail of the MF, even though spherical
collapse is asymptotically recovered. In particular, the ellipsoidal model curve is
quite similar to the PS 1.5 curve (PS curve with $\delta_c=1.5$). It is however to be stressed that
this similarity, while encouraging, has to be taken with care, as it
is not clear how the collapsed objects predicted by this theory are
related to the N-body clumps.  As a technical remark, this ellipsoidal model curve
has been computed by using the full $b_c$ curves; in
Monaco (1995) only the overdense curve was considered, and the Zel'dovich
behavior was forced at large $x$ and $y$ values.

\subsubsection{\bf MF and $n(\Lambda)$}

The passage from the resolution to the mass
variable requires knowledge of how collapsed matter gathers in clumps,
an element which is missing in the excursion set approach; 
The volume of the excursion sets as a function of resolution,
and then the mass of structures, is obtained by means of the
%reasonable ``golden rule'' 
rule: 
%given in Eq.  (\ref{eq:ps_plain})
%, here
%again reported:
\be M n(M) dM = \bar{\varrho} \left|\frac{d\Omega}{dM}\right| dM = 
\bar{\varrho} n(\mres) \left| \frac{d\Lambda}{dM}\right| dM. 
\label{eq:golden_rule} \ee
%the MF has
%been found by means of the usual resolution-mass relation (the
%so-called golden rule, Eq. \ref{eq:golden_rule}), and the consequences
%of this approximation have been discussed.

A more realistic resolution-mass relation would predict a whole
distribution of masses to form at a given resolution:

\be \mres \rightarrow p(M;\mres). \label{eq:mass_res} \ee

\noindent
The $p(M;\mres)$ function gives the probability that a mass $M$ is
formed at a resolution \res; its mean value will be of order:

\be \int_0^\infty M p(M;\mres) dM \sim \bar{\varrho}R(\mres)^3,
\label{eq:mean_mass} \ee

\noindent
as the smoothing length $R(\mres)$ is the relevant scale in this case;
the proportionality constant, of order one, will in general depend on
the shape of the filter, on the power spectrum and on the resolution.
The MF would then be given by:

\be M n(M)dM = \bar{\rho} \left( \int_0^\infty n(\mres)
p(M,\mres) d\mres \right)\, dM \; ,\label{eq:no_golden_rule} \ee 

\noindent 
i.e., by the \dmfr\ curve convolved with the distribution
$p(M,\mres)$. 

Such a $p$
distribution is expected, at small resolutions, to be peaked at its
mean value, while its shape at large resolutions is expected be more
complex, probably influenced by the details of the prescription chosen
to fragment the collapsed medium (M98). Then, with respect to using the
simple golden rule, the introduction of the $p$ distribution is
expected not to influence dramatically the large mass tail of the MF,
while it is likely to influence the small-mass part, which is
confirmed to be a not robust prediction of this kind of MF theories.
Finally, differences in the $p$ distributions for different filter
shapes could in principle at least attenuate the differences between
the MFs found with SKS and Gaussian filtering.

\subsubsection{\bf Conclusions}

%In this Chapter it has been shown that the excursion set formalism can
%be extended to the case in which the $F$ process, a non-linear and
%non-local functional of the initial potential, is considered. 

As shown in M98, a PS-like procedure is enough to obtain the main features of the MF. 
If SKS filtering is used, it has been demonstrated that the diffusion
formalism can be extended to the non-Gaussian $F$ process, by
considering it as a diffusion process. The problem can then be
transformed to the diffusion of a Wiener process with a moving
absorbing barrier. For Gaussian smoothing, the simple approximation
proposed by Peacock \& Heavens (1991) has been shown very successful
in finding the MF.

%The main conclusions are the following:

The following conclusions
%, which will be fully confirmed below, 
can be drawn:
%
%\begin{enumerate}
%\item 
%realistic collapse predictions cause an enhanced production of
%large-mass clumps; the resulting MF is similar to a PS one with
%$\delta_c=1.5$;
%\item 
%the fact that spherical collapse is asymptotically recovered for the
%strongest overdensities does not guarantee that the ``spherical''
%canonical PS MF, with \dc=1.69, is recovered at large masses.
%\end{enumerate}
%

\begin{enumerate}
\item 
A larger number of large-mass objects is expected to form with respect
to the canonical PS prediction with \dc=1.69.
\item 
The fact that spherical collapse is asymptotically recovered for the
strongest overdensities does not guarantee that the ``spherical''
canonical PS MF, with \dc=1.69, is recovered at large masses.
\item
The large-mass part of the MF is considered robust with respect to the
dynamical prediction. The ELL prediction tends to give more objects
than the 3RD one in the large-mass tail; this is due to the fact that
3rd-order Lagrangian theory slightly underestimates spherical
collapse; thus the ELL prediction is considered more believable in
that range.
\item
PS-like 
%and Gaussian-smoothing MFs 
gives fewer objects than the
SKS one, by roughly a factor of 2, in the large- and intermediate-mass
ranges.
\item
PS-like is very similar to the PS one with a $\delta_c
\simeq1.5$ parameter (indicated as PS 1.5).
\end{enumerate}

The small-mass part of the MF is not considered a robust prediction of
the theory, for at least three reasons:

\begin{enumerate}
\item
The definition of collapse, which is based on the concept of orbit
crossing, is not expected to reproduce common small-mass structures
like virialized halos. OC regions rather represent those large-scale
collapsed environments in which the virialized halos are embedded.
\item
All the dynamical predictions used are considered good as long as the
inverse collapse time is not small. Thus, the small-mass part of the
MF is based on non-robust dynamical predictions.
\item
The $p$ distribution of the forming masses, at a given resolution, is
likely to significantly affect the small-mass part of the MF.
\end{enumerate}

The prediction of more large-mass objects, caused by the improved
dynamical description of collapse, confirms some previous claims, for
example by Lucchin \& Matarrese (1988) and Porciani et al.  (1996),
who introduced non-Gaussianity of dynamical origin in the PS or
diffusion approaches. This increase can be seen as the effect
of tides on the dynamics of the mass element (Bertschinger \& Jain
1994).  Besides, even spherical collapse, when the global
interpretation of collapse times is assumed (\S 2.3.4), leads to the
prediction of more large-mass objects (Blanchard et al. 1992; Yano et
al. 1996; Betancort-Rijo \& Lopez-Corredoira 1996).

The similarity of the (Gaussian-smoothed) dynamical MF with a PS one,
with \dc=1.5, makes the dynamical MF be consistent with many numerical
simulation (\S 2.2), even though this agreement is not a real proof of
validity as (i) OC regions are not the halos extracted from
simulations, and (ii) the resolution-mass relation is still treated in
a simplified way. However, the dynamical MF theory does not predict
the trend of lower \dc\ values at higher redshifts, observed in recent
N-body simulations (see references in \S 2.2). An explanation of this
behavior was proposed by Monaco (1995): the presence of small-scale
power can effectively slow down gravitational collapse, through
dynamical events of the kind of previrialization, proposed by Peebles
(1990; see also Lokas et al. 1996; Bouchet 1996), or through dynamical
friction with those possible particles which evaporate out of the
collapsed structures (Antonuccio \& Colafrancesco 1994). Such events
could be effective for power spectra with $n>-1$ (Lokas et al. 1996).
In CDM-like spectra, the effective spectral index at $M_*$ is smaller
than $-1$ at high redshifts, where \dc=1.5 (or even less) is found,
but becomes larger at lower redshifts, where \dc\ starts to increase to
1.7 or even larger values. However, such a behavior could also be due
to a dependence of the $p$ distribution on the power spectrum.

\subsection{\bf More recent developments using the excursion set model: the moving barrier model and the multiplicity function}

%DEFINIRE ST, ST1, SMT, J01, YNY

As reported in the introduction, the PS model when compared to numerical simulations gives a smaller number of high-mass halos while 
giving a larger number of low-mass halos. The quoted discrepancy, that lead to research new expressions for the mass function, 
is not surprising since the PS model, as any other analytical model, should make several assumptions to get simple 
analytical predictions. The main assumptions that the PS model combines are the simple physics 
of the spherical collapse model with the assumption that the initial fluctuations were Gaussian and small. 
On average, initially denser regions collapse before less dense ones, which means that, at any given epoch, 
there is a critical density, $\delta_c(z)$, which must be exceeded if collapse is to occur. 
In the spherical collapse model, this critical density does not depend on the mass of the collapsed object, while taking
account of the effects of asphericity and tidal interaction with neighbors, it is possible to show that it is mass dependent
(Del Popolo \& Gambera 1998, SMT).
%
%%Taking account of the effects of asphericity and tidal interaction with neighbors, Del Popolo \& Gambera (1998) and SMT,
%%using a parameterization of the ellipsoidal collapse, showed that the threshold is mass dependent, and in particular that %%of the set of 
%%objects that collapse at the same time, the less massive ones must initially have been denser than the more massive, since %%the less massive ones would have had to hold themselves together against stronger tidal forces. 
%
%
%In the 
%quoted model, a region collapse at a redshift $z$ if the inner overdensity exceeds a critical value, $\delta_c(z)$    
%given by the spherical collapse model, which is constant and independent of the initial size of the region.
%
In the second hand, the Gaussian nature of the fluctuation field means that a good approximation to the number density 
of bound objects that have mass $m$ at time $z$ is given by considering the barrier crossing statistics of many independent and uncorrelated random walks, where the barrier shape $B(m,z)$, is connected to the collapse threshold. 
Simply changing the barrier shape, SMT showed that it is possible to incorporate the ``quoted effects" \footnote{Namely that in the case of objects collapsing at the same time, the less massive regions must initially have been denser than the more massive ones.}
%
%%, since the less massive ones would have had to hold themselves together against stonger tidal forces) 
%
in the excursion set approach. 
Moreover, using the shape of the modified barrier in the excursion set approach, it is possible to obtain a good fit to the universal halo mass function \footnote{Note that at present there is no good numerical test of analytic predictions for the low mass tail of the mass function.}. 
The excursion set approach allows one to calculate good approximations to several 
important quantities, such as the ``unconditional" and ``conditional" mass functions. 
ST1 provided formulas to calculate these last quantities starting from the shape of the 
barrier.

%They also showed that while the ``unconditional" and ``conditional" mass function is in good agreement with results from numerical simulations, neither the 
%constant nor the moving barrier models (barrier obtained from non-spherical collapse) were able to describe the simulations results at small lookback times, in the case of the %rescaled (in terms of $\nu$) ``conditional" mass function. The reason for this discrepancy is probably due to the excursion set approach's neglect of correlations between %scales (Peacock \& Heavens 1990; Bond et al. 1991; ST) or to the too simple parametrization of the ellipsoidal collapse outlined in SMT. 
In the following, I'll use an improved version of the barrier obtained in Del Popolo \& Gambera (1998) to get the multiplicity functions, which shall 
be compared with those obtained by PS, ST, J01, YNY, 
and with numerical simulations of YNY. 

In order to calculate the barrier shape, it is possible to follow Del Popolo \& Gambera (1998) model.
The equation governing the collapse of a density perturbation taking account 
angular momentum acquisition by protostructures 
can be obtained using a model due to Peebles (Peebles 1993) (see also Del Popolo \& Gambera 1998, 1999).\\
Let's consider an ensemble of gravitationally growing mass concentrations 
and suppose that the material in each system collects within the
same potential well
with inward pointing acceleration given by $g(r)$ (see Del Popolo \& Gambera 1998). We
indicate with $dP=f(L,r v_r,t)dL dv_r dr$ the probability that a particle
can be found  in the proper radius range $r$, $r+dr$, in the radial
velocity range $v_r={\dot r}$, $v_r+d v_r$ and with angular momentum
$L=r v_\theta$ in the range $dL$.
%then from Liouville's theorem it follows that
%the distribution function, $f$,
%satisfies the collisionless Boltzmann equation:
%\begin{equation}
%\frac{\partial f}{\partial t} + v_{r}
%\frac{\partial f}{\partial r} + \frac{\partial f}{\partial v_{r}}
%\cdot [ \frac{L_{2}}{r^{3}} - g(r)] = 0
%\end{equation}
The radial
acceleration of the particle is: 
\begin{equation}
\frac{dv_r}{dt}=\frac{L^2(r)}{M^2r^3}-g(r)=\frac{L^2(r)}{M^{2} r^3}-\frac{G M }{r^2}  
\label{eq:col}
\end{equation}
%%where $g(r)$ is the acceleration.
%and $g(r)$ the acceleration.
Eq. (\ref{eq:col}) can be derived from a potential
and then from Liouville's theorem it follows that
the distribution function, $f$,
satisfies the collisionless Boltzmann equation:
\begin{equation}
\frac{\partial f}{\partial t} + v_{r}
\frac{\partial f}{\partial r} + \frac{\partial f}{\partial v_{r}}
\cdot \left[ \frac{L_{2}}{r^{3}} - g(r) \right] = 0
\end{equation}
%The equation governing the collapse of a density perturbation 

Assuming a non-zero cosmological constant Eq. (\ref{eq:col}) 
becomes:
%taking account of a cosmological constant and 
%the 
%interaction of the quadrupole moment of the system with the tidal field of
%the matter of the neighbouring proto-clusters is given by: 
\begin{equation}
\frac{dv_r}{dt}=-\frac{G M }{r^2}+\frac{L^2(r)}{M^{2} r^3}+\frac{\Lambda}{3} r \label{eq:coll}
\end{equation}
(Peebles 1993; Bartlett \& Silk 1993; Lahav 1991; Del Popolo \& Gambera 1998, 1999).
%where, $L(r)$ is the angular momentum.
Integrating Eq. (\ref{eq:coll}) we have: 
\begin{equation}
\frac{1}{2}\left( \frac{dr}{dt}\right) ^{2}=\frac{GM}{r}+\int 
\frac{L^{2}}{M^{2}r^{3}}dr+\frac{\Lambda }{6}r^{2}+\epsilon
\label{eq:coll1}
\end{equation}
where the value of the specific binding energy of the shell, $\epsilon$, can be obtained using the condition for turn-around, $\frac{dr}{dt}=0$.

In turn the binding energy of a growing mode solution is uniquely given
by the linear overdensity, $\delta_{i}$, at time $t_{i}$.
From this overdensity, using the linear theory, we may obtain that of
the turn-around epoch and then that of the collapse.
%which is given by:
We find the binding energy of the shell, $C$, using the
relation between $v$ and $\delta_{i}$ for the growing mode
(Peebles 1980) in Eq. (\ref{eq:coll1}) and finally the
linear overdensity at the time of collapse is given by:
%Using
%a technique similar to that by Bartlett \& Silk (1993) 
%it is possible to obtain the overdensity at the time of collapse:
\begin{equation}
\delta _{\rm c}=\delta _{\rm co}\left[ 1+
\int_{r_{\rm i}}^{r_{\rm ta}}  \frac{r_{\rm ta} L^2 \cdot {\rm d}r}{G M^3 r^3}+\Lambda \frac{r_{\rm ta} r^2}{6 G M}
\right] \simeq \delta _{\rm co} \left[ 
1+\frac{\beta_1}{\nu^{\alpha_1}}+\frac{\Omega_{\Lambda} \beta_2}{\nu^{\alpha_2}}
\right]
\label{eq:maa} 
\end{equation}
where $\alpha_1=0.585$, $\beta_1=0.46$, $\alpha_2=0.4$ and $\beta_2=0.02$, 
where $\delta _{\rm co}=1.68$ is the critical threshold for a spherical model,
$r_{\rm i}$ is the initial radius, $r_{\rm ta}$ is the turn-around radius, 
$L$ the angular momentum. 
%(for $\nu>0.1)$. 
The angular momentum appearing in Eq. ~(\ref{eq:maa}) is the total angular momentum acquired by the proto-structure during evolution. In order to calculate $L$, it is possible to use the same model
as described in Del Popolo \& Gambera (1998, 1999) (more hints on
the model and some of the model limits can be found in
Del Popolo, Ercan \& Gambera 2001).

The CDM spectrum used to calculate the mass function plotted in Figs. 17-19 is that of BBKS (equation~(G3)), with transfer function:
\begin{equation}
T(k) = \frac{[\ln \left( 1+2.34 q\right)]}{2.34 q}
\cdot [1+3.89q+
(16.1 q)^2+(5.46 q)^3+(6.71)^4]^{-1/4}
%
%T^2(k) &=& [\ln \left( 1+4.164k\right)]^2 \cdot (192.9+1340k+ \nonumber \\
%& + &  1.599\cdot 10^5k^2+1.78\cdot 10^5k^3+3.995\cdot
%10^6k^4)^{-1/2}
%
\label{eq:ma5}
\end{equation}
(where 
%$ A$ is the normalizing constant and 
$q=\frac{k\theta^{1/2}}{\Omega_{\rm X} h^2 {\rm Mpc^{-1}}}$.
Here $\theta=\rho_{\rm er}/(1.68 \rho_{\rm \gamma})$
represents the ratio of the energy density in relativistic particles to
that in photons ($\theta=1$ corresponds to photons and three flavors of
relativistic neutrinos).
The power spectrum was normalized to reproduce the observed abundance of rich 
cluster of galaxies (e.g., Bahcal \& Fan 1998).

In the excursion set approach, the average comoving number density of haloes of mass $m$ 
%the often called 
the universal or ``unconditional" mass function, $n(m,z)$, is given by:
\begin{equation}
n(m,z)=\frac{\overline{\rho}}{m^{2}}\frac{d\log{\nu }}{d\log m}\nu f(\nu )
\label{eq:universal}
\end{equation}
(BCEK), where $\overline{\rho}$ is the background density, $\nu=\left(\frac{\delta_{\rm c}(z)}{\sigma(m)}\right)^2$ is 
the ratio between the critical overdensity required for collapse in the spherical model, $\delta_{\rm c}(z)$, to the r.m.s. density fluctuation $\sigma(m)$, on the scale $r$ of the initial size of the object $m$. The function $\nu f(\nu)$ is obtained by computing the distribution of first crossings, $f(\nu) d \nu$, of a barrier $B(\nu)$, by independent, uncorrelated Brownian motion random walks. The mass function can be thus calculated once a shape for the barrier is given and the power spectrum is known. In the case of spherical collapse, characterized by a constant barrier (for all $\nu$), PS and BCEK obtained:
\begin{equation}
\nu f(\nu )=\left( \frac{\nu }{2\pi }\right) ^{\frac{1}{2}}\exp (-\frac{\nu}{2})
\label{eq:bond}
\end{equation}   
In the case of a nonspherical collapse, the shape of the barrier is no longer a constant and moreover 
%(for elipsoidal collapse) 
it depends on mass (Del Popolo \& Gambera 1998; SMT). As shown by ST1, for a given barrier shape, $B(S)$, where $S\equiv S_{\ast }\left( \frac{\sigma }{\sigma _{\ast }}\right)^{2}=\frac{S_{\ast }}{\nu}$ and $\sigma _{\ast }=\sqrt{S_{\ast }}=\delta _{co}$, the first crossing distribution is well approximated by:
\begin{equation}
f(S)dS=|T(S)|\exp (-\frac{B(S)^{2}}{2S})\frac{dS/S}{\sqrt{2\pi S}}
\label{eq:distrib}
\end{equation}   
where $T(S)$ is the sum of the first few terms in the Taylor expansion of $B(S)$:
\begin{equation}
T(S)=\sum_{n=0}^{5}\frac{(-S)^{n}}{n!}\frac{\partial ^{n}B(S)}{\partial S^{n}}
\label{eq:expans}
\end{equation}
The quantity $Sf(S,t)$ is a function of the variable $\nu$ alone, where $\nu\equiv (\delta_c(t)/\sigma(M))^2$. Since $\delta_c$ and $\sigma$ evolve
with time in the same way, the quantity $Sf(S,t)$ is independent on time. Setting $2Sf(S,t)=\nu f(\nu)$, one obtains the so-called
multiplicity function $f(\nu)$. 
%The multiplicity function is the distribution of first crossings of  a barrier $B(\nu)$ by 
%independent uncorrelated Brownian
%random walks (Bond et al. 1991). 
That's why the shape of the barrier influences the form of the multiplicity function.
   
%The previous Eq. ~(\ref{eq:distrib}),(\ref{eq:expans}), reduce to Eq. ~(\ref{eq:bond}) for constant  
%barriers. 
In the case of the ellipsoidal barrier shape given in ST:
\begin{equation}
B(\sigma ^{2},z)=\sqrt{a}\delta _{c}(z)\left[ 1+\frac{\beta }{\left( a\nu \right) ^{\alpha }}\right] 
\end{equation}
the Eqs. ~(\ref{eq:distrib}),(\ref{eq:expans}), give, after truncating the expansion at $n=5$ (see ST):
\begin{equation}
\nu f(\nu)=\sqrt{a \nu / 2 \pi}[1+\beta(a
{\nu}^2)^{-\alpha}g(\alpha)]\exp\left(-0.5a\nu^2[1+\beta(a\nu^2)^{-\alpha}]^2\right)
\label{eq:sstt}
\end{equation}
%\begin{eqnarray}
%f(\nu)&=&\sqrt{a \nu / 2 \pi}[1+\beta(a
%{\nu}^2)^{-\alpha}g(\alpha)]\exp\left(-0.5a\nu^2[1+\beta(a\nu^2)^{-\alpha}]^2\right)
%\nonumber \\
%& &
%\simeq A \left( 1+\frac{0.094}{\left( a\nu \right) ^{0.6}}\right) \sqrt{\frac{a\nu }{2\pi }}
%\exp{\{-a\nu \left[ 1+\frac{0.5}{\left( a\nu \right) ^{0.6}}\right] ^{2}/2\}}
%\label{eq:sstt}
%\end{eqnarray}
where
\begin{equation}
g(\alpha)=
\mid 1-\alpha +\frac{\alpha (\alpha
-1)}{2!}-...-\frac{\alpha(\alpha-1)\cdot \cdot \cdot
(\alpha-4)}{5!} \mid
\end{equation}

Using the values for $\beta$ and $\alpha$ of ST ($a=0.707$, $\delta_{\rm c}(z)=1.686 (1+z)$, $\beta \simeq 0.485$ and $\alpha \simeq 0.615$) in Eq. (\ref{eq:sstt}), one gets (ST1):

\begin{equation}
\nu f(\nu) \simeq A_1 \left( 1+\frac{0.094}{\left( a\nu \right) ^{0.6}}\right) \sqrt{\frac{a\nu }{2\pi }}\exp{\{-a\nu \left[ 1+\frac{0.5}{\left( a\nu \right) ^{0.6}}\right] ^{2}/2\}}
\label{eq:sstt1}
\end{equation}
%\begin{equation}
%f(\nu )d\nu=A \left( 1+\frac{0.094}{\left( a\nu \right) ^{0.6}}\right) \sqrt{\frac{a\nu }{2\pi }}\exp{\{-a\nu \left[ 1+\frac{0.5}{\left( a\nu \right) ^{0.6}}\right] ^{2}/2\}}
%\label{eq:sstt}
%\end{equation}
with $A_1 \simeq 1$.
This last result is in good agreement with the fit of the simulated first crossing distribution (ST):
\begin{equation}
\nu f(\nu )d\nu =A_2\left( 1+\frac{1}{\left( a\nu \right) ^{p}}\right) \sqrt{\frac{a\nu }{2\pi }}\exp (-a\nu /2)
\label{eq:ssttt}
\end{equation}
where $p=0.3$, and $a=0.707$. 
%or
%\begin{equation}
%\nu f(\nu)=A_4(1+\nu'^{-2p})\sqrt{2/\pi}\nu'exp(-\nu'^2/2)
%\end{equation}
%where $\nu' =\nu\sqrt{a}$.
%
%and the values of the constants
%are: $A=0.322$, $p=0.3$ and $a=0.707$.

The normalizatition factor $A_2$ has to satisfy the constraint:
\begin{equation}
\int_0^{\infty} f(\nu) d \nu=1
\end{equation}
and as a consequence it is not an independent parameter, but is expressed in the form:
\begin{equation}
A_2=\left[1+2^{-p} \pi^{-1/2} \Gamma(1/2-p)\right]^{-1}=0.3222
\footnote{
%Note that ST used Eq. \ref{eq:sstt} to compare model and data. 
Note, that Eq. \ref{eq:ssttt} gives a better fit to Eq. \ref{eq:sstt} if $A \simeq 0.3$ and $a \simeq 0.79$. Vice versa a smaller value of $a$ ($a \simeq 0.63$) and $A=1.08$ in Eq. \ref{eq:sstt} gives a better fit to Eq. \ref{eq:ssttt} (with $A_1=0.3222$ and $a=0.707$), which was the one ST used to compare model and data.}.
\end{equation}
%ST gave the best-fit parameter values as $A=0.322$, $p=0.3$ and $a=0.707$ 

%%%%%LE EQUAZIONI SOPRA, 6,7 devono avere le p uguali ecc, e sono diverse. CORREGGERE.

%Thus, given Eqs. ~(\ref{eq:distrib})-(\ref{eq:expans}), it is possible to obtain 
%mass function, if the barrier shape and the power spectrum are given. 

%Putting Eq. (\ref{eq:ma7}) into Eqs. ~(\ref{eq:distrib})-(\ref{eq:expans}) and truncating the %expansion at $n=5$, 

If the barrier takes account of the cosmological constant, like in Eq. (\ref{eq:maa}), using the same method 
that lead to Eq. (\ref{eq:sstt}), we have that: 
\begin{equation}
%f(\nu )d\nu \simeq 1.1\left( 1+\frac{0.073}{\left( a\nu \right) ^{0.585}}\right) \sqrt{\frac{a\nu }{2\pi }}\exp{\{-a\nu 
%\left[ 1+\frac{0.52}{\left( a\nu \right) ^{0.585}}\right] ^{2}/2\}}
%f(\nu )d\nu \simeq A _1 \left( 1+\frac{0.1218}{\left( a\nu \right) ^{0.585}}\right) \sqrt{\frac{a\nu }{2\pi }}\exp{\{-
%0.4019 a\nu \left[ 1+\frac{0.5526}{\left( a\nu \right) ^{0.585}}\right] ^{2}\}}
\nu f(\nu )=A _1 \left( 1+\frac{\beta_1 g(\alpha_1)}{\left( a\nu \right) ^{\alpha_1}}
+\frac{\beta_2 g(\alpha_2)}{\left( a\nu \right) ^{\alpha_2}}
\right) \sqrt{\frac{a\nu }{2\pi }}\exp{\{-a \nu \left[ 1+\frac{\beta_1}{\left( a\nu \right) ^{\alpha_1}}
+\frac{\beta_2}{\left( a\nu \right) ^{\alpha_2}}
\right] ^{2}/2\}}
%\label{eq:mia}
%f(\nu )d\nu \simeq 1.21\left( 1+\frac{0.06}{\left( a\nu \right) ^{0.585}}\right) \sqrt{\frac{a\nu }{2\pi }}\exp{\{-a\nu 
%\left[ 1+\frac{0.57}{\left( a\nu \right) ^{0.585}}\right] ^{2}/2\}}
\label{eq:mia}
\end{equation}

In the case of the barrier given in Eq. (\ref{eq:maa}) with $\Lambda=0$, the ``unconditional" multiplicity function can be approximated by:
\begin{equation}
%f(\nu )d\nu \simeq 1.1\left( 1+\frac{0.073}{\left( a\nu \right) ^{0.585}}\right) \sqrt{\frac{a\nu }{2\pi }}\exp{\{-a\nu 
%\left[ 1+\frac{0.52}{\left( a\nu \right) ^{0.585}}\right] ^{2}/2\}}
%f(\nu )d\nu \simeq A _1 \left( 1+\frac{0.1218}{\left( a\nu \right) ^{0.585}}\right) \sqrt{\frac{a\nu }{2\pi }}\exp{\{-
%0.4019 a\nu \left[ 1+\frac{0.5526}{\left( a\nu \right) ^{0.585}}\right] ^{2}\}}
\nu f(\nu ) \simeq A _4 \left( 1+\frac{b}{\left( a\nu \right) ^{0.585}}\right) \sqrt{\frac{a\nu }{2\pi }}\exp{\{-a c\nu \left[ 1+\frac{d}{\left( a\nu \right) ^{0.585}}\right] ^{2}\}}
\label{eq:mia}
%f(\nu )d\nu \simeq 1.21\left( 1+\frac{0.06}{\left( a\nu \right) ^{0.585}}\right) \sqrt{\frac{a\nu }{2\pi }}\exp{\{-a\nu 
%\left[ 1+\frac{0.57}{\left( a\nu \right) ^{0.585}}\right] ^{2}/2\}}
\label{eq:miaa}
\end{equation}
where $a=0.707$, $b=0.1218$, $c=0.4019$, $d=0.5526$ and $A_4 \simeq 1.75$ is obtained from the normalization condition.
%IMPORTANTE TOLTO
%%\begin{equation}
%%\nu f(\nu )=\frac{A}{2}\left( 1+\frac{1}{\nu _{1}^{\alpha_2 }}+\frac{1}{\nu _{2}^{\beta_2 }}\right) \sqrt{\frac{\nu %%_{1}}{2\pi }}\exp (-\nu _{3}/2)
%%\end{equation}
%where $A \simeq 0.395$, $\nu _{1}=0.83 \nu$, $\nu _{2}=0.12 \nu$, $\nu _{3}=0.73 \nu$, $\alpha_2=0.3$ and $\beta_2=0.12$.  
%%where $A \simeq 0.395$, $\nu _{1}=0.83 \nu$, $\nu _{2}=0.12 \nu$, $\nu _{3}=0.8 \nu$, $\alpha_2=0.355$ and %%$\beta_2=0.028$.  

In the case of the barrier with non-zero cosmological constant, Eq. (\ref{eq:maa}), a good approximation to the multiplicity function is given by:
\begin{equation}
\nu f(\nu ) \simeq A _5 \left( 1+\frac{0.1218}{\left( a\nu \right) ^{0.585}}
+\frac{0.0079}{\left( a\nu \right) ^{0.4}}
\right) \sqrt{\frac{a\nu }{2\pi }}\exp{\{-0.4019 a \nu \left[ 1+\frac{0.5526}{\left( a\nu \right) ^{0.585}}
+\frac{0.02}{\left( a\nu \right) ^{0.4}}
\right] ^{2}\}}
%\label{eq:mia}
\label{eq:mia1}
\end{equation}
where $A_5=1.75$.
As previously reported, for matter of completeness, to the previous functions, namely PS, ST, Eq. (\ref{eq:mia1}) 
we have to add J01, which satisfies the equation:
\begin{equation}
\nu f(\nu)=0.315 exp(-\mid 0.61+ln[\sigma^{-1}(M)]\mid^{3.8})
\end{equation}
In order to express the above relation as a function of $\nu$, one
substitutes $\sigma^{-1}(M)=\nu/\delta_c$ and I assume a
constant value of $\delta_c$, that of the Einstein-de Sitter Universe namely
$\delta_c=1.686$. The above formula is valid for $0.5 \leq \nu \leq  4.8$.\

\begin{figure}
\centerline{\hbox{
\hspace{4cm} 
\psfig{file=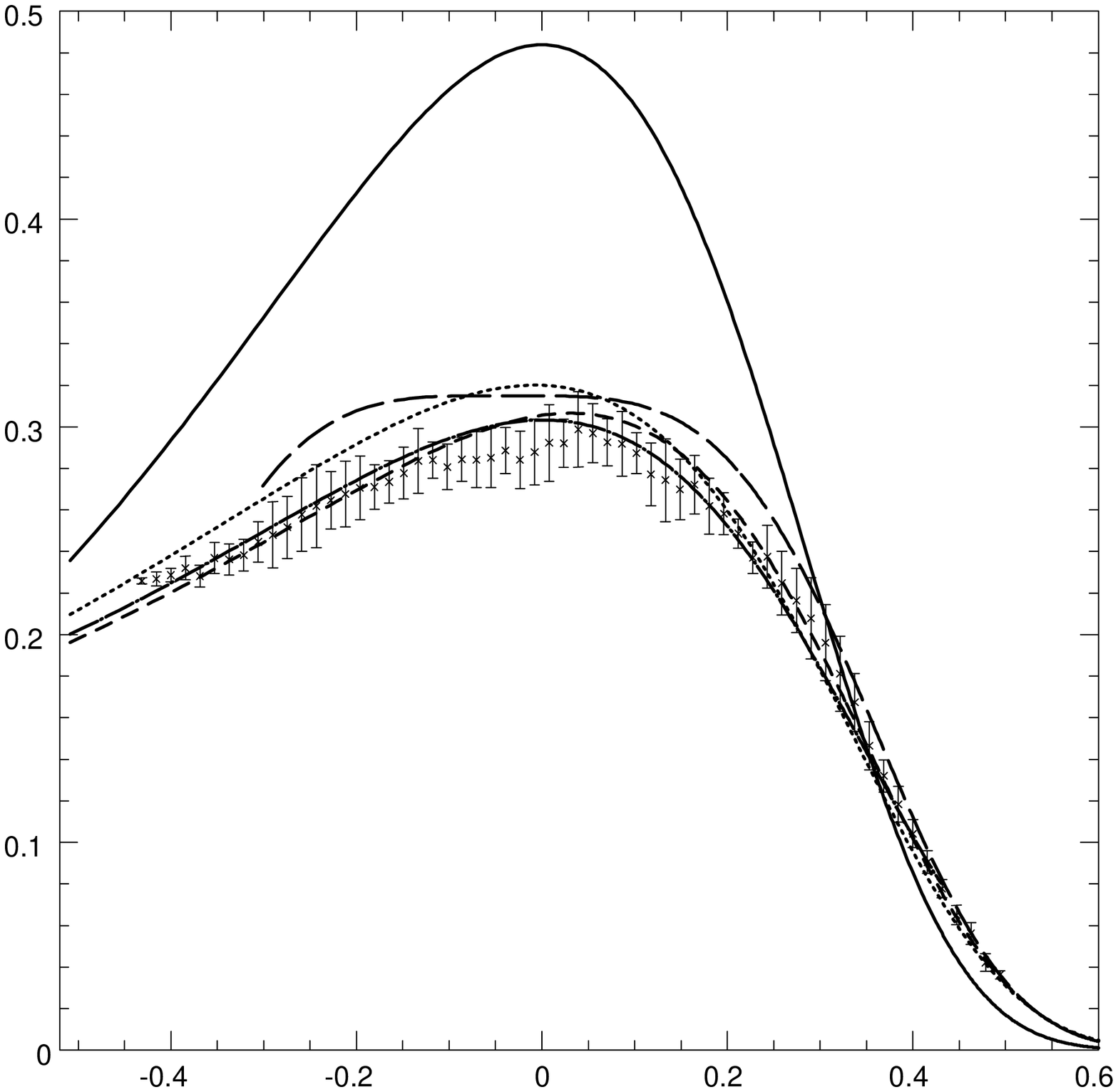,width=10cm} 1a 
\psfig{file=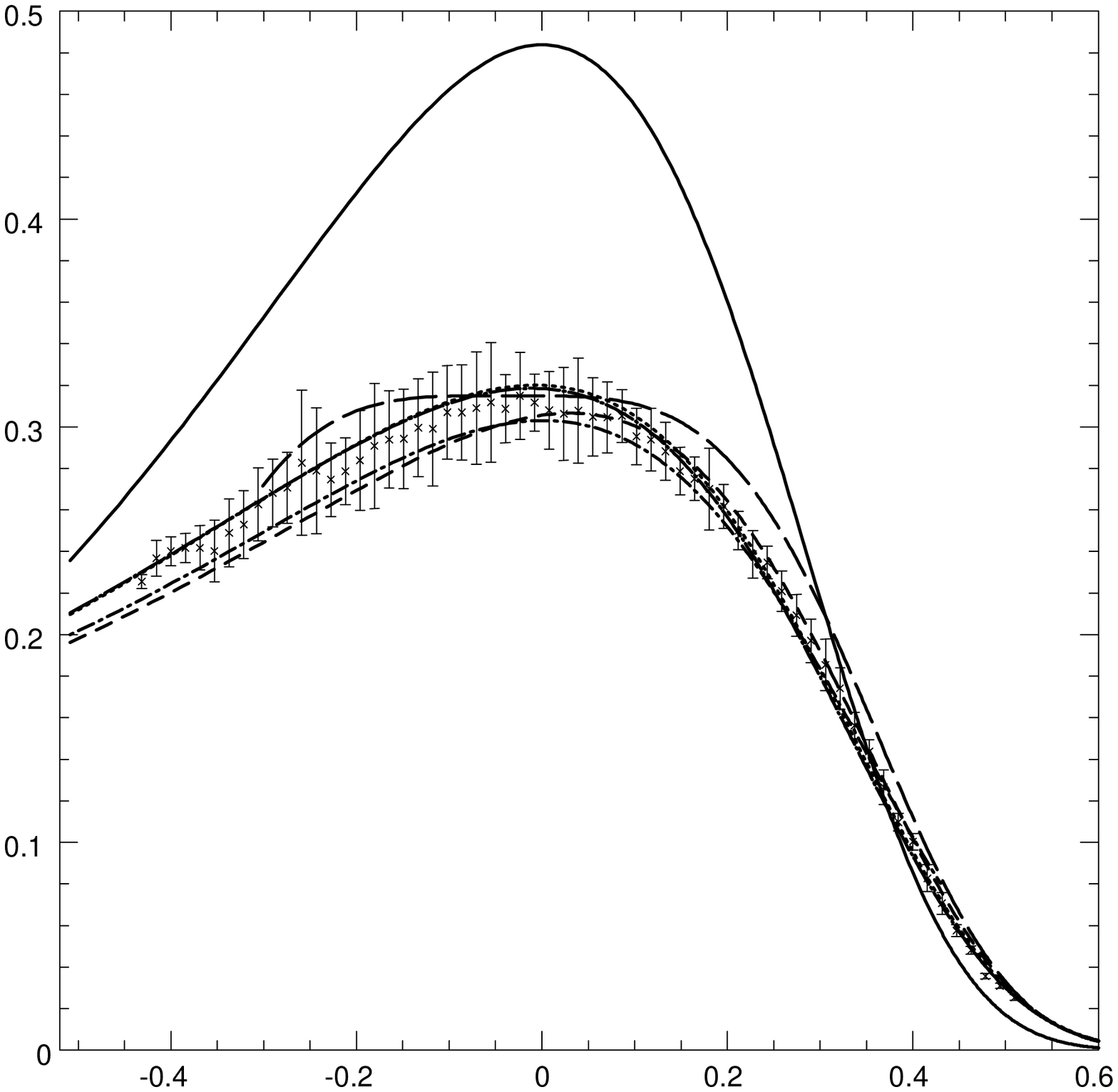,width=10cm} 1b
}}
%\caption[]{P}
%\end{figure*}
%\begin{figure}
\centerline{\hbox{
\hspace{4cm} 
\psfig{file=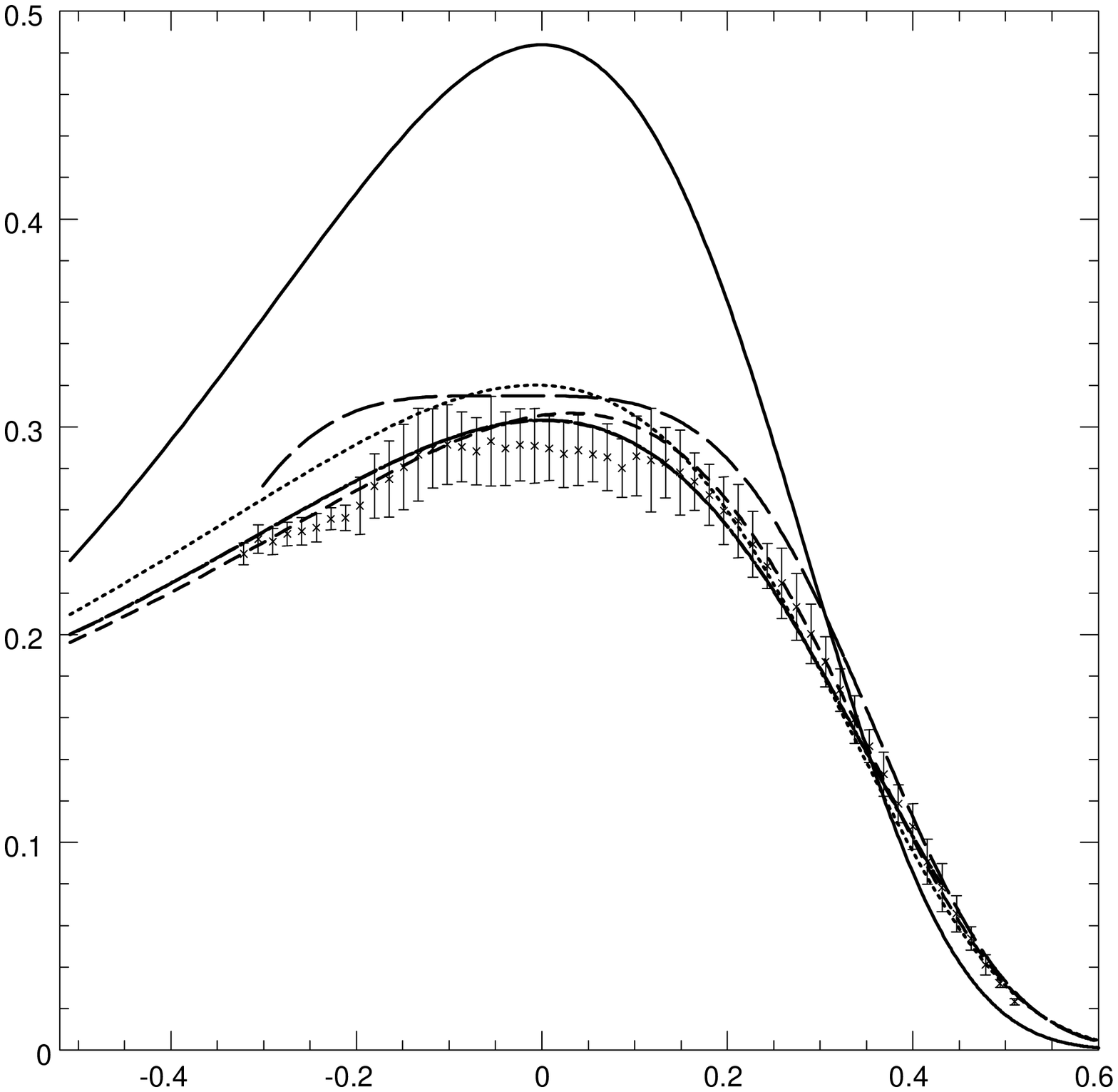,width=10cm} 1c
\psfig{file=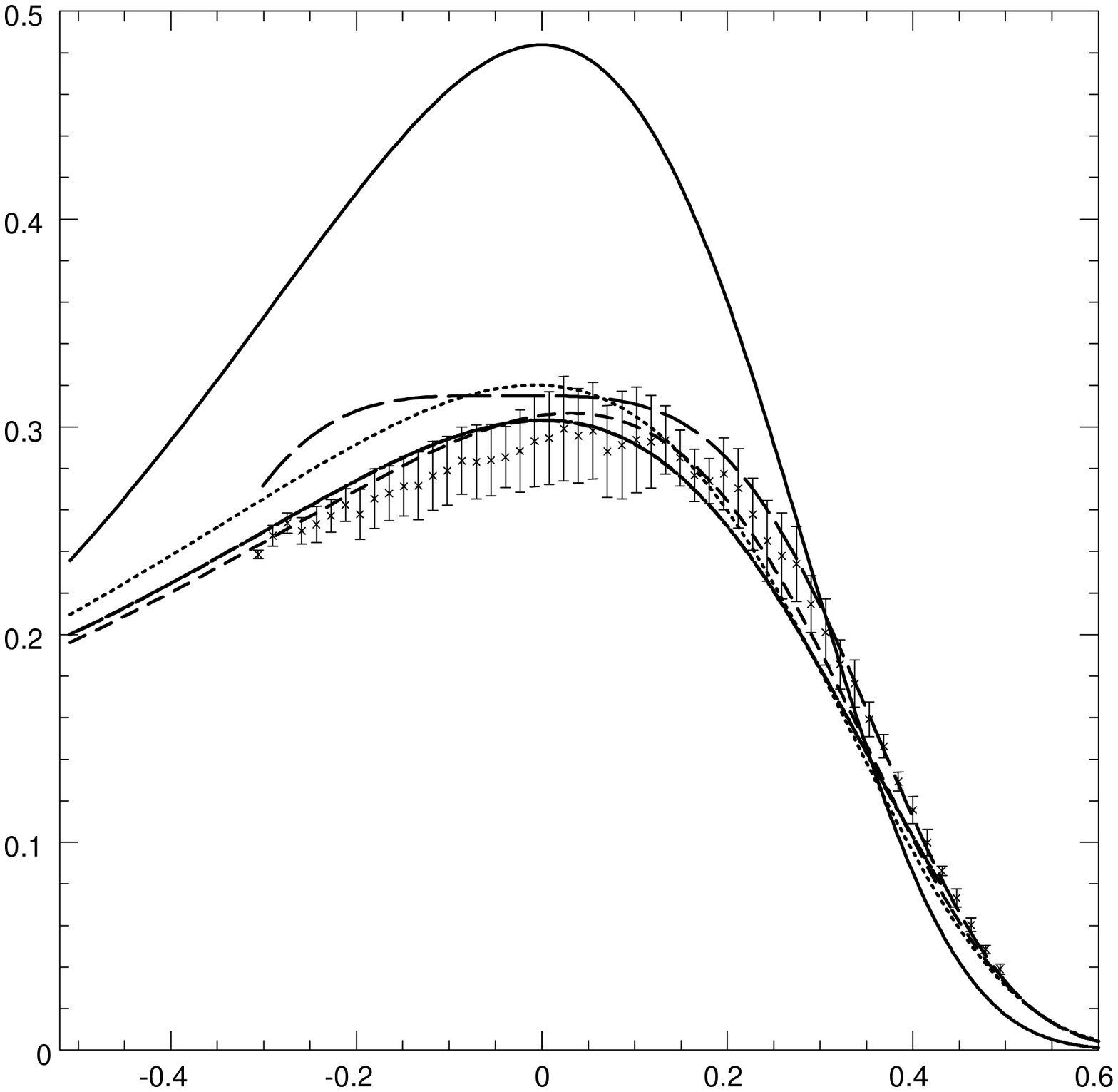,width=10cm} 1d
}}
%\caption[]{P}
\end{figure}

\begin{figure}
\centerline{\hbox{
\hspace{4cm} 
\psfig{file=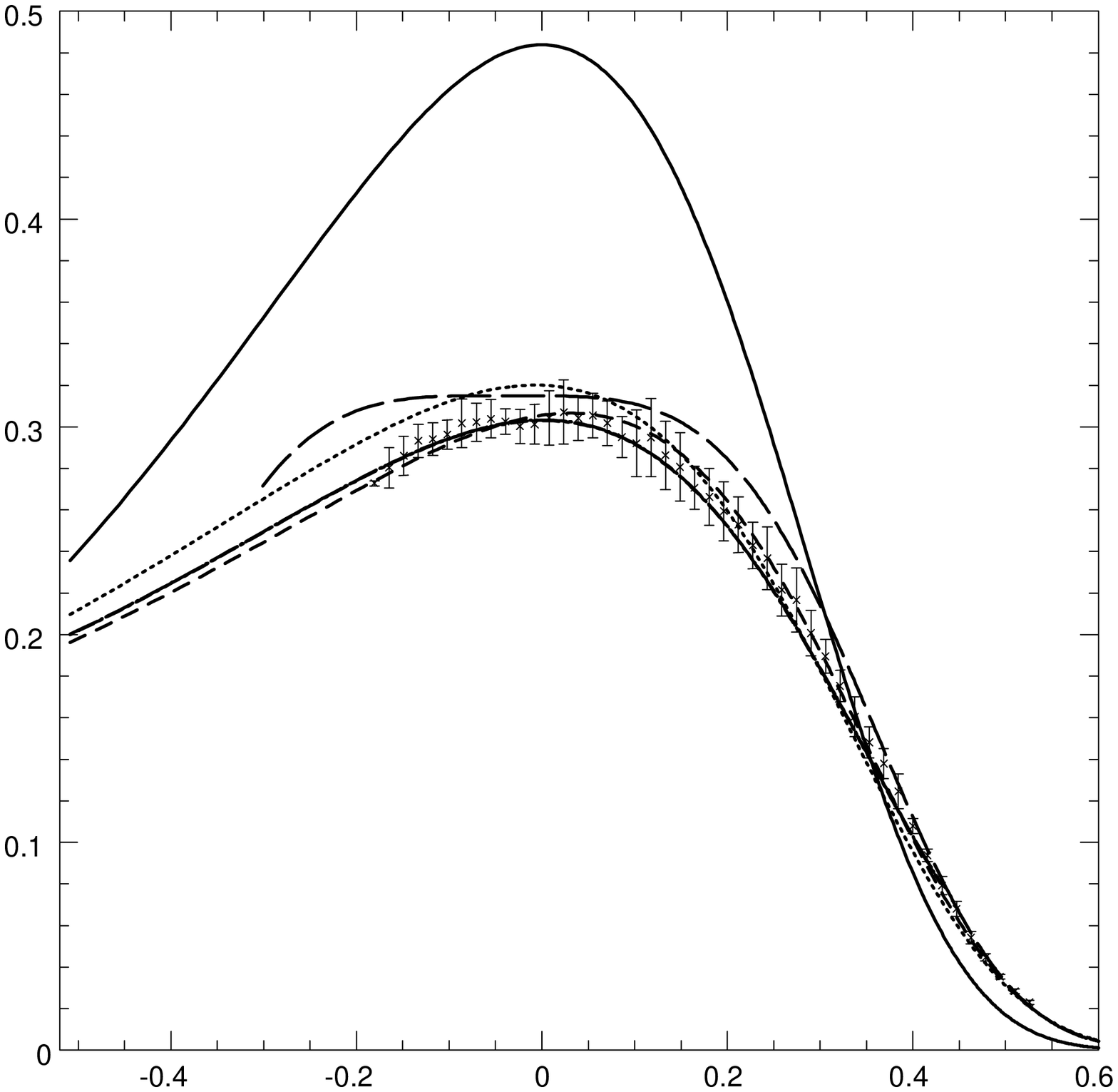,width=10cm} 1e
\psfig{file=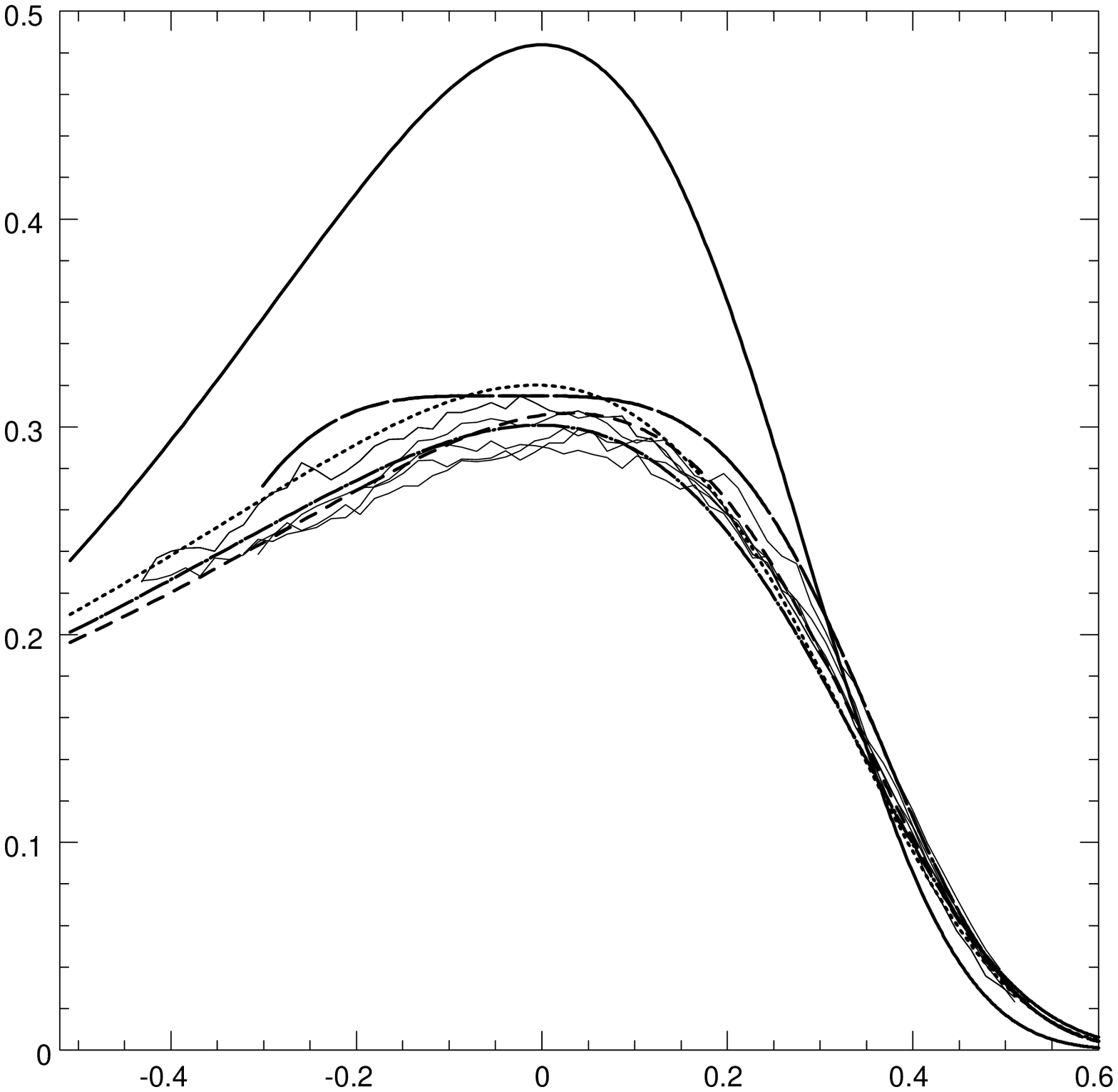,width=10cm} 1f
}}
%%\centerline{\hbox{(3)
%%\psfig{file=ltrel1.ps,width=8cm,angle=-90}
%}}
\caption[]{
The multiplicity functions from five runs of YNY simulations ({crosses with error bars}) are shown in the panels ($a$-$e$),
except for the panel ($f$) which shows the results from all the runs
({thin lines}).
In each panel are the solid line represents PS multiplicity
function, the dotted line the ST multiplicity function, 
the long-dashed line the J01 multiplicity function, the short-dashed 
line YNY7, and the dot-dashed line the one calculated  by Del Popolo (2005).
Figure taken from Del Popolo (2005). 

%The J01 multiplicity function, originally given as
%a function of $\sigma$ \citep{j01}, is expressed here in terms of $\nu
%= \delta_c / \sigma$, assuming that $\delta_c$ is constant although it
%varies slightly in the $\Lambda$CDM universe. In the high-mass range
%($\nu>$1), the numerical multiplicity functions reside
%between the ST and J01 functions, and its maximum value at
%$\nu \sim 1$ is below those of the ST and J01 functions.
%\label{fig:mltpl}}
}
\end{figure}

YNY (Eq. 7, hereafter YNY7) proposed the following function to fit the numerical multiplicity function:
\begin{eqnarray}
%\nu f(\nu) = A [1+(B \nu / \sqrt{2})^C] \nu^D \exp[-(B \nu)^2/2]
\nu f(\nu) = A_6 [1+(B \nu / \sqrt{2})^C] \nu^D \exp[-(B \nu/\sqrt{2})^2],
\label{eq:4fit}
\end{eqnarray}
where, $A_6$ is a normalization factor to satisfy the unity constraint,
$\int_0^{\infty}f(\nu)d\nu=1$, therefore
\begin{eqnarray}
A_6=2 (B/\sqrt{2})^D\{\Gamma[D/2] + \Gamma[(C+D)/2]\}^{-1}.
\end{eqnarray}
The best-fit parameters are given
as $B$=0.893, $C$=1.39, and $D$=0.408, and from these parameters, $A_6$
is constrained so that $A_6=0.298$.  

This best-fit, function from Eq. (\ref{eq:4fit}), is shown in
Figs. 17-19 and is only valid at $0.3 \leq \nu \leq 3$.

\begin{figure}
\centerline{\hbox{
\psfig{file=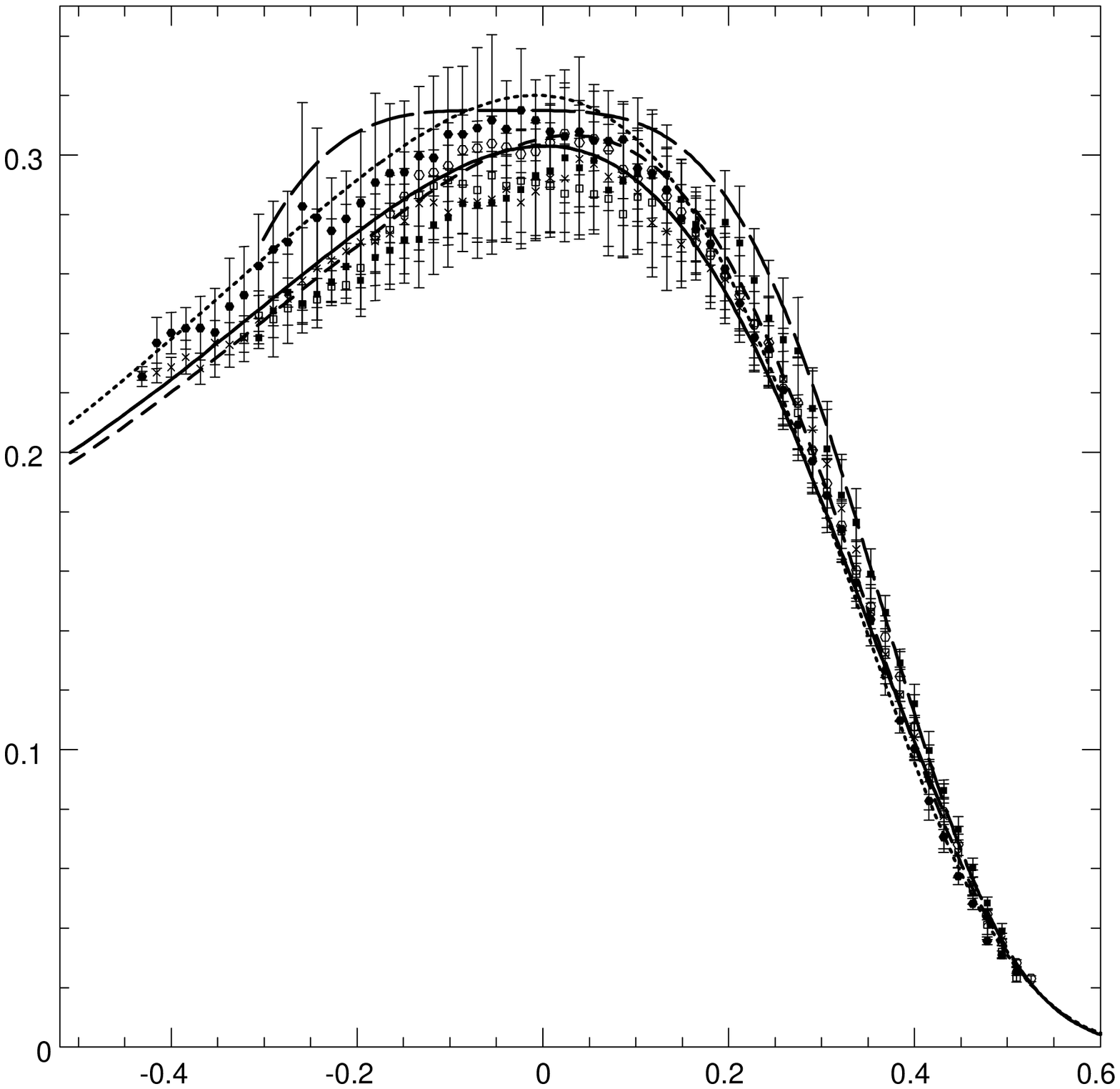,width=16cm}
}}
\caption[]{The best-fit multiplicity function. 
%equation \ref{eq:4fit} 
%({\it solid line}).  
In the plot the solid line represents the multiplicity function
obtained in Del Popolo (2005), the short-dashed 
line YNY7, the dotted line the ST multiplicity function, 
the long-dashed line the J01 multiplicity function. The errorbars with open 
circles represents the run 140 of YNY, those with filled squares the case 70b, 
those with open squares the case 70a, those with filled circles the case 35b, 
those with crosses the case 35a. Figure taken by Del Popolo (2005). 
}
\end{figure}

\begin{figure}
\centerline{\hbox{
\psfig{file=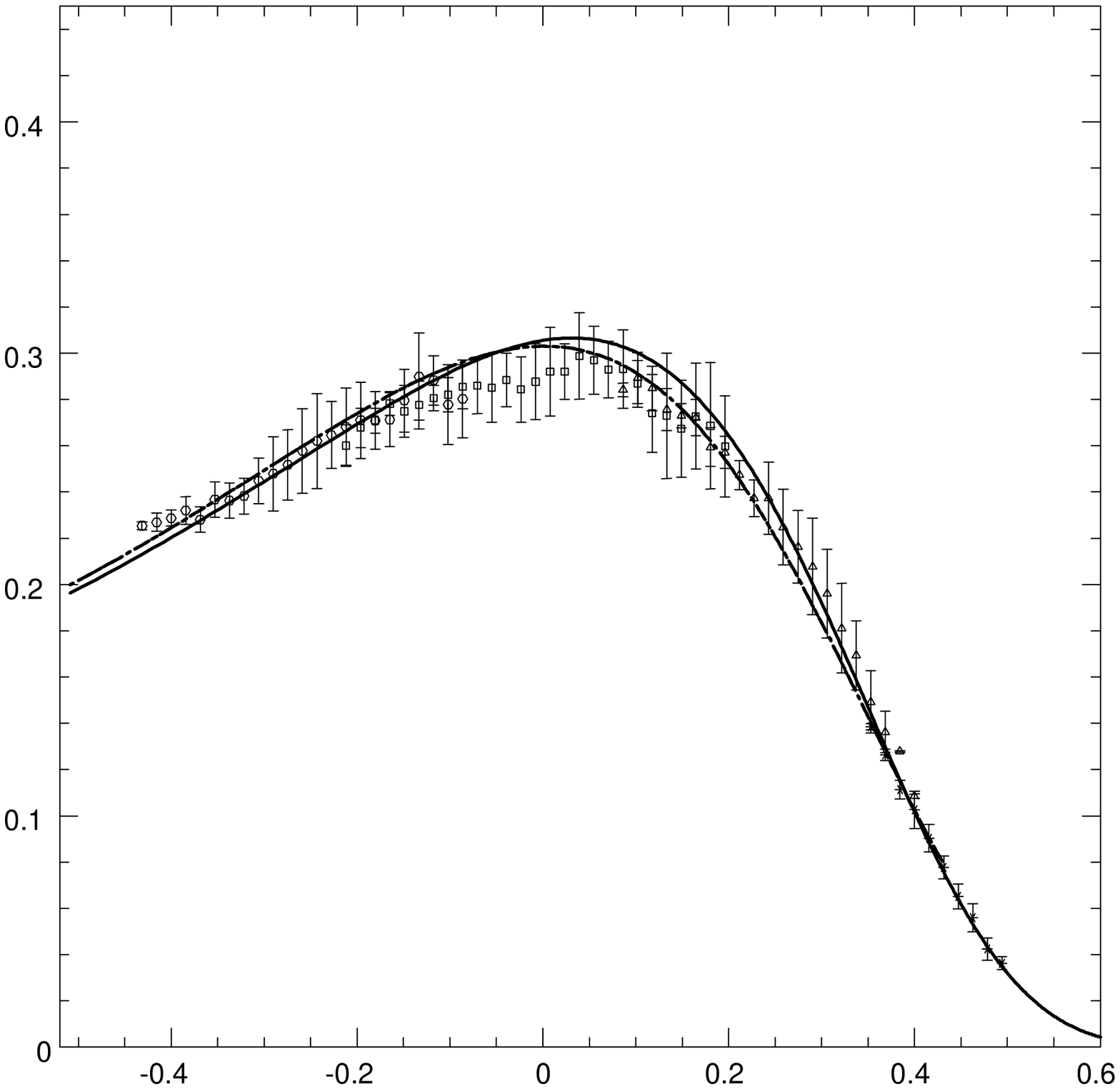,width=16cm}
}}
\caption[]{Time dependence of the multiplicity function
%. Theshows the multiplicity
%function 
from the 35a run, for four redshift ranges of 
$0 \leq z < 1$ ({open circles}), $1 \leq z < 3$ ({open squares}),
$3 \leq z < 6$ ({open triangles}), and $z \geq 6$,    ({crosses}).
Also shown are 
%the best fit function for all the runs using the ST
%functional form
%({\it dot-dashed line}) 
YNY7 (solid line)
%and  equation \ref{eq:4fit} ({\it solid line}).
and the model of Del Popolo (2005) (dot-dashed line). Figure taken by Del Popolo (2005).}
\end{figure}

%RISULTATI, ECC.

\subsubsection{\bf Results}

The analytic multiplicity functions of PS, ST, J01, YNY7, and Eq. (\ref{eq:mia1}),  are compared with the numerical simulations
of YNY.
Those simulations adopt the $\Lambda$CDM cosmological parameters of $\Omega_m=0.3$,
$\Omega_\lambda=0.7$, $h=0.7$, and $\sigma_8=1.0$, using
$512^3$ particles in common (see YNY for details).
 
The numerical multiplicity functions are shown
by crosses with errorbars in five panels of Fig. 17.
All the data from the initial redshift to the present $z=0$ is
compiled to draw the average curves (crosses) with 
error bars indicating the epoch to epoch variation.
In the panel (f), all the numerical multiplicity
functions are shown by thin lines.  

Four analytic multiplicity functions described in the previous section are also
shown in this figure, that is PS ({solid line}), ST ({dotted
line}), J01 ({long-dashed line}), YNY7 (dashed line), and the one of the Del Popolo (2005) described by Eq. (\ref{eq:mia1}) ({ dot-dashed line}).
%%%%%The best-fit functions based on the ST functional form
%%%%%are also shown by short-dashed lines. The best-fit parameters for the different runs of simlations are given in Tab. 1 of YNY.
%
%and 
%are very close to those of
%\citetalias{wht02}.  
Since the data are available only in the region at
$\nu \leq 3$, these functions could be erroneous at $\nu \geq 3$.

Note that the comparison of the above curves, except  for the PS model, with the results of N-body simulations show a very good agreement. 
However, there are some discrepancies between the YNY
multiplicity function and other model functions (except the one in Del Popolo 2005).  
First, the multiplicity function of the Del Popolo (2005), similarly to that of YNY,
in the low-$\nu$ region of $\nu \leq 1$, systematically falls below the ST and
the J01 functions. In this region the multiplicity function of the Del Popolo (2005) is very close to that of YNY.

As seen in Fig. 17, and in agreement with YNY, the numerical multiplicity functions reside between the ST and J01 multiplicity functions at 
$2 \leq \nu \leq 3$ (except for the run 35b). 
Additionally, the numerical multiplicity functions have an apparent peak at $\nu \sim 1$  instead of the plateau
that is seen in the J01 function.

On the other hand, in the high-$\nu$ region, where
$\nu$ is significantly larger
than unity, the multiplicity function of the Del Popolo (2005) 
like YNY takes values between ST
and J01 functions. These differences between numerical multiplicity functions and analytic ones, like ST, ST1 and J01, 
are within 1 $\sigma$ error bars, and they are possibly due
to the different box sizes adopted (see YNY for a discussion). To be more precise, throughout the peak range of $0.3 \leq \nu \leq 3$, the ST multiplicity function is in disagreement with the high mass resolution $N$-body simulations of YNY and that of Del Popolo (2005). As shown by YNY the ST functional form provides a good fit to them only choosing  parameter values of $a=0.664, p=0.321$, and
$A_2=0.301$. The multiplicity function obtained in the Del Popolo (2005) has a peak at $\nu \sim 1$ as in the ST function, 
and YNY numerical multiplicity function and YNY7, instead of a plateau as in the J01 function.

Moreover, the functional form proposed in YNY, namely YNY7, provides a better fit when compared
with the ST functional form but 
%we need an analytic function based on a
it is not based on theoretical background. The function obtained in this paper, similarly to YNY7 provides a better fit 
to simulations 
than the ST functional form, and at the same time has been obtained from solid physical, theoretical, arguments.
%which fits the numerical multiplicity function
%even better.
The better agreement observed between the multiplicity function of Del Popolo (2005) and YNY simulations, when compared with the ST, 
is connected to the shape of the barrier ($\delta_{\rm c}$).
As reported in Sec. 2, 
%In the spherical collapse model, this critical density does not depend on the mass of the collapsed object. 
taking account of the effects of asphericity and tidal interaction with neighbors, Del Popolo \& Gambera (1998),
%using a parametrization of the ellipsoidal collapse, 
showed that the threshold is mass dependent, and in particular that of the set of 
objects that collapse at the same time, the less massive ones must initially have been denser than the more massive, 
since the less massive ones would have had to hold themselves together against stronger tidal forces. 

The shape of the barrier given in Eq. (\ref{eq:maa}) is a direct consequence of the angular momentum acquired by the proto-structure during evolution and 
%while Eq. (\ref{eq:ma8}) introduces 
the effects of the cosmological constant.
%The good agreement between Sheth \& Tormen (1999) model and that of the present paper is due to the similitude of the 
%barriers of the two papers. 
%In both two, 

Similarly to ST, the barrier increases with $S$ (decrease with mass, M) differently from other models (see Monaco 1997a, b). 
It is interesting to note that the increase of the barrier with $S$ has
several important consequences and these models have a richer structure
than the constant barrier model. 
%In the case of non-spherical collapse
%with  
%increasing 
%barrier, a small fraction of 
%the mass in the universe remains unbound, while for the spherical dynamics, at the given time, all the mass is bound 
%up in collapsed objects. Moreover, incorporating the non-spherical collapse 
%with increasing 
%barrier in the excursion set approach results in a model in which
%fragmentation and mergers may occur (ST). If the barrier decreases with
%$S$ (Monaco 1997 a,b), this implies that all walks are guaranteed
%to cross it and so there is no fragmentation associated with this
%barrier shape.

%In other words, this 

%{\bf The barrier given in Eqs. ~(\ref{eq:ma7})-(\ref{eq:ma8}), differently from that of the spherical collapse is mass %dependent. 
%Eqs. ~(\ref{eq:ma7})-(\ref{eq:ma8}) show 
%that the threshold
%for collapse decreases with mass, or similarly it increases with $\sigma$ since this quantity is a decreasing function of %mass.
%%
%%In other words, this means that, in order to form
%%structure, peaks in more dense
%%regions must have a lower value of the threshold, $\delta_c(nu)$, with respect
%%to those of under-dense regions.}
%}
The decrease of the barrier with mass means that, in order to form structure, more massive peaks must
cross a lower threshold, $\delta_c(\nu,z)$, with respect to under-dense ones.
At the same time, since the
probability to find high peaks is larger in more dense regions, 
this means that, statistically, in order to form structure, 
peaks in more dense
regions may have a lower value of the threshold, $\delta_c(\nu,z)$, with respect
to those of under-dense regions.
This is due to
the fact that less massive objects are more influenced by external tides, and
consequently they must be more overdense to collapse by a given time.
In fact, the angular momentum acquired by a shell centred on a peak
in the CDM density distribution is anti-correlated with density: high-density
peaks acquire less angular momentum than low-density peaks
(Hoffman 1986; Ryden 1988).
A larger amount of angular momentum acquired by low-density peaks
(with respect to the high-density ones)
implies that these peaks can more easily resist gravitational collapse and consequently 
it is more difficult for them to form structure.
%
%%This is in agreement with Audit et al. (1997), Peebles (1990) and Del Popolo \& Gambera (1998), 
%%which pointed out that the gravitational collapse is slowed down by the  effect of the shear
%%rather than fastened by it (as substained by other authors).
%
Therefore, on small scales, where the shear is statistically greater,
structures need, on average, a higher  density contrast to collapse.

It is evident that the effect of a non-zero cosmological
constant adds to that 
%is that of reducing the effect 
of L. 
%\footnote{I studied in a previous paper the effect of a non-zero cosmological constant on evolution 
%of some relations like the M-T relation (Del Popolo 2002).  
%The evolution of the M-T relation is more rapid in models
%with L =0. 
The effect of a non-zero cosmological constant is that of
slightly changing the evolution of the multiplicity function with respect to
open models with the same value of $\Omega_0$. This is caused by the fact that
in a flat universe with $\Omega_{\Lambda}>0$, the density of the universe remains close
to the critical value later in time, promoting perturbation growth
at lower redshift. The evolution is more rapid for larger values (in
absolute value) of the spectral index, n.
%}. 
%and that the CDM
%cosmology, with 	0 =0.3, 	 =0.7, is in better agreement with
%the observed bend than the 	0 =0.3 OCDM model.

As previously reported, the ST model gives a better fit to simulations than PS model, but 
it has some discrepancies with simulations.
ST model was introduced at the beginning (Sheth \& Tormen 1999) as a fit to the GIF simulations and in  a subsequent paper (SMT) was recognized the importance of aspherical collapse in the  
functional form of the mass function. The effects of asphericity were taken into account by changing the functional form of the critical overdensity (barrier) by means of a simple intuitive parameterization of elliptical collapse of isolated spheroids. The model proposed in Del Popolo (2005) has several similitudes with ST and ST1 models, namely it uses the excursion 
set approach as extended by ST1 to calculate the multiplicity function, but at the same time it differs from ST and ST1 for the way the barrier was calculated and for the fact that takes account of angular momentum acquisition, and a non-zero cosmological constant, things 
that are not taken into account into ST and ST1. These differences gives rise to a multiplicity function in better agreement with simulations. 
This shows the importance of the form of the barrier.
%: in the case of the PS model it was constant, describing a spherical collapse. In ST, it was 
%mass dependent and taking account the effects of ellipsoidal collapse, with a noteworthy 
%improvement in the multiplicity function. 
%The quoted discrepancy (PS model versus N-body simulations and ST model) 
%is not surprising since the PS model, as any other analytical model, should make several 
%assumptions to get simple 
%analytical predictions. As previously reported, the main assumptions that the PS model combines %are the simple physics 
%of the spherical collapse model with the assumption that the initial fluctuations were Gaussian %and small. 
%
%The above considerations show that it is possible to get accurate predictions for a number of statistical quantities 
%associated with the formation and clustering of dark matter haloes by incorporating a non-spherical collapse in the   
%excursion set approach. 
%
The improvement of the multiplicity function of Del Popolo (2005) and ST with respect to PS is probably connected also to the fact that incorporating the non-spherical collapse with increasing barrier in the excursion set approach results in a model in which fragmentation and mergers may occur, effects important in structure formation. 

In the case of
non-spherical collapse with increasing barrier, a small fraction of
the mass in the Universe remains unbound, while for the spherical
dynamics, at the given time, all the mass is bound up in collapsed
objects. Moreover, incorporating the non-spherical collapse with
increasing barrier in the excursion set approach results in a model
in which fragmentation and mergers may occur (ST). If the barrier
decreases with S (Monaco 1997a,b), this implies that all walks are
guaranteed to cross it and so there is no fragmentation associated
with this barrier shape.

In other words, the excursion set approach with a barrier taking account effects of physics of 
structure formation gives rise to good approximations to the numerical multiplicity function: the  approximation goodness increases with a more improved form of the barrier (taking account more and more physical effects: angular momentum acquisition, non zero cosmological constant, etc). 
Another important aspect of the quoted method is its noteworthy versatility: for example it is very easy to take account of the presence of a non zero 
cosmological constant englobing it in the barrier. I recall that the YNY numerical multiplicity function assumes a non zero cosmological constant while the theoretical models (ST,ST1, J01) does not take this into account. 
%Following the same method of Del Popolo \& Gambera (1998) it is easy to englobe the 
%cosmological constant effects. We shall show this in a future paper. 

%In most cases, the numerical multiplicity functions and the best-fit
%functions to them are consistent with the ST and J01 multiplicity
%functions at $\nu \gtrsim 3$.  However, each of the numerical multiplicity
%functions reside between the ST and J01 functions at $1.5
%\lesssim \nu \lesssim 3$, and is below the ST
%function at $\nu \lesssim 1$ except for the 35b run.  The
%numerical multiplicity functions have an apparent peak at $\nu
%\sim 1$, instead of a plateau as seen in the J01 function.  

%Finally I checked the time dependence of the multiplicity function.
Fig. 19 shows the multiplicity
function from the 35a run, for four redshift ranges of 
$0 \leq z < 1$ ({open circles}),
$1 \leq z < 3$ ({open squares}),
$3 \leq z < 6$ ({open triangles}), and
$z \geq 6$,    ({crosses}).
%The dashed line represents YNY7 while the solid line Eq. (\ref{eq:mia}) of te present paper.
At high redshifts, high-$\nu$ halos in the exponential part of the
%best-fit ST function and equation \ref{eq:4fit} 
YNY7 (solid line) function and Eq. (\ref{eq:mia1}) (dot-dashed line of Del Popolo 2005)
are probed.  As redshift decreases,
the probe window moves to the lower-$\nu$ region.
Fig. 19 shows that the multiplicity function of Del Popolo (2005), Eq. (\ref{eq:mia1}), and YNY7 both gives a good fit to the numerical simulations. For small values of $\nu$, Eq. (\ref{eq:mia1}) is a slightly better fit to data, and at large values of $\nu$ the two functions decays in the same way. 

\begin{flushleft}
{\it 3.5.2  Multiplicity function evolution}
\end{flushleft}

In this section, I compare the analytic mass function of the present paper with that of ST, and with \cite{reed} (R03) simulation results at several
redshifts. In Fig. 7, I compare the mass function of the present paper with ST, and with R03 simulation results at several
redshifts. In the figure, the solid line represents the ST mass function at $z=0, 5, 8, 15$, going from right to left, respectively. The dashed line the mass function of the present paper for the same values of the redshift, the errorbars with open squares, crosses, open triangles and solid triangles
represents R03 at $z=0, 5, 8, 15$.
Fig. 7 shows that the ST function provides a good fit to R03 data, except at very high redshifts, where it significantly overpredicts the halo abundance. At all
redshifts up to z$=$10, the difference is $\simeq 10 \%$ for each of our well sampled mass bins.  However, the ST  function begins to
overpredict the number of haloes increasingly with redshift for z$>$10, up to $\simeq 50$\% by z$=$15.  The simulation mass
functions appear to be generally steeper than the ST function, especially at high redshifts. This is in agreement with the theoretical mass function calculated in the present paper which gives a better description of the R03 mass function for higher values of $z$ for which the ST mass function overpredicts the simulation results. 

In Fig. 8, I plot the mass function for all of our outputs in the $f(\sigma)- \ln(\sigma^{-1})$ plane.  Large values of $\ln\sigma^{-1}$
correspond to rare haloes of high redshift and/or high  mass, while
small values of $\ln\sigma^{-1}$ describe haloes of low mass and
redshift  combinations.  
The solid line is the ST mass function while the dashed line the one obtained in the present paper and the dotted line represents a crude multiplicative
factor to the ST function as follows, with $\delta_{\rm co}$ $=$ 1.686:
% and FOF $ll=$0.2:
\begin{equation}
%\label{sheth_tormen mod}
f(\sigma) = f(\sigma; {\rm S{\rm}T})\bigg[exp[-0.7/(\sigma
[\cosh(2\sigma)]^5)]\bigg],
\label{eq:reed}
\end{equation}
valid over the range of -1.7 $\leq
\ln\sigma^{-1} \leq$ 0.9. 
%Eq. (\ref{eq:reed}). 
The ST and the mass function of the present paper differs more in the high mass region, where the mass function of the present paper is steeper than ST
and in better agreement with numerical simulations data than ST mass function. 
The ST function fits the simulated mass function to better than 10$\%$ over the range of -1.7 $\leq
\ln\sigma^{-1} \leq$ 0.5 while it  
appears to significantly overpredict haloes for $\ln\sigma^{-1} \geq$ 0.5.  
The magnitude of the ST overprediction at
high values of ln$\sigma^{-1}$ is consistent with being a function
purely of ln$\sigma^{-1}$ rather than redshift, a natural consequence
of the fact that  the mass function is self similar in time (e.g.
\cite{ef7}; \cite{la1}; \cite{jen}).
The empirical adjustment to the ST mass function (Eq. (\ref{eq:reed})), dotted line, describes much better numerical simulations data: for -1.7 $\leq \ln\sigma^{-1} \leq$
0.5, Eq. (\ref{eq:reed}) matches R03 data to better than 10$\%$ for well-sampled
bins, while  for 0.5 $\leq \ln\sigma^{-1} \leq$ 0.9, where Poisson
errors are larger, data is matched  to roughly 20$\%$. 

YNY and \cite{warre} made a similar choice, namely they introduced an empirical mass function obtained from a fit to their simulations 
that gives a better fit to simulations than ST model.
It is important to stress that even if   
the functional forms proposed in R03, YNY and \cite{warre} provide a better fit to simulations when compared
with the ST functional form, they are   
not based on theoretical background. 
The function obtained in the present paper, similarly, for example, to R03 provides a better fit 
to simulations than the ST functional form, and at the same time has been obtained from solid physical, theoretical, arguments.
The better agreement observed between the mass function of the present paper and R03 simulations, when compared with the ST, 
is connected to the shape of the barrier ($\delta_{\rm c}$).

In other words, the excursion set approach with a barrier taking account effects of physics of 
structure formation gives rise to good approximations to the numerical multiplicity function: the  approximation goodness increases with a more improved form of the barrier (taking account more and more physical effects: angular momentum acquisition, non zero cosmological constant, etc). 
Another important aspect of the quoted method is its noteworthy versatility: for example it is very easy to take account of the presence of a non zero 
cosmological constant englobing it in the barrier. I recall that the YNY numerical multiplicity function assumes a non zero cosmological constant while the theoretical models (ST, \cite{pre}, \cite{jen}) does not take this into account. 
\begin{figure}
\centerline{\hbox{
\psfig{file=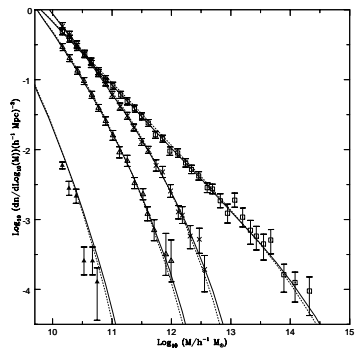,width=5cm}
}}
\caption[]{Comparison of the mass function evolution calculated in the present paper with ST mass function and R03 simulations.  
Solid curves are the Sheth \& Tormen function at z$=$0, 5, 8, \& 15 (from right to left).  
The dashed line the mass function of the present paper for the same values of the redshift, the errorbars with open squares, crosses, open triangles and solid triangles
represents R03 result at the same redshift. 
}
\end{figure}

\begin{figure}
\centerline{\hbox{
\psfig{file=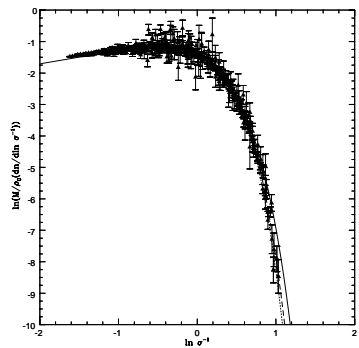,width=5cm}
}}
\caption[]{Mass function plotted in redshift independent form for all of
R03 outputs: redshifts used are 0, 1., 2., 3., 4., 5., 6.2,
7.8, 10., 12.1, 14.5. The solid line is ST prediction while the dashed and dotted line represent the result of the present paper and 
Eq. (\ref{eq:reed}), respectively.
}
\end{figure}

%\section{\Large Prospects and Conclusions}

\begin{flushleft}
{4.  PROSPECTS AND CONCLUSIONS
%Prospects and Conclusions
}
\end{flushleft}

In the previous sections, we have seen that from the seminal paper of PS to last results, large progress has been made 
in the field of the cosmological mass function. If the PS model, explained quite well the 
N-body simulations of 80' and first part of 90', its limits have been shown by more recent simulations (e.g. \cite{gov1}; YNY).
Its theoretical limits, for example the lack of the "fudge factor of 2", has been solved in some papers (e.g. \cite{bon1}). 
We have reviewed the excursion set model, the relation between peak model and the MF, one approach (PS-like approach) that takes into account the dynamics 
in the MF theory. All the previous results, give a MF in better agreement with simulations but only in more recent years several works have shown how it is possible to obtain MF functions in very good agreement with simulations (\cite{del}, \cite{del1}, ST, \cite{pre}, \cite{sh2}, \cite{jen}, YNY).

Even if the last models for the mass function gives much better results than the previous ones, it is necessary to recall which are the 
limits of the theory and the problems that must be attacked.

%From the theoretical point of view, 
The main problems are the
following:

%\begin{itemize}
%\item
---It is necessary to add more physics to the dynamical MF theory, to
describe events such as the aggregation and fragmentation of already
collapsed structures. In this way, the dynamics inside structures
could be resolved; this is necessary to describe objects such as
galaxies inside larger structures, as groups or clusters. 
The theories presented in the last subsection are somehow able to take account of some of the quoted effects, like fragmentation,
angular momentum acquisition, presence of a non zero cosmological constant. 
Changing the shape of the barrier it is possible to take into account more physical effects to the dynamical MF theory. \\
%The kinetic
%theories presented in \S 2.5.3 go in this direction.
%%%\item
---It is necessary to take in full account the geometry of collapsed
regions in Lagrangian space, going beyond the usual golden rule
described in previous sections. In the realistic case in which the stochastic
process on which the MF is based is non Gaussian, this investigation
can be performed through Monte Carlo simulations of initial fields. \\
%\item
---It is necessary to understand the role of filtering in the MF problem,
as different kinds of filters (sharp $k$-space, Gaussian) lead to
different and not equivalent formulations of the problem.\\
%\item
---It is necessary to understand in detail what is the total mass of a
structure, and to find a rigorous, well-posed and easily interpretable
definition for it, suitable to help in the interpretation of
observations.  
%\end{itemize}

Another important point to remark is connected to N-body simulations. For example, as shown by YNY 
there is a discrepancy
in the numerical multiplicity functions from various
simulation runs. There are three strategies to resolve this discrepancy.
The first is to run simulations having still higher
mass dynamic range free from the box size effect.
The second is to increase the number of realizations, because
there is a scatter from the runs using the same box size. The
third is to run simulations whose box size is smaller than that
of the present work, although it might sound contradictorily.
From simulations with smaller box size, one obtain the information
on the conditional multiplicity function which coincides
with the unconditional multiplicity function at $\nu <<1$.
Comparing the unconditional multiplicity function from simulations
with a large box size and the conditional multiplicity
function from those of a small box size will offer not only
the clues to resolve the above mentioned discrepancies, but
also insights into the mechanism how the PS ansatz works to
reproduce the numerical multiplicity function.

\begin{flushleft}
{ACKNOWLEWDGMENTS}
\end{flushleft}

The author thanks the referee M. V. Sazhin for his helpful comments.

\end{document}